\documentclass{IEEEtran}
\usepackage[noadjust]{cite}
\usepackage{srcltx}
\usepackage{graphicx}
\usepackage{epstopdf}
\usepackage{psfrag}
\usepackage{epsfig}
\usepackage[tbtags,sumlimits,nointlimits,reqno]{amsmath}
\usepackage{amssymb}
\usepackage{xcolor}
\usepackage{float}
\usepackage{subfigure}
\usepackage[most]{tcolorbox}
\usepackage{soul}
\usepackage{footnote}
\usepackage{amsmath}
\usepackage{color}
\usepackage{graphicx}  
\usepackage{dcolumn}   
\usepackage{bm}        
\usepackage{verbatim}   
\usepackage{array}
\usepackage{psfrag}
\usepackage{booktabs}
\usepackage{mathtools}
\usepackage{cuted}
\usepackage[utf8]{inputenc}
\def\BibTeX{{\rm B\kern-.05em{\sc i\kern-.025em b}\kern-.08em
		T\kern-.1667em\lower.7ex\hbox{E}\kern-.125emX}}
\usepackage[framemethod=TikZ]{mdframed}
\usepackage{lipsum}
\mdfdefinestyle{MyFrame}{%
	linecolor=blue,
	outerlinewidth=2pt,
	roundcorner=20pt,
	innertopmargin=\baselineskip,
	innerbottommargin=\baselineskip,
	innerrightmargin=20pt,
	innerleftmargin=20pt,
	backgroundcolor=gray!50!white}

\begin{document}

\title{Space-Time Modulation:\\Principles and Applications}
\author{Sajjad Taravati and Ahmed A. Kishk, \IEEEmembership{Fellow, IEEE}
		\thanks{Manuscript accepted for publication in the IEEE Microwave Magazine October 19, 2018.}
	\thanks{Sajjad Taravati (sajjad.taravati@utoronto.ca) is with the Edward S. Rogers Sr. Department of Electrical and Computer Engineering, University of Toronto, Toronto, Ontario M5S 3H7, Canada.}
	\thanks{Ahmed A. Kishk is with the Department of Electrical and Computer Engineering, Concordia University, Montr\'{e}al, QC H3G 2W1, Canada. 
		(e-mail:sajjad.taravati@concordia.ca)}
	}

\markboth{Accepted and to appear in the IEEE Microwave Magazine}%
{*** \MakeLowercase{\textit{et al.}}: Bare Demo of IEEEtran.cls for IEEE Journals}

\maketitle

\section{Introduction}

The ever-increasing demand for high data rate wireless systems has led to a crowded electromagnetic spectrum. This demand has successively spurred the development of versatile microwave and millimeter-wave integrated components possessing high selectivity, multi-functionalities and enhanced efficiency. These components require a class of nonreciprocal structures endowed with extra functionalities, e.g., frequency generation, wave amplification and full-duplex communication. Space-time (ST) modulation has been shown to be a perfect candidate for high data transmission given its extraordinary capability for electromagnetic wave engineering. Space-time-modulated (STM) media are dynamic and directional electromagnetic structures whose constitutive parameters vary in both space and time. Recently, STM media have received substantial attention in the scientific and engineering communities. The unique and exotic properties of STM media have led to the development of original physics concepts, and novel devices in acoustics, microwave, terahertz and optic areas. STM media were initially studied in the context of traveling wave parametric amplifiers in the 50s and 60s~\cite{Cullen_NAT_1958,Tien_JAP_1958,cullen1960theory,Oliner_PIEEE_1963,Peng1969,Chu1969}. In that era, magnet-based nonreciprocity was the dominant approach to the realization of isolators, circulators and other nonreciprocal systems. However, magnet-based nonreciprocity is plagued with bulkiness, nonintegrability, heaviness and incompatibility with high frequency techniques. 

Recently, ST modulation has experienced a surge in scientific attention thanks to its extraordinary and unique nonreciprocity. It eliminates issues of conventional nonreciprocity techniques, such as bulkiness, heaviness and incompatibility with integrated circuit technology associated with magnet-based nonreciprocity, power restrictions of nonlinear-based nonreciprocity, and frequency limitations and low power handling of transistor-based nonreciprocity. It provides asymmetric interband photonic transitions~\cite{Fan_PRB_1999,dong2008inducing,Fan_NPH_2009,Fan_PRL_109_2012,Taravati_PRB_2017,Taravati_PhNt_Equi_2018,Taravati_PRApplied_2018,Taravati_Kishk_PRB_2018}, subluminal and superluminal phase velocities, and asymmetric dispersion diagrams~\cite{Oliner_PIEEE_1963,Taravati_PRB_2017,Taravati_PRApplied_2018,Taravati_Kishk_PRB_2018}, and holds potential for energy accumulation~\cite{lurie2017energy}. Various enhanced-efficiency magnet-free microwave and optical components have been recently realized by taking advantage of the unique properties of ST modulation, including isolators~\cite{bhandare2005novel,Fan_NPH_2009,lira2012electrically,Taravati_PRB_2017,Taravati_PRB_SB_2017,serranomagnetic}, circulators~\cite{qin2014nonreciprocal,estep2016magnetless,reiskarimian2017cmos,reiskarimian2018integrated}, pure frequency mixer~\cite{Taravati_PRB_Mixer_2018} metasurfaces~\cite{Shalaev_OME_2015,hadad2015space,shi2016dynamic,salary2018electrically}, one-way beam splitters~\cite{Taravati_PhNt_Equi_2018,Taravati_Kishk_PRB_2018}, nonreciprocal antennas~\cite{Taravati_APS_2015,Alu_PNAS_2016,ramaccia2018non}, and advanced wave engineering~\cite{Taravati_Kishk_TAP_2018}.

\begin{figure*}
	\begin{center}
			\includegraphics[width=1.8\columnwidth]{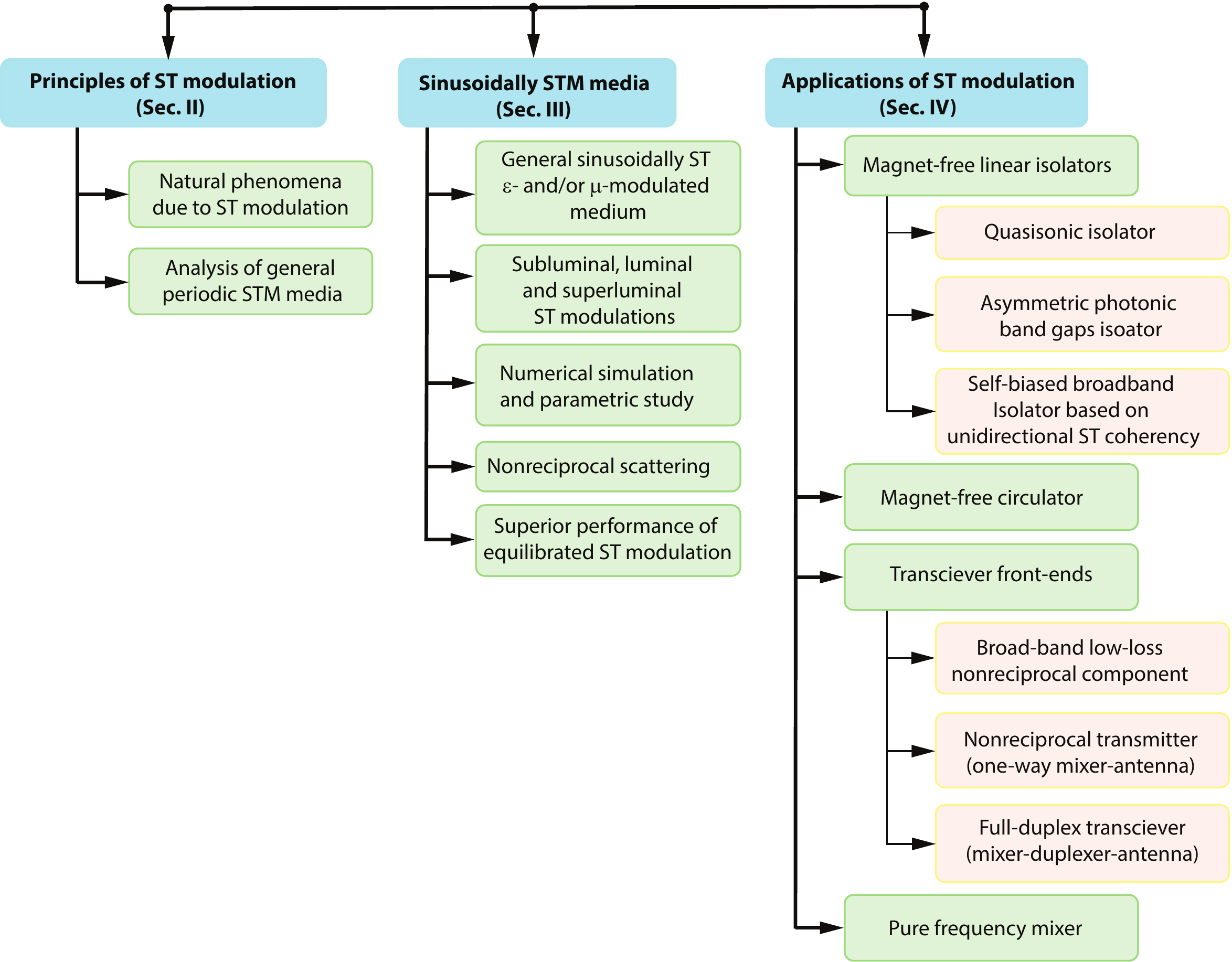} 
		\caption{Organization of the paper.} 
		\label{Fig:org}
	\end{center}
\end{figure*}

This paper overviews the principles, theoretical analysis and numerical simulation of STM media and their applications in communication systems. Fig.~\ref{Fig:org} presents an overview of the organization of the paper. We first present a global perspective on the principles of ST modulation, naturally and artificially created STM media and their applications in communication systems. Next, the interaction of artificial STM media with electromagnetic waves is analyzed, and numerical simulation of electromagnetic wave propagation and scattering from these media is provided. Next, the dispersion diagram of the STM medium is studied for the characterization of subluminal, luminal and superluminal ST modulations. In addition, we investigate the effect of length, modulation velocity, modulation frequency, and equilibrium on wave scattering from the STM medium. The extraordinary interaction of the STM medium with electromagnetic waves leads to unidirectional frequency generation and amplification, which are leveraged for the realization of various enhanced-efficiency broadband microwave components that are integrable with circuit technology, lightweight, tunable, and low cost. These applications include, but are not limited to, magnet-free isolators, magnet-free circulators, transceiver front-ends, and pure frequency mixers.

\section{Principles of ST Modulation}
\subsection{Natural Phenomena Due to ST Modulation}
\begin{figure*}
	\begin{center} 
		\includegraphics[width=1.4\columnwidth]{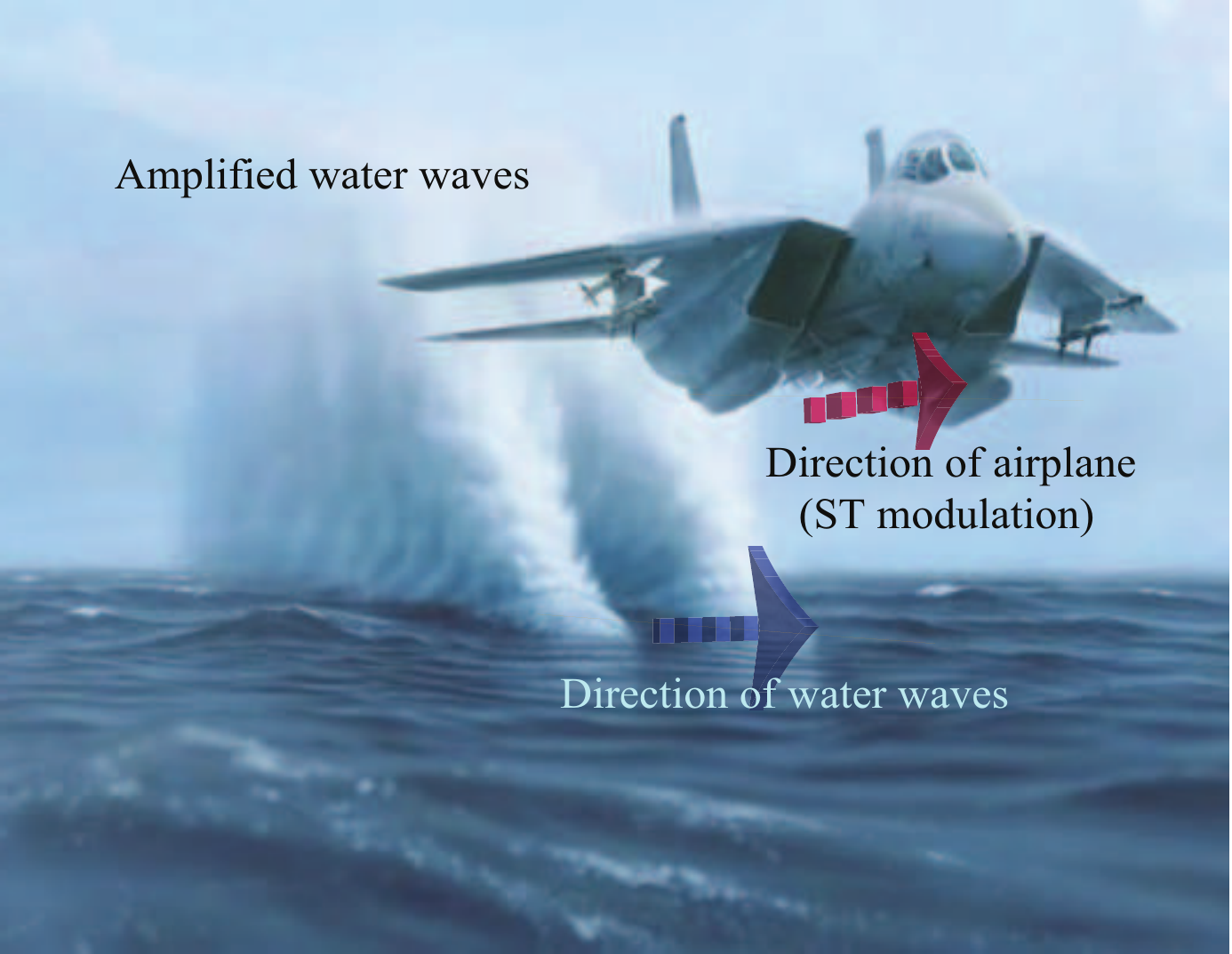} 
		\caption{An airplane flying on top of the ocean, spatiotemporally modulates the water waves, resulting in tremendous amplification of the waves~\cite{sonic_website}.} 
		\label{Fig:airplane_water}
	\end{center}
\end{figure*}

Space and time represent general forms of the existence of matter. Space possesses three dimensions:length, width and height, whereas time is represented by one dimension--from the past through the present to the future--which is unrepeatable and irreversible. ST modulation leads to various natural phenomena. For instance, consider Fig.~\ref{Fig:airplane_water}. The water waves in the ocean are unidirectional and possess very long wavelengths with small amplitude. However, flying an airplane on top of the ocean along the same direction as the water waves leads to ST modulation of the water. This, in turn, yields tremendous amplification of the water waves. The same phenomenon occurs in the ocean when a strong wind blows along the same path as the water waves and leads to huge water waves at the ocean's surface. 

\begin{figure*}
\begin{center}
\includegraphics[width=1.4\columnwidth]{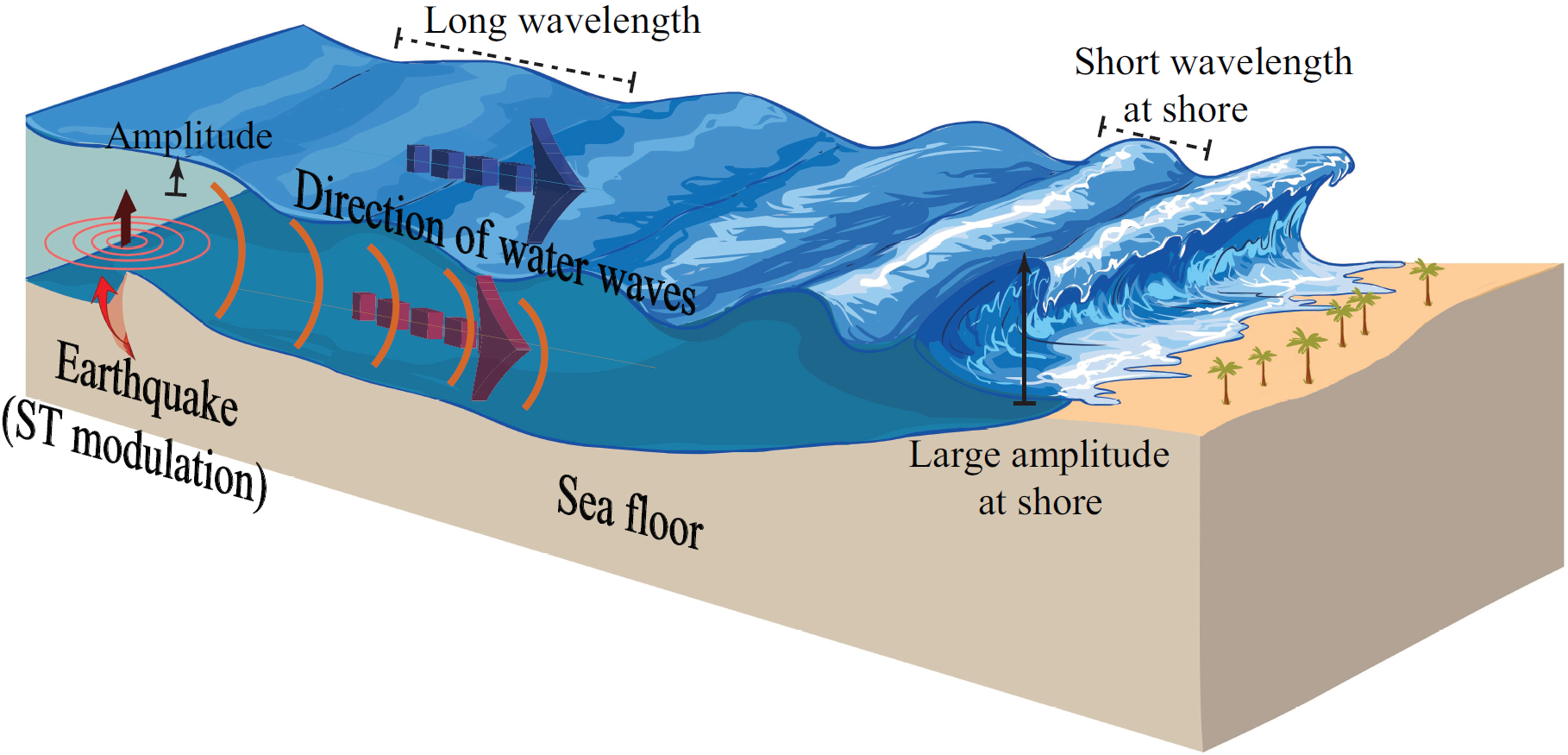}  
\caption{A tsunami as a natural phenomenon resulting from ST modulation of water waves via an underwater earthquake, so that the water waves at the shore acquire short wavelength but strong amplitude.} 
		\label{Fig:Tsunami}
\end{center}
\end{figure*}

A tsunami is another natural phenomenon caused by ST modulation of water waves (Fig.~\ref{Fig:Tsunami}). Tsunamis, or seismic sea waves, are giant waves caused by earthquakes, specifically subduction zone earthquakes, occurring under the ocean. The movement of the tectonic plates due to the earthquake also results in the movement of the sea water. A major earthquake under the ocean's floor creates shock waves which push the water up, leading to ST modulation of the water. As a consequence, at the shore, the height of the water waves increases enormously as the speed decreases, causing the tsunami. The velocity of a tsunami depends on the depth of the ocean over which it is traveling, as well as the depth of the earthquake. If the earthquake occurs more than $100$ km below the ocean floor, a tsunami will not occur because of insufficient vertical displacement of the water.

\subsection{Analysis of General Periodic STM Media}\label{sec:gen_sol}
\begin{figure*}
	\begin{center}
\includegraphics[width=1.4\columnwidth]{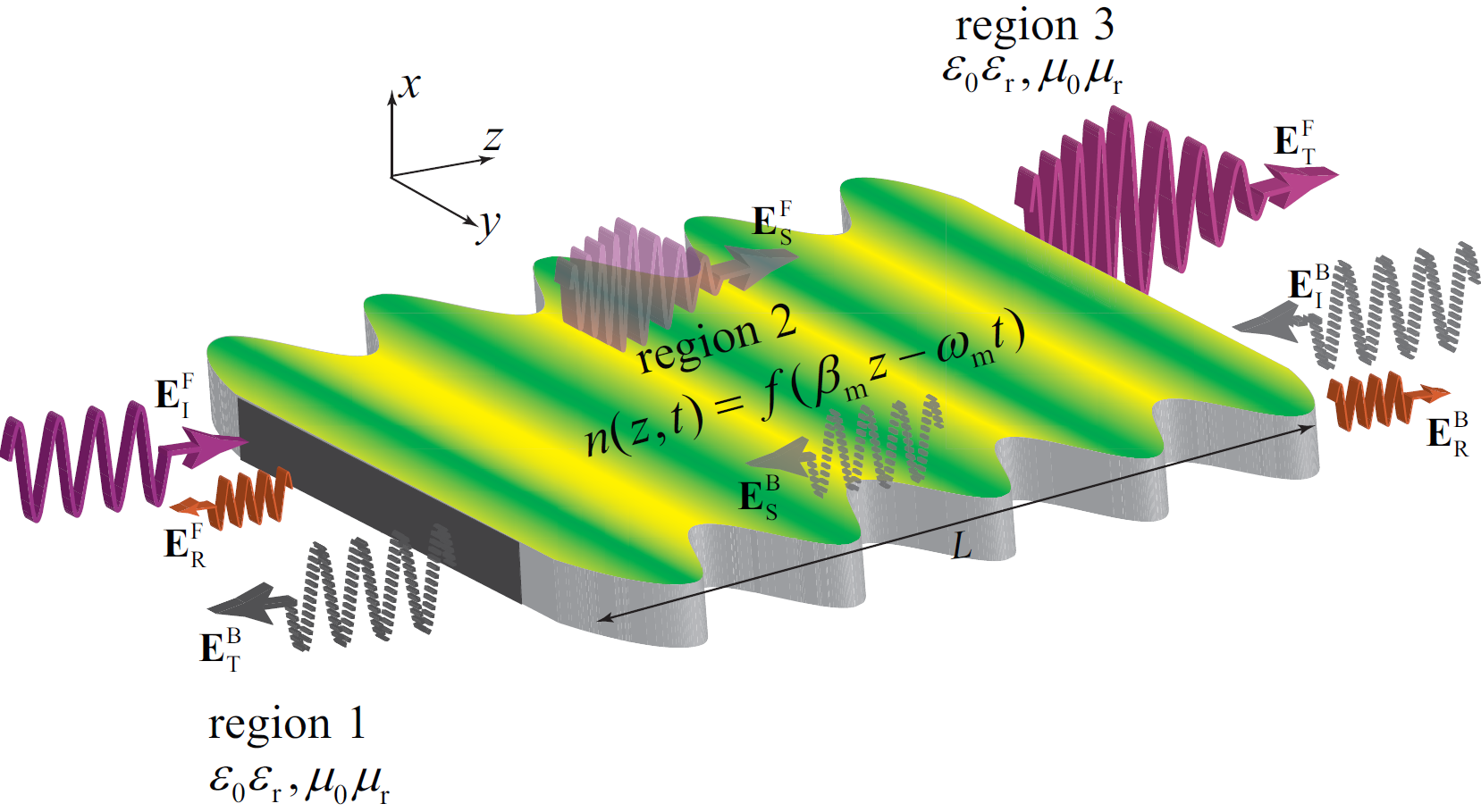}
\caption{Electromagnetic scattering from a periodically STM slab (region~2) sandwiched between two semi-infinite unmodulated media (regions~1 and~3). Due to the unidirectionality of the modulation, $\epsilon(z,t)=f_\text{per}(\beta_\text{m}z-\omega_\text{m}t)$, the system is nonreciprocal, with different temporal and spatial frequencies scattered in the two directions~\cite{Taravati_PRB_2017,Taravati_PRApplied_2018}.}
		\label{Fig:S-T_slab}
	\end{center}
\end{figure*}

In physics and engineering, ST modulation is created by spatiotemporal modulation of the constitutive parameters of media, e.g., electric permittivity~\cite{Oliner_PIEEE_1963,Fan_NPH_2009,Taravati_TAP_65_2017,Taravati_PRB_2017}, magnetic permeability~\cite{Tien_JAP_1958,tien1958traveling,Taravati_PRApplied_2018}, or electrical conductivity~\cite{reiskarimian2017cmos,salary2018time,reiskarimian2018integrated}. Such ST modulation has been realized by a time-varying modulation signal that propagates along the medium and spatiotemporally modulates variable capacitances/inductances which are added to the intrinsic capacitance/inductance of the background medium~\cite{Taravati_TAP_65_2017,Taravati_PRB_2017}. A general STM slab is depicted in Fig.~\ref{Fig:S-T_slab}~\cite{Taravati_PRB_2017,Taravati_PRApplied_2018}. Such a general ST medium is represented by a unidirectional refractive-index
\begin{equation}
n(z,t)= f(\beta_\text{m}z-\omega_\text{m}t),
\label{eqa:Gen_ref}
\end{equation}
\noindent where $f(.)$ is arbitrary periodic functions of the ST phase variable \mbox{$\xi=\beta_\text{m}z-\omega_\text{m}t$}, and $\beta_\text{m}$ and $\omega_\text{m}$ are respectively the spatial and temporal modulation frequencies. Since the refractive-index of the STM slab is unidirectional, we investigate the electromagnetic wave transmission for the wave incidence from the left and right sides of the STM slab separately, called the forward (``F'') and backward (``B'') problems, throughout this paper. The forward problem is characterized by an incident wave impinging on the slab from the left side $\mathbf{E}_\text{I}^\text{F}=\mathbf{\hat{x}} A_{0} e^{-i(k_0 z-\omega_\text{0}t)}$, and the transmitted wave from the opposite side of the slab, $\mathbf{E}_\text{T}^\text{F}$; while the backward problem is represented by the incident wave impinging on the slab from the right side $\mathbf{E}_\text{I}^\text{B}=\mathbf{\hat{x}} A_{0} e^{i(k_0 z+\omega_\text{0}t)}$, and the transmitted wave from the opposite side of the slab $\mathbf{E}_\text{T}^\text{B}$. 

The spatial modulation frequency $\beta_\text{m}$ is related to the temporal modulation frequency $\omega_\text{m}$, as 
\begin{equation}
\beta_\text{m}
=\frac{\omega_\text{m}}{v_\text{m}}=\frac{\omega_\text{m}}{\gamma v_\text{b}},
\label{eqa:vm_omegam_betam}
\end{equation}
\noindent where $v_\text{m}$ is the phase velocity of the ST modulation which may be smaller or greater than the phase velocity of the background medium $v_\text{b}=c/\sqrt{\epsilon_\text{r}\mu_\text{r}}$, with $c=1/\sqrt{\mu_0 \epsilon_\text{0}}$ being the speed of light in vacuum, and with $\epsilon_\text{r}$ and $\mu_\text{r}$ denoting the relative permittivity and permeability common to regions~1 and~3. In Eq.~\eqref{eqa:vm_omegam_betam}, $\gamma$ is the ratio between the modulation and background phase velocities which is called the \emph{ST velocity ratio}. The limit $\gamma=0$ corresponds to a purely space-modulated medium, whereas the limit $\gamma=\infty$ yields a purely time-modulated medium~\cite{Kalluri_2010,Taravati_PRB_2017}, and $\gamma=1$ corresponds to the STM medium where the modulation propagates exactly at the same velocity as a wave in the background medium. 

The general periodic STM refractive-index $n(z,t)$ in~\eqref{eqa:vm_omegam_betam} may be represented in terms of general periodic STM electric permittivity and magnetic permeability, i.e., $\epsilon(z,t)$ and $\mu(z,t)$. Due to the spatial and temporal periodicity of the permittivity and permeability, they may be expressed in terms of ST Fourier series expansion, i.e.,
\begin{subequations}\label{eqa:Fourier_per}
\begin{equation}
\epsilon (z,t) = \sum\limits_{k =  - \infty }^\infty  \tilde{\epsilon}_k {e^{-jk(\beta_\text{m}z-\omega_\text{m}t)}},
\label{eqa:Fourier_perm}
\end{equation}
\begin{equation}
\mu (z,t) = \sum\limits_{k =  - \infty }^\infty  \tilde{\mu}_k {e^{-jk(\beta_\text{m}z-\omega_\text{m}t)}},
\label{eqa:Fourier_perme}
\end{equation}
\end{subequations}
where $\tilde{\epsilon}_k$ and $\tilde{\mu}_k$ are the coefficients of the $k^\text{th}$ term, and $\tilde{\epsilon}_0$ and $\tilde{\mu}_0$ are the average permittivity and permeability of the STM slab. Assuming $\text{TM}_{yz}$ or $E_x$ polarization, the electromagnetic fields inside the slab, i.e., $\mathbf{E}_\text{S} (z,t)$ and $\mathbf{H}_\text{S}(z,t)$, may be represented in the double ST Bloch-Floquet form
\begin{subequations}\label{eqa:EH_anzatz}
	\begin{equation}\label{eqa:E_anzatz}
	\mathbf{E}_\text{S} (z,t)=\mathbf{\hat{x}} e^{-i(\beta_\text{0} z-\omega_{0}t)} \sum_{n=-\infty}^{\infty} E_{n} e^{-in(\beta_\text{m} z-\omega_\text{m}t)},
	\end{equation}
	\begin{equation}\label{eqa:H_anzatz}
	\mathbf{H}_\text{S}(z,t)=\mathbf{\hat{y}} e^{-i(\beta_\text{0} z-\omega_{0}t)} \sum_{n=-\infty}^{\infty} H_{n} e^{-in(\beta_\text{m} z-\omega_\text{m}t)},
	\end{equation}
\end{subequations}
where $\beta_{0}$ and $\omega_0$ represent, respectively, the spatial and temporal frequencies of the fundamental harmonic, $n=0$.

In Eq.~\eqref{eqa:EH_anzatz}, the given parameters are the amplitude of the incident field (fundamental harmonic) $A_0$, the spatial and temporal frequencies of the incident field ($k_0$ and $\omega_0$), the spatial and temporal modulation frequencies ($\beta_\text{m}$ and $\omega_\text{m}$), the periodic function $f(.)$, and length of the slab $L$. We achieve the dispersion relation $\beta_\text{0}(\omega_{0})$ and the electromagnetic field solutions inside ($E_{n}$ and $H_{n}$) the STM slab in Fig.~\ref{Fig:S-T_slab} by satisfying Maxwell's equations. Since the slab assumes no variation in the $x$- and $y$-directions, ${\partial\textbf{E}_\text{S}}/{\partial x}=0$ and ${\partial\textbf{E}_\text{S}}/{\partial y}=0$. Hence, $\nabla\times\textbf{E}_\text{S}=\mathbf{\hat{y}} \partial E_x(z,t)/\partial z$ and $\nabla\times\textbf{H}_\text{S}=-\mathbf{\hat{x}} \partial H_y(z,t)/\partial z$. As a result, the sourceless Maxwell equations read
\begin{subequations}
	\begin{equation}\label{eqa:Max1}
	\dfrac{\partial E_x(z,t)}{\partial z}   =- \dfrac{\partial  \left[\mu(z,t) H_y(z,t) \right]}{\partial t},
	\end{equation}
	\begin{equation}\label{eqa:Max2}
	\dfrac{\partial H_y(z,t)}{\partial z}=-\dfrac{\partial \left[\epsilon (z,t) E_x(z,t) \right]}{\partial t}.
	\end{equation}
\end{subequations}

\begin{figure}
	\begin{center}
		\includegraphics[width=1\columnwidth]{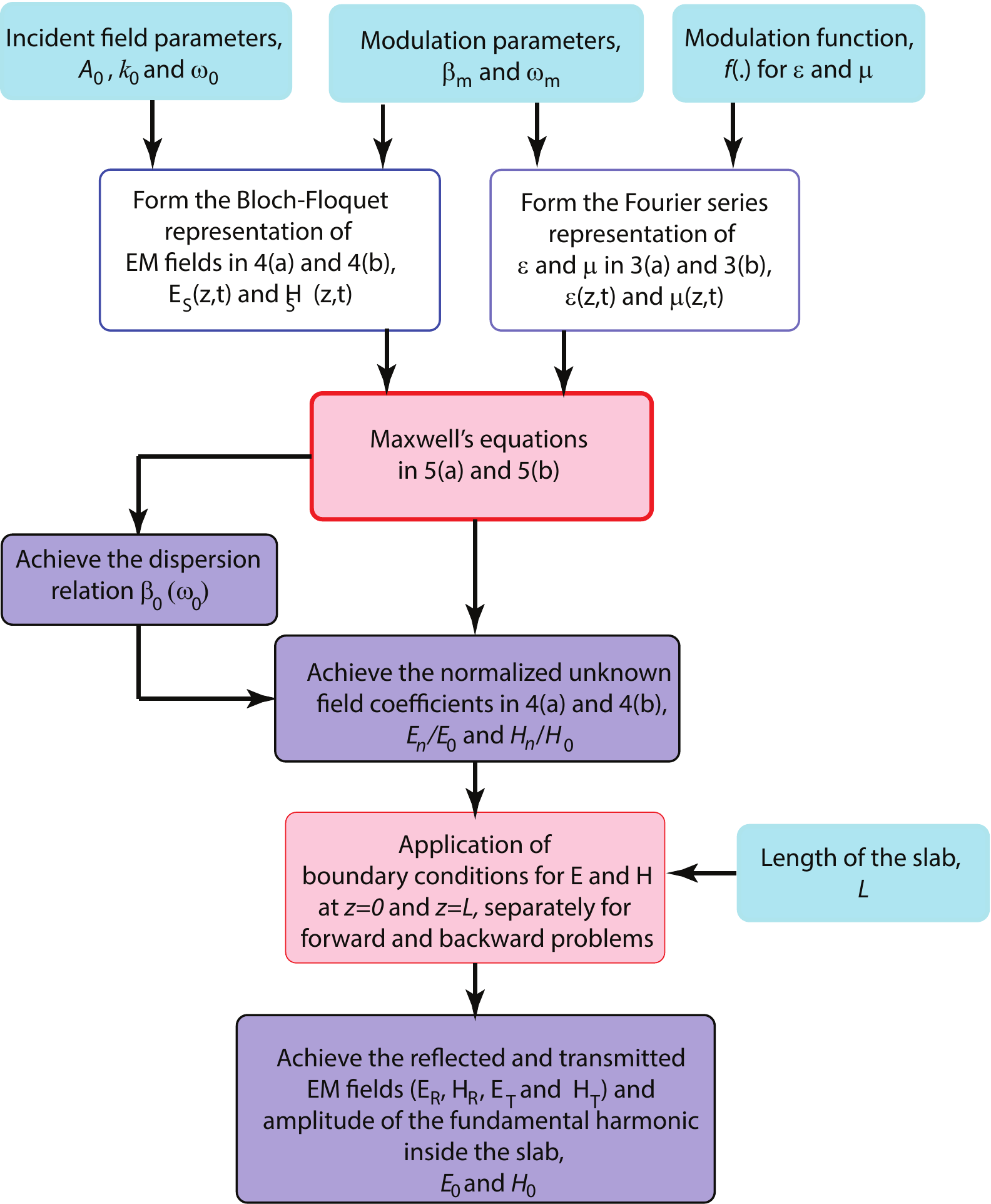} 
		\caption{Procedure for achieving the scattered electromagnetic (EM) field solutions in the STM slab in Fig.~\ref{Fig:S-T_slab}.} 
		\label{Fig:filed_sol}
	\end{center}
\end{figure}

The scattered electromagnetic fields, i.e., the reflected fields $\mathbf{E}_\text{R}^\text{F,B}$ and $\mathbf{H}_\text{R}^\text{F,B}$ and the transmitted fields $\mathbf{E}_\text{T}^\text{F,B}$ and $\mathbf{H}_\text{T}^\text{F,B}$ are achieved by applying the boundary conditions for the continuity of electromagnetic fields at $z=0$ and $z=L$~\cite{Taravati_PRB_2017,Taravati_PRApplied_2018}. Fig.~\ref{Fig:filed_sol} presents the procedure for achieving unknown field amplitudes and dispersion relations.

\section{Sinusoidally STM Media}
\subsection{General Sinusoidally ST Permittivity- and/or Permeability-Modulated Medium}
Sinusoidal ST modulation is the most common and applicable form of ST modulation. Such a sinusoidal and unidirectional ST modulation is achieved using variable capacitances and variable inductances, distributed at a subwavelength level on a transmission line, which are spatiotemporally modulated by a $+z$-propagating harmonic wave generator with temporal frequency $\omega_\text{m}$~\cite{Tien_JAP_1958,tien1958traveling,Taravati_TAP_65_2017,Taravati_PRB_2017,Taravati_PRApplied_2018}. Practical realization of variable inductors (permeability) has been extensively discussed in~\cite{Taravati_PRApplied_2018}. It has been shown that voltage-controlled variable inductors may be realized using varactors and thyristors~\cite{turgul2013study,amirpour2015varactor,yoshihara2005wide,vroubel2004integrated,james2005variable}. In addition, spatiotemporal modulation of the permeability has been experimentally demonstrated in~\cite{tien1958traveling}. Here, we consider the general case of a sinusoidal STM transmission line (Fig.~\ref{Fig:S-T_slab}), represented by the STM permittivity and permeability 
\begin{subequations}\label{eqa:sin_perm}
	\begin{equation}\label{eqa:STpermit}
	\epsilon(z,t)=\epsilon_\text{av} \left[1+\delta_{\epsilon} \cos(\beta_\text{m} z-\omega_\text{m} t) \right],
	\end{equation}
	\begin{equation}\label{eqa:STpermeab}
	\mu(z,t)=\mu_\text{av}\left[1+\delta_{\mu} \cos(\beta_\text{m} z-\omega_\text{m} t) \right],
	\end{equation}
\end{subequations}
\noindent interfaced with two semi-infinite unmodulated regions. In~\eqref{eqa:sin_perm}, $\epsilon_\text{av}$ and $\mu_\text{av}$ respectively denote the average electric permittivity and average magnetic permeability of the transmission line, and $\delta_\epsilon$ and $\delta_\mu$ respectively represent the amplitudes of the permittivity and permeability ST modulations. The sinusoidal ST modulation in~\eqref{eqa:sin_perm} may be decomposed into its ST Fourier components in~\eqref{eqa:Fourier_per}, i.e., $\tilde{\epsilon}_{-1}=\tilde{\epsilon}_{+1}= \epsilon_\text{av} \delta_{\epsilon}/2$,$\quad\tilde{\mu}_{-1}=\tilde{\mu}_{+1}=  \mu_\text{av}\delta_{\mu} /2$, $\tilde{\epsilon}_0= \epsilon_\text{av}$ and $\tilde{\mu}_0=\mu_\text{av}$.
\subsection{Subluminal, Luminal and Superluminal ST Modulations}
\begin{figure*}
	\begin{center}
       \subfigure[]{\label{Fig:subson}
			\includegraphics[width=0.64\columnwidth]{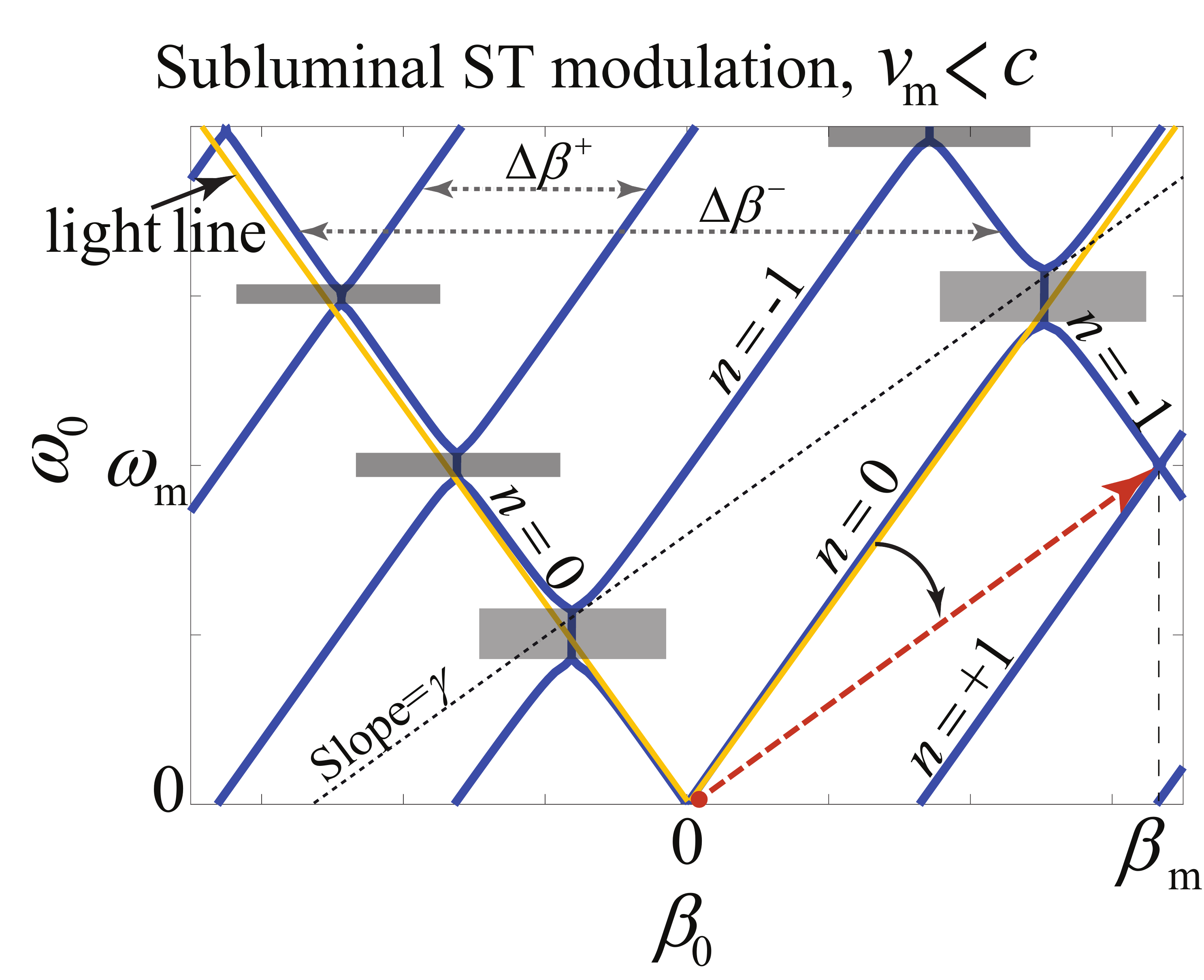}	}
		\subfigure[]{\label{Fig:son}
			\includegraphics[width=0.64\columnwidth]{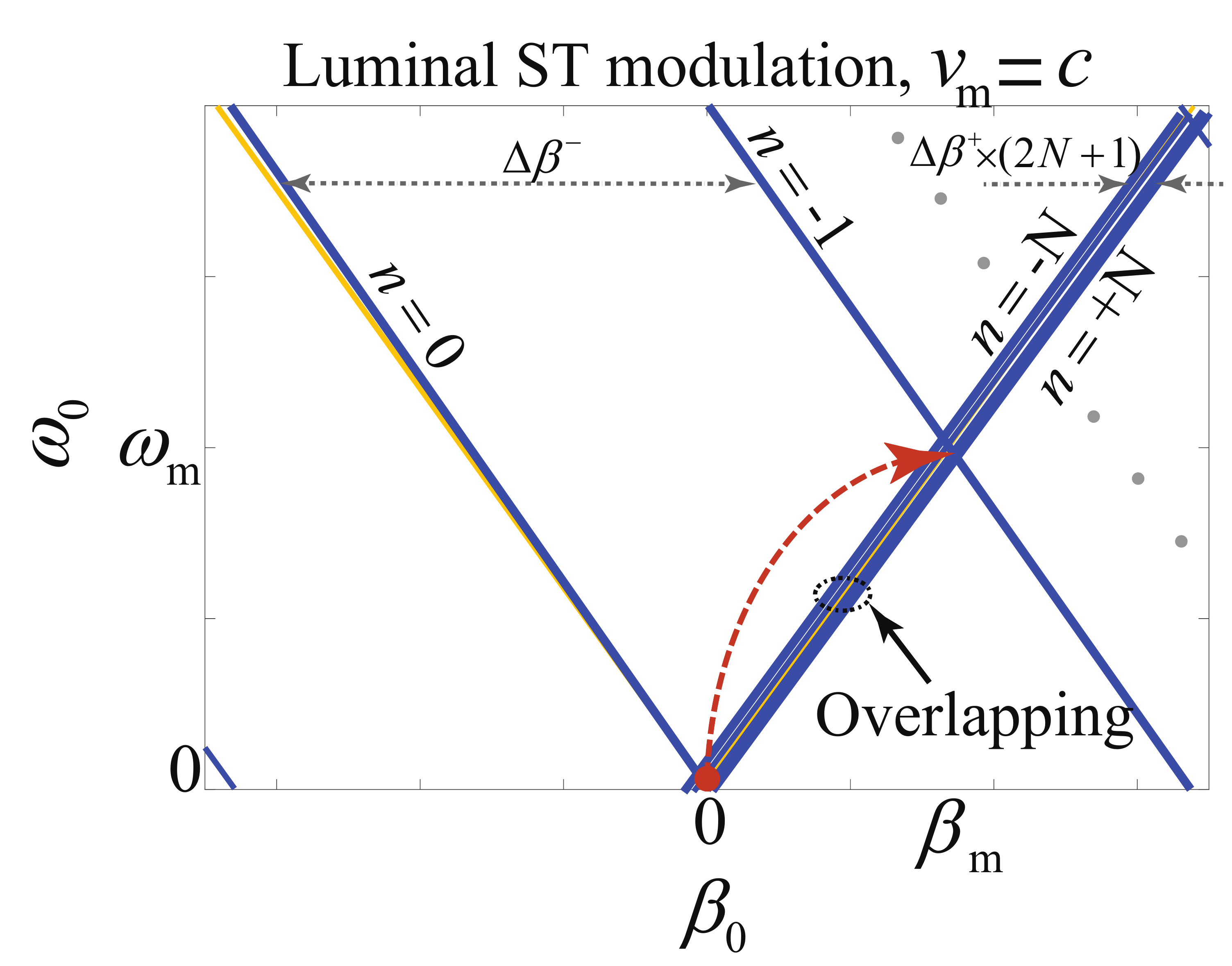}		}	
		\subfigure[]{\label{Fig:superson}
			\includegraphics[width=0.64\columnwidth]{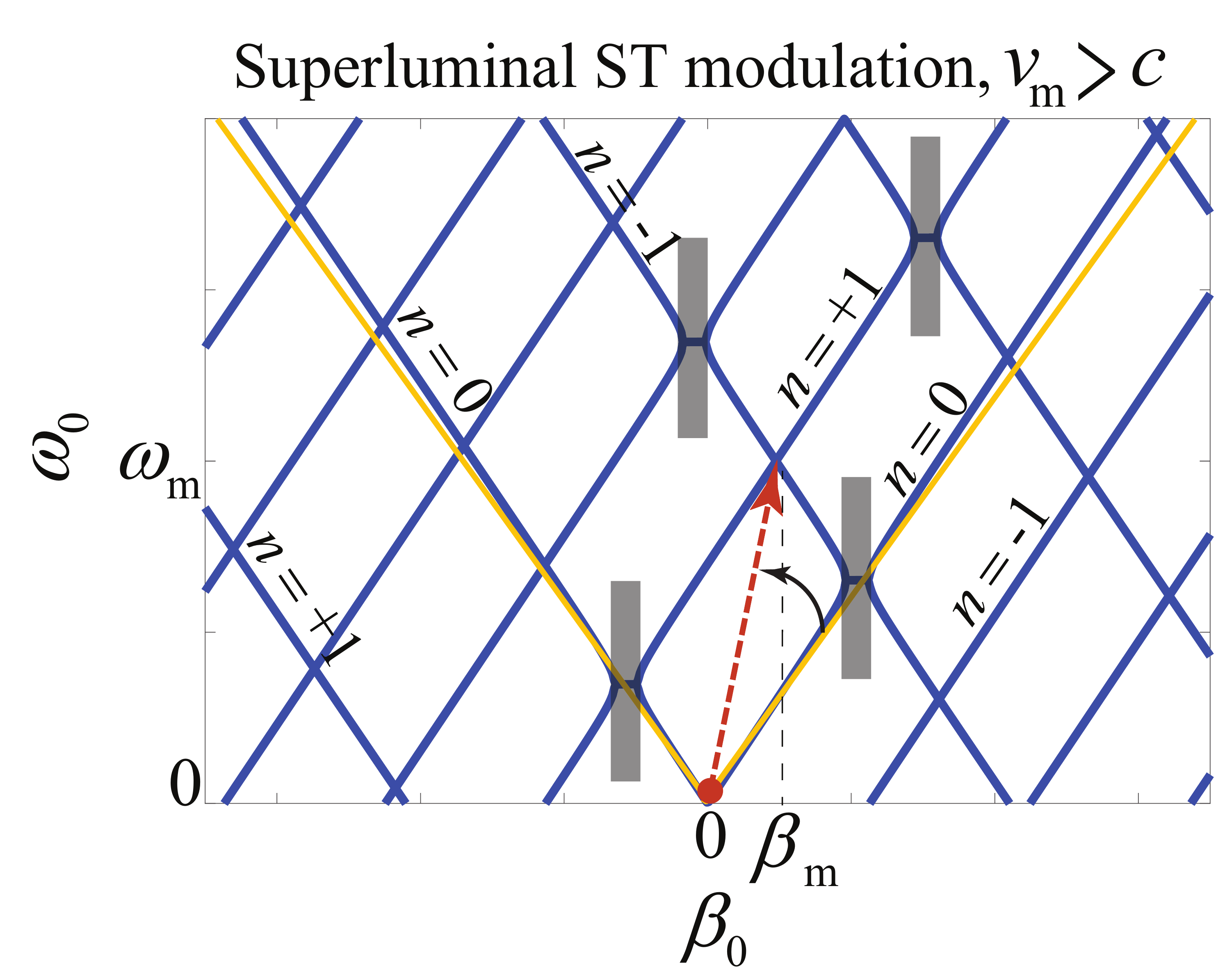}   }  
			\subfigure[]{\label{Fig:airplane_sup}
				\includegraphics[width=1.95\columnwidth]{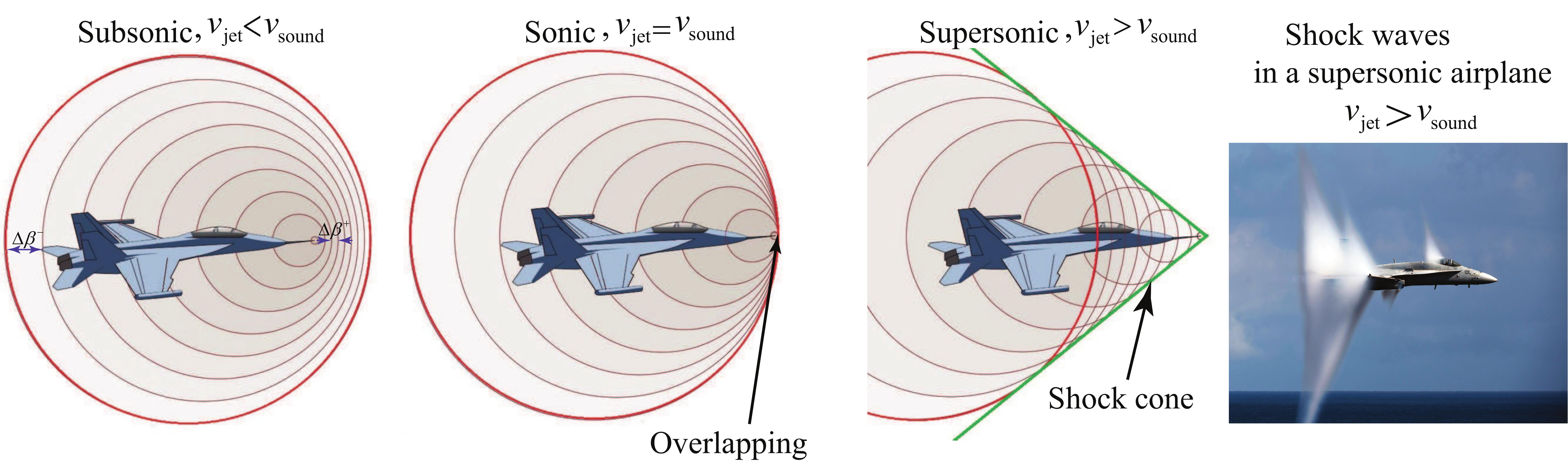}   } 
		\caption{Analogy between a sonic airplane and an STM medium. (a)-(c)~Normal incidence dispersion diagrams for the sinusoidally STM (unbounded) slab medium. (a)~Subluminal ST modulation with $\delta_{\epsilon}=0.4$, $\delta_{\mu}=0$ and $\gamma=0.5$. (b)~Same as (a) except for the vanishingly small permittivity modulation amplitude $\delta_{\epsilon}=\delta_{\mu}\rightarrow 0$ and the subsonic ST velocity ratio $\gamma\rightarrow 1$. (c) Superluminal ST modulation with $\delta_{\epsilon}=0.4$, $\delta_{\mu}=0$ and superluminal ST velocity ratio $\gamma=3.5$. (d)~Sonic airplane travels at three different speeds, i.e., subsonic, sonic and supersonic~\cite{sonic_website}. 
		 }
		\label{Fig:sonic}
	\end{center}
\end{figure*}

The wave propagation phenomenology within the STM slab may also be studied by virtue of the dispersion diagrams of the corresponding unbounded medium. The dispersion relation is represented by $\beta_0(\omega_0)$ which is achieved following the procedure given in Fig.~\ref{Fig:filed_sol}, that is, by injecting the Bloch-Floquet decomposed electromagnetic fields into Maxwell's equations and taking the corresponding space and time derivatives~\cite{Oliner_PIEEE_1963,Taravati_TAP_65_2017,Taravati_PRApplied_2018}. In the limiting case of a vanishingly small modulation amplitude, $\delta_\epsilon=\delta_\mu \rightarrow 0$, the dispersion relation for the sinusoidal ST modulation will be simplified to the closed-form dispersion relation~\cite{Taravati_PRB_2017,Taravati_PRApplied_2018}
\begin{equation}
	\frac{\beta_0^\pm }{\beta_\text{m}}
	= \gamma \frac{\omega_0}{\omega_\text{m}} +n \left( \gamma \mp  1\right).
	\label{eqa:KKKK3}
\end{equation}
where $\beta_0^+$ and $\beta_0^-$ represent the spatial frequency of the fundamental space-time harmonic (STH) ($n=0$) for forward and backward, respectively. The ST modulation may be classified in three different categories, i.e., subluminal ($v_\text{m}<v_\text{b}$ or $0<\gamma <1$), luminal ($v_\text{m}=v_\text{b}$ or $\gamma =1$) and superluminal ($v_\text{m}>v_\text{b}$ or $\gamma >1$) ST modulations, where $v_\text{b}=c$ if the background medium is air. Figs.~\ref{Fig:subson},~\ref{Fig:son} and~\ref{Fig:superson} plot the analytical dispersion diagrams of the STM medium of the aforementioned three ST modulations. The dispersion diagram of periodic STM media is constituted from an infinite periodic set of $\beta_0^\pm/\beta_\text{m}-\omega_0/\omega_\text{m}$ straight lines, labeled $n$~\cite{Taravati_PRB_2017,Taravati_PRApplied_2018}. Each curve represents a mode excited at the incident frequency $\omega_0$, as well as the oblique STH of another mode, excited at another frequency. For a homogeneous non-periodic unmodulated medium with $\omega_\text{m}=\beta_\text{m}=0$, $n=0$ is the only remaining curve, and for a vanishingly small pumping strength, $\delta_\epsilon=\delta_\mu\rightarrow 0$, the medium is quasi homogeneous, where most of the energy resides in the fundamental forward and backward STH $n=0$.

Fig.~\ref{Fig:subson} plots the dispersion diagram of the subluminal (space-like) ST modulation, with $\delta_{\epsilon}=0.4$, $\delta_{\mu}=0$ and $\gamma=0.5$. It may be seen from this figure that the forward and backward STHs acquire different horizontal distances, i.e., $\Delta\beta^\pm=\beta_{n+1}^\pm-\beta_n^\pm$ so that increasing $\gamma$ yields an increase in the $\Delta\beta^-$ and a decrease in the $\Delta\beta^+$. For a static medium, $v_\text{m}=0$ and $\Delta\beta^+=\Delta\beta^-$. However, as $v_\text{m}$ increases, the forward and backward harmonics experience different velocities relative to the modulation wave, i.e. $v^+=v_\text{b} - v_\text{m}$ and $v^-=v_\text{b}+ v_\text{m}$, respectively~\cite{Taravati_PRB_2017,Taravati_PRApplied_2018}. As a consequence, the ST transitions (exchange of energy and momentum between STHs) for the forward STHs are stronger than those of the backward STHs, revealing a nonreciprocal wave transmission for forward and backward wave incidences. It may be observed from Fig.~\ref{Fig:subson} that for a nonzero modulation strength, i.e., $\delta_\epsilon\neq\delta_\mu \geq 0$, the subluminal ST modulation provides vertical $\omega$-band-gaps at the synchronization points between forward and backward harmonics.

Fig.~\ref{Fig:son} plots the dispersion diagram except for the luminal ST modulation, i.e., $\gamma\rightarrow1$ and $\delta_\epsilon=\delta_\mu\rightarrow 0$. In this case, the forward harmonics acquire the minimum distance of $\Delta\beta^+\rightarrow0$, while the backward ST harmonics acquire the maximal distance of $\Delta\beta^-\rightarrow2$. Hence, a strong nonreciprocal transmission for forward and backward incident fields is provided by the STM medium. For the superluminal (time-like) ST modulation in Fig.~\ref{Fig:superson}, similar to the subluminal ST modulation in Fig.~\ref{Fig:subson}, the forward and backward STHs acquire different horizontal distances $\Delta\beta^\pm=\beta_{n+1}^\pm-\beta_n^\pm$, but increasing $\gamma$ leads to a decrease in the $\Delta\beta^-$ and an increase in the $\Delta\beta^+$. In contrast to the subluminal ST modulation, the superluminal ST modulation with a non-zero modulation strength $\delta_\epsilon\neq\delta_\mu \geq 0$ exhibits horizontal $\beta$-band-gaps, which occur at the synchronization points between the forward and backward STHs.

One may draw an analogy between a wave traveling in an STM medium and an airplane traveling through the air. Fig.~\ref{Fig:airplane_sup} demonstrates the occurrence of three different cases when the velocity of an airplane changes with respect to the sound velocity. In the sonic ($\gamma=1$) and supersonic ($\gamma>1$) regimes, the velocity of the airplane is equal and larger than the velocity of the sound, respectively. In these cases, shock waves start at the nose and end at the tail of the aircraft. The waves extend in different radial directions which result in a cone similar to that of the plane's vapor trail, called a Mach cone, which is shown in the photo on the right side of Fig.~\ref{Fig:airplane_sup}. When the leading parabolic edge of the cone contacts the earth, it creates a huge amount of sound energy, where the air particles are pushed aside with a huge force, and which is experienced on land as a sonic boom. The pressure waves generated in front of and behind the airplane travel at the speed of sound and vary by altitude and temperature. Hence, as the airplane travels faster and faster, the pressure waves cannot get out of the way of each other; instead they build up and compress together and eventually form a single shock wave at the speed of sound. The shape of the shock wave is called a Mach cone and the opening angle of the cone is represented by $\sin(\theta_\text{cone}) = v_\text{sound}/v_\text{object}$. A shock wave of light can also be generated. Light travels at the speed of $3\times10^8$ m/s in the air but it slows down in natural materials by the factor of the refractive index of the material. Hence, a very fast particle can exceed the speed of light in a material with a refractive index of larger than unity. For instance, a visible shock wave of light called Cherenkov radiation can be produced by accelerating electrons to a very high velocity and firing them into a piece of glass or plastic.

It can be shown that the ST Bloch-Floquet decomposition in~\eqref{eqa:EH_anzatz} is valid for subluminal ($v_\text{m}<v_\text{b}$) and superluminal ($v_\text{m}>v_\text{b}$) ST modulations. However, the ST Bloch-Floquet decomposition is not convergent in the sonic interval~\cite{Oliner_PIEEE_1963,Taravati_PRB_2017,Taravati_PRApplied_2018}, i.e.,
\begin{equation}\label{eqa:sonic}
\begin{split}
\gamma_\text{s,min} =&\sqrt{\frac{\epsilon_\text{r}\mu_\text{r}}{\tilde{\epsilon}_0 \tilde{\mu}_0 (1+\delta_\epsilon) (1+\delta_\mu)}}
\leq\gamma \\
& \qquad \qquad \leq
\sqrt{\frac{\epsilon_\text{r} \mu_\text{r}}{\tilde{\epsilon}_0 \tilde{\mu}_0 (1-\delta_\epsilon) (1-\delta_\mu)}}=\gamma_\text{s,max}.
\end{split}
\end{equation}
where $\gamma_\text{s,min}$ and $\gamma_\text{s,max}$ are the lower and upper sides of the sonic interval, respectively. The sonic condition corresponds to the luminal ST modulation where the modulation and background velocities are close. This is due to the fact that none of the amplitudes of the infinite number of ST harmonics are small in the sonic condition~\cite{Oliner_PIEEE_1963,Taravati_PRB_2017,Taravati_PRApplied_2018}.
\subsection{Numerical Simulation}
\begin{figure}
	\begin{center}
		\subfigure[]{\label{Fig:FDTD}
			\includegraphics[width=0.85\columnwidth]{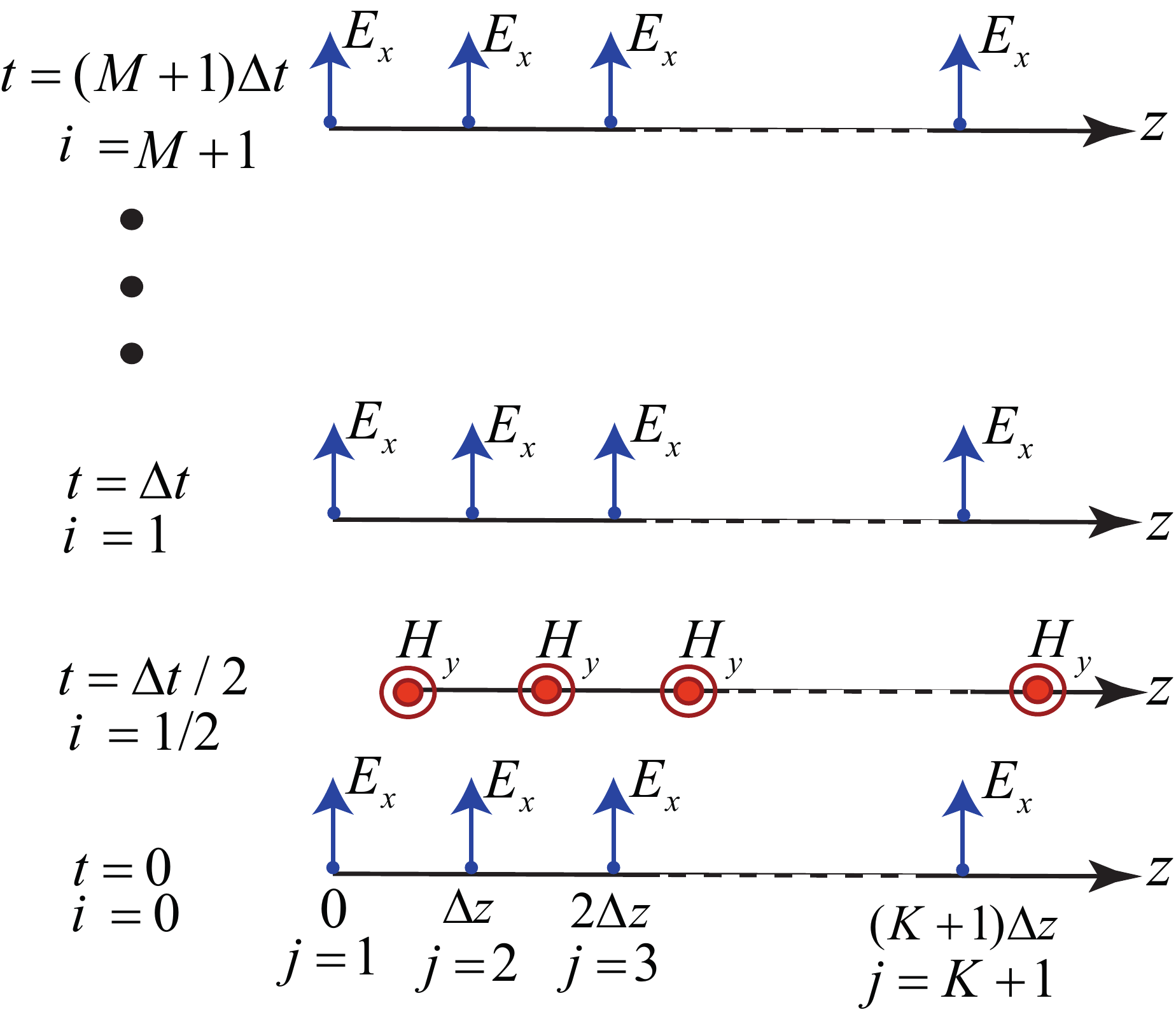} }
		\subfigure[]{\label{Fig:var_gamma}
			\includegraphics[width=0.89\columnwidth]{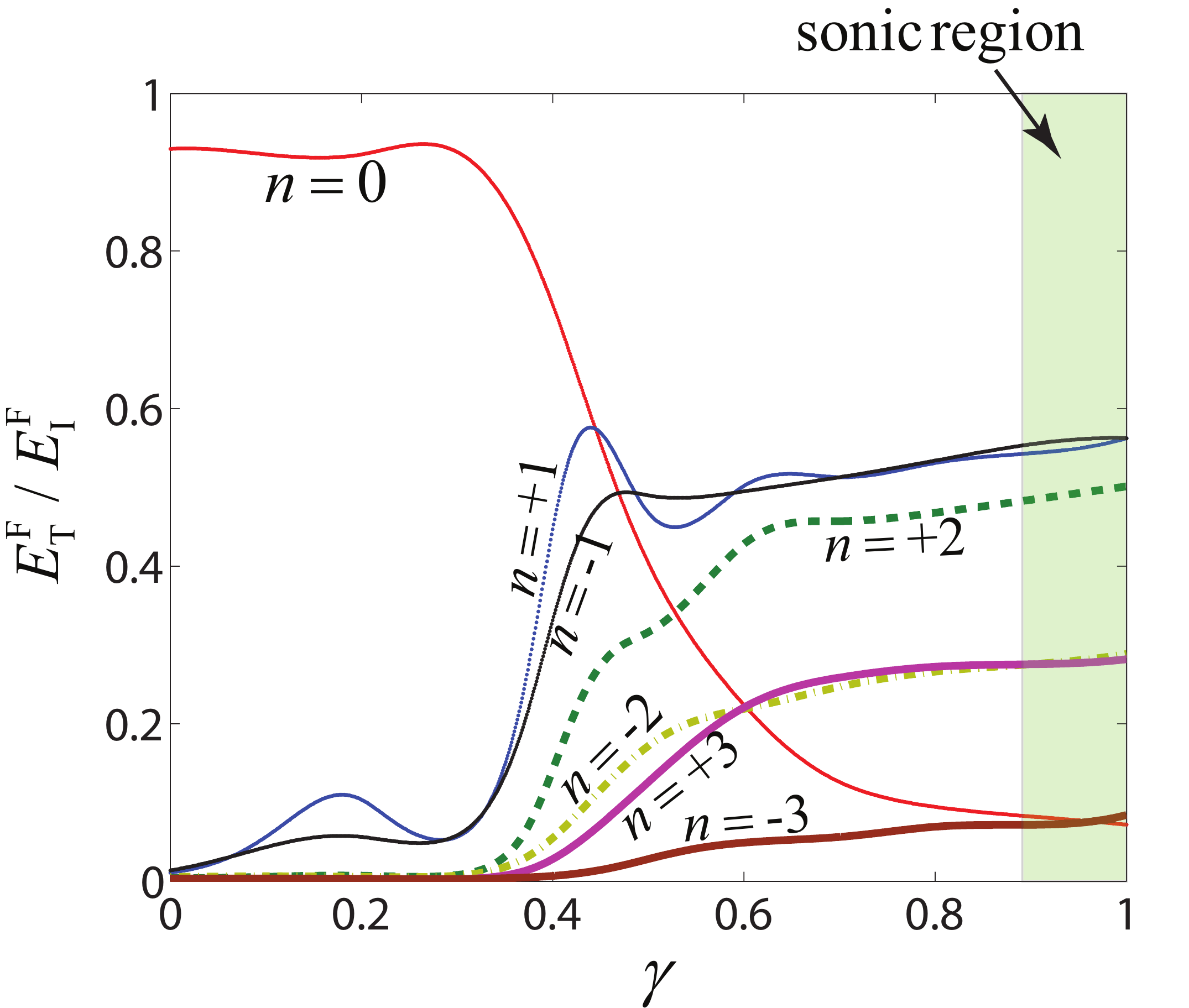}}
		\subfigure[]{\label{Fig:var_L}
			\includegraphics[width=0.89\columnwidth]{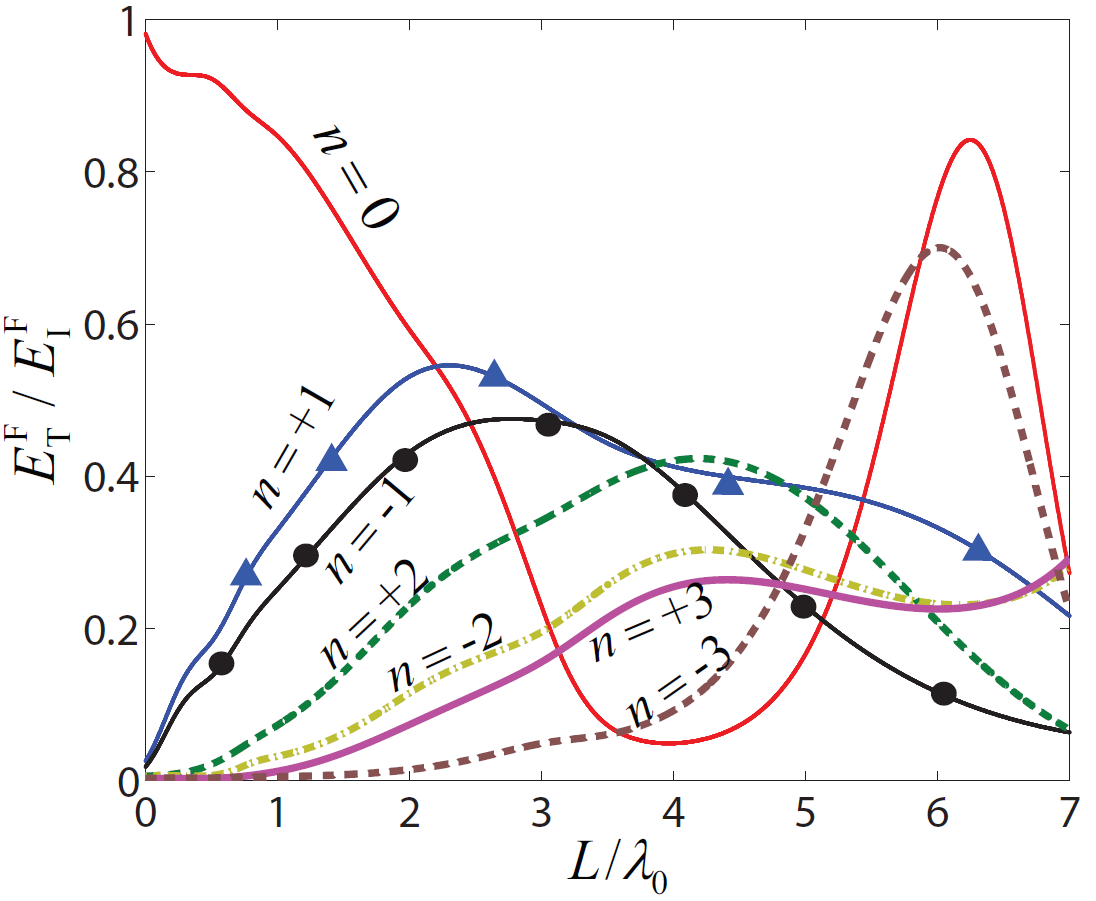}}
		\caption{Finite-difference time-domain numerical simulation of the ST permittivity- and permeability-modulated media. (a)~Space-time sampling scheme~\cite{Taravati_PRApplied_2018}. (b) and (c) Numerical simulation results for harmonic generation in non-equilibrated STM medium with parameters $\delta_{\epsilon}=0.22$, $\delta_{\mu}=0$, $\omega_\text{m}=2\pi \times 0.2$~GHz and $\omega_\text{0}=2\pi \times 1.5$~GHz versus (b) the velocity ratio $\gamma$ and (c) the length of the slab for $\gamma=1$~\cite{Taravati_PRB_2017}.}
		\label{Fig:effect}
	\end{center}
\end{figure}

Numerical simulation of an STM medium is an excellent tool for the parametric investigation of wave propagation and transmission through such a medium, especially in the sonic interval where the Block-Floquet solution does not converge. Fig.~\ref{Fig:FDTD} sketches the implemented finite-difference time-domain (FDTD) scheme for numerical simulation of general ST permittivity- and permeability-modulated media~\cite{Taravati_PRApplied_2018}. In this scheme, the STM medium is sampled with $K+1$ spatial steps and $M+1$ temporal steps, with the steps of $\Delta z$ and $\Delta t$, respectively. Then, the finite-difference discretized form of the electric and magnetic fields may be achieved by satisfying the sourceless Maxwell equations at each ST sample, as~\cite{Taravati_PRApplied_2018}

\begin{subequations}
	\begin{equation}\label{eqa:num_Max1c}
	\begin{split}
	H_y\lvert_{j+1/2}^{i+1/2}=&\left(1-\Delta t \dfrac{\mu'\lvert_{j+1/2}^{i-1/2}}{\mu \lvert_{j+1/2}^{i}}  \right)	H_y\lvert_{j+1/2}^{i-1/2} \\
 & \qquad \qquad-  \dfrac{\Delta t/\Delta z}{ \mu \lvert_{j+1/2}^{i}} \left( E_x\lvert_{j+1}^{i}-E_x\lvert_{j}^{i} \right) 
    \end{split}
	\end{equation}
	\begin{equation}\label{eqa:num_Max2c}
	E_x \lvert_{j}^{i+1}= \left(1- \dfrac{\Delta t\epsilon'\lvert_{j}^{i}}{\epsilon \lvert_{j}^{i+1/2}}  \right) E_x \lvert_{j}^{i}
- \dfrac{\Delta t/\Delta z}{ \epsilon \lvert_{j}^{i+1/2}} \left( H_y \lvert_{j+1/2}^{i+1/2}-H_y\lvert_{j-1/2}^{i+1/2}\right) 
	\end{equation}
\end{subequations}
\noindent where $\mu'=\partial \mu(z,t)/\partial t=\omega_\text{m} \mu_0 \mu_\text{av}\delta_{\mu} \sin(\beta_\text{m} z-\omega_\text{m} t)$ and $\epsilon'=\partial \epsilon(z,t)/\partial t=\omega_\text{m} \epsilon_0 \epsilon_\text{av}\delta_{\epsilon} \sin(\beta_\text{m} z-\omega_\text{m} t)$.

We next investigate the effect of ST velocity ratio $\gamma=1$ and the effect of the length of the STM slab $L$ on the forward transmission. It should be noted that, due to the negligible effect of the ST modulation on the backward transmission, varying the $\gamma=1$ and $L$ leads to minor change on the backward problem. Fig.~\ref{Fig:var_gamma} plots the FDTD numerical results for the amplitude of the electric field of the STHs at the output of the STM slab versus the ST velocity ratio $\gamma$. It may be observed from this figure that, in the subsonic interval $\gamma~\rightarrow~0$, the energy resides mainly inside the fundamental harmonic ($n=0$). However, in the sonic condition, a strong energy transition occurs from the fundamental harmonic to the higher STHs. 

Fig.~\ref{Fig:var_L} plots the amplitude of the transmitted harmonics versus the length of the STM slab. The structure operates in the middle of the sonic condition, i.e. $\gamma=1$. As the length of the slab is increased, the input power is more efficiently coupled to the ST harmonics $\omega_0 \pm n\omega_\text{m}$, $n\ge1$. The incident wave gradually couples its energy to these harmonics as it propagates through the slab. Therefore, a longer slab exhibits a more efficient power transition but in a quasi-periodic manner. It may be seen from Fig.~\ref{Fig:var_L} that a quasi-periodic transition of power between the fundamental harmonic ($n=0$) and higher-order harmonics occurs. One may derive a closed form solution for the period of the wave transition between the fundamental STH and the first higher (or lower) STH by ignoring the effect of all other STHs and considering a pure transition from $n=0$ to $n=1$~\cite{Fan_NPH_2009}. Thereby, the coherency length $l_\text{c}$ is achieved, which depends on the modulation parameters, i.e., $\omega_\text{m}$, $\omega_0$, $\gamma$, $\delta_{\epsilon}$ and $\delta_{\mu}$~\cite{Fan_NPH_2009}.  

\begin{figure}
	\begin{center}
			\includegraphics[width=1\columnwidth]{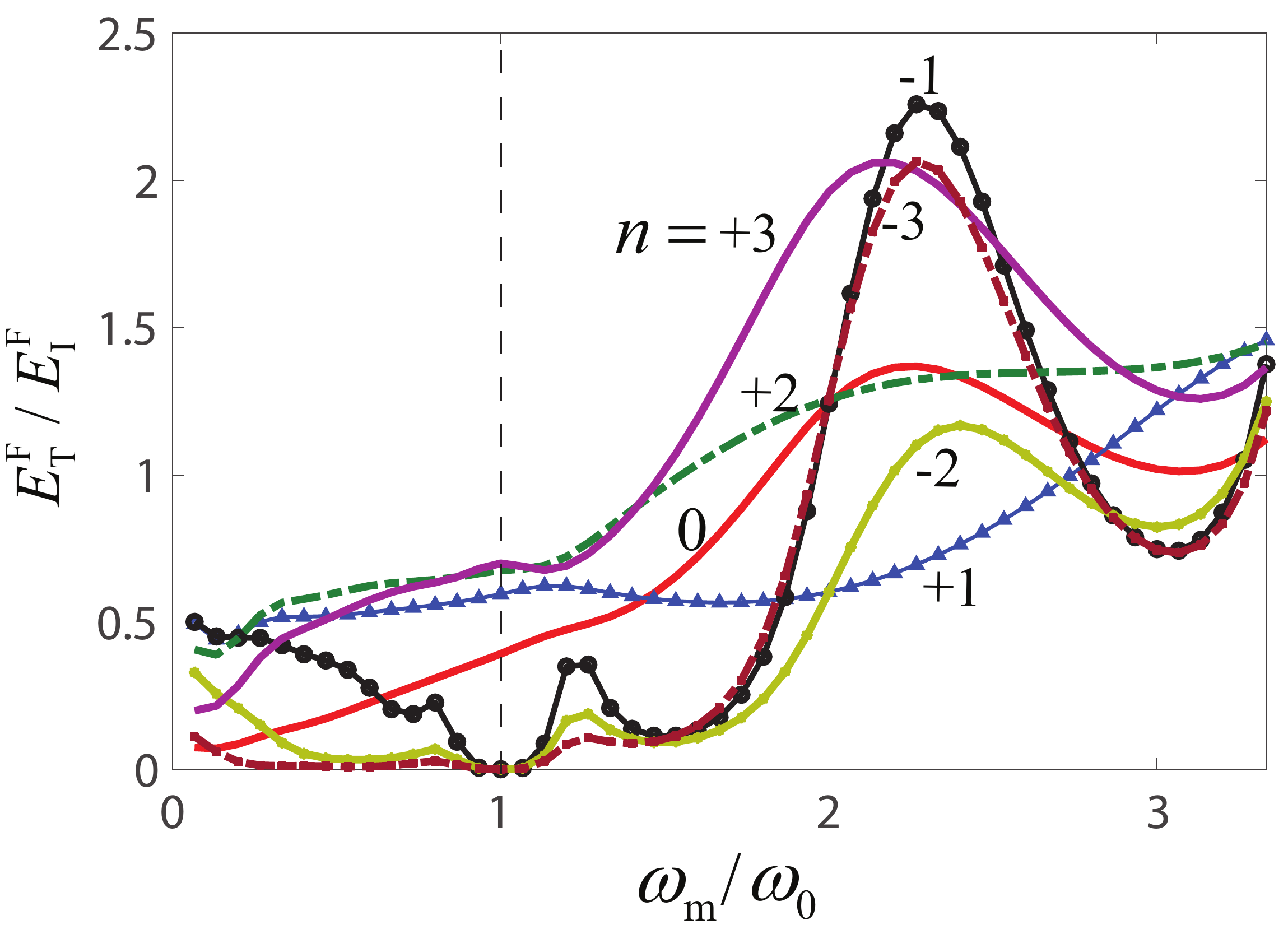}  
		\caption{Investigation of effect of the modulation temporal frequency $\omega_\text{m}/\omega_\text{0}$ on the transmission gain, based on the numerical simulation results, for $\delta_{\epsilon}=0.22$, $\delta_{\mu}=0$, $L=3\lambda$ and $\gamma=1$.} 
		\label{Fig:omegam}
	\end{center}
\end{figure}

The factor $\omega_\text{m}/\omega_0$ specifies the frequency of the generated STHs, and therefore, it is a crucial parameter. However, it also directly affects the amplitude of the generated and transmitted STHs. Fig.~\ref{Fig:omegam} presents the numerical simulation results that investigate the effect of the modulation temporal frequency $\omega_\text{m}/\omega_\text{0}$ on the transmission gain, where $\delta_{\epsilon}=0.22$, $\delta_{\mu}=0$, $L=3\lambda$ and $\gamma=1$. It may be seen from this figure that a transmission gain is achieved for $\omega_\text{m}/\omega_0>1$. This is consistent with the results presented in~\cite{Wang_TMTT_2014}. Moreover, in a space-time-varying system, the ratio of the power of the generated harmonics is $P_n/P_0=|\beta_n/\beta_0|$~\cite{Taravati_TAP_65_2017}, where $P_n$ and $P_0$ are powers of the $n$th and the fundamental ($n=0$) STHs, respectively~\cite{Taravati_TAP_65_2017}. This phenomenon has also been reported in the Manley-Rowe relation for a time-varying medium~\cite{Manley_Rowe_1956}.
\subsection{Nonreciprocal Scattering}
\begin{figure*}
	\begin{center}
		\subfigure[]{\label{Fig:Nonrec_trans_a}
			\includegraphics[width=0.9\columnwidth]{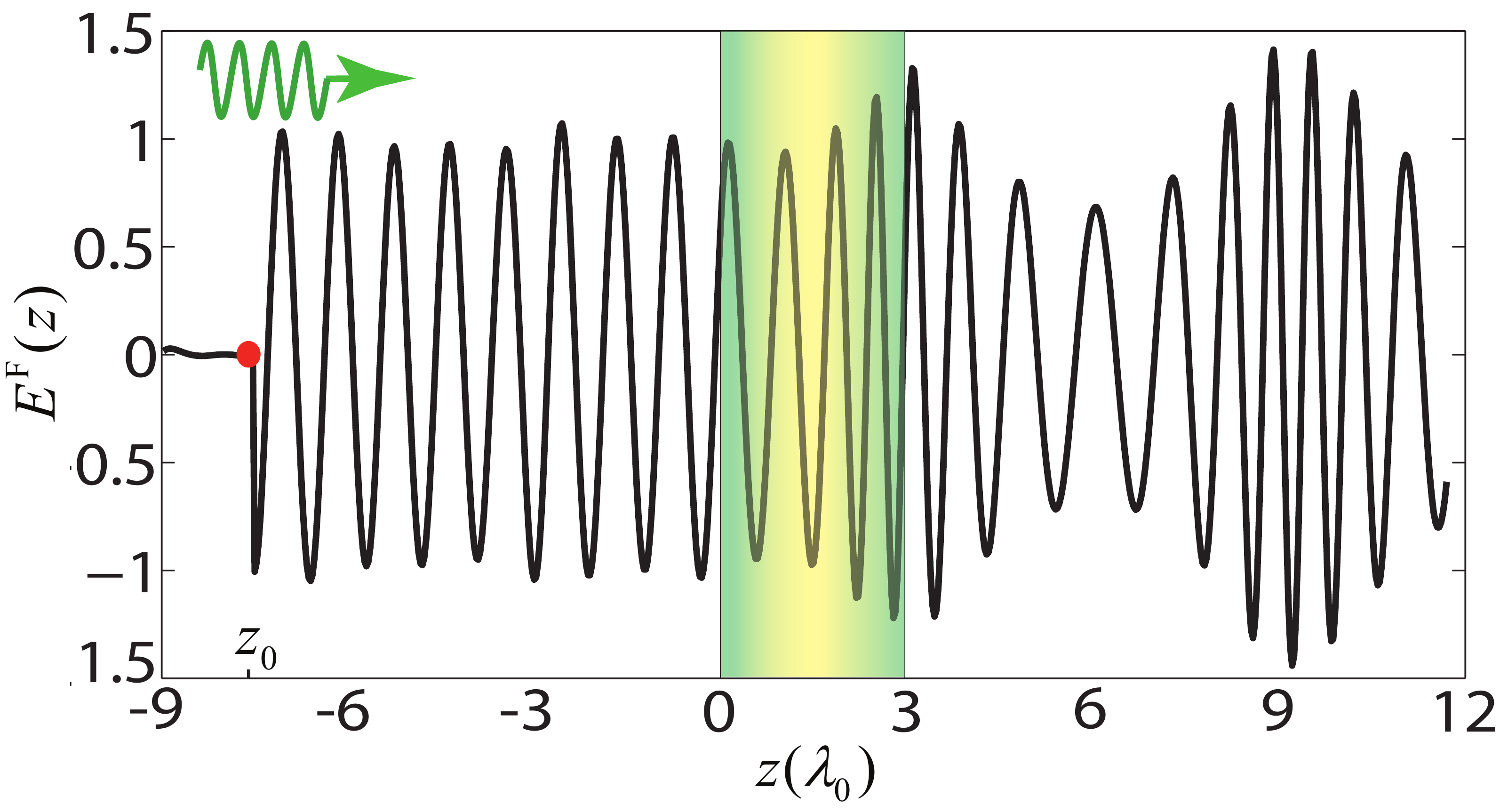}}
				\subfigure[]{\label{Fig:Nonrec_trans_c}
			\includegraphics[width=0.9\columnwidth]{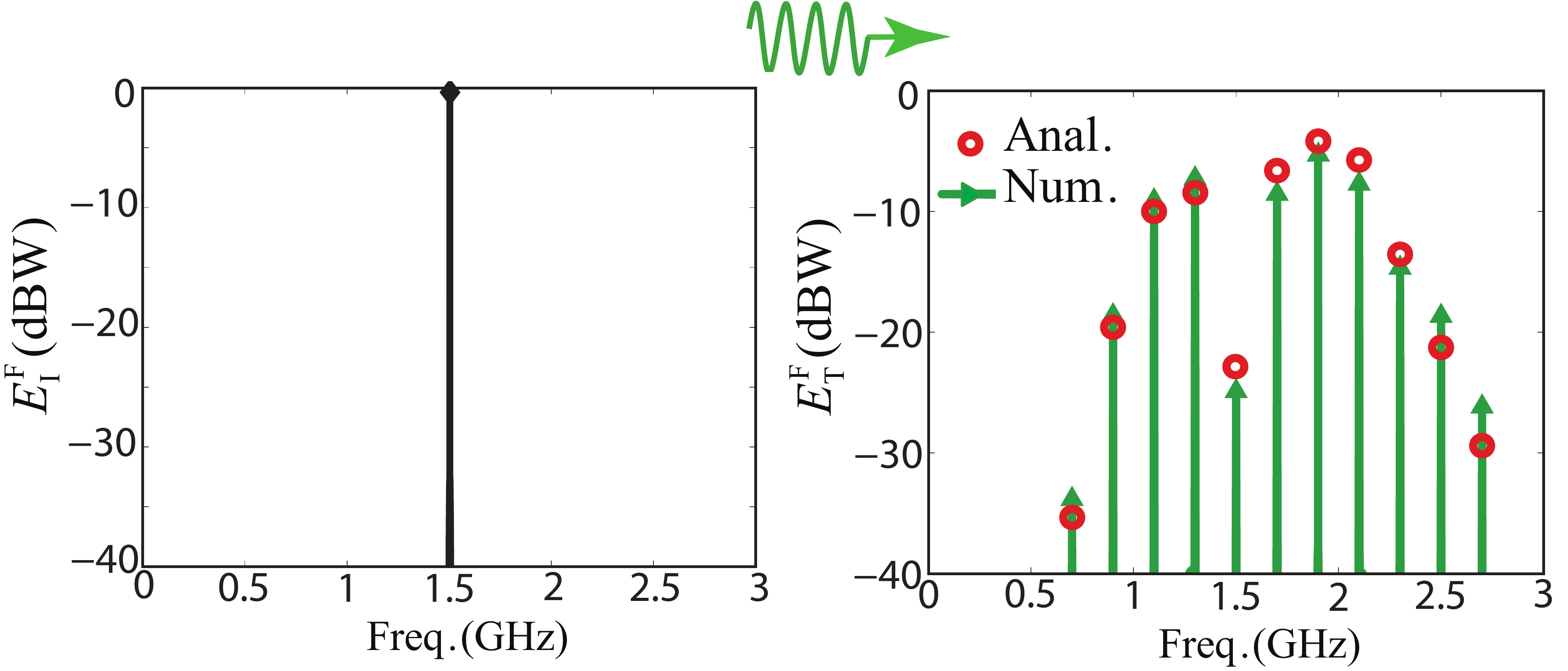}}
		\subfigure[]{\label{Fig:Nonrec_trans_b}
			\includegraphics[width=0.9\columnwidth]{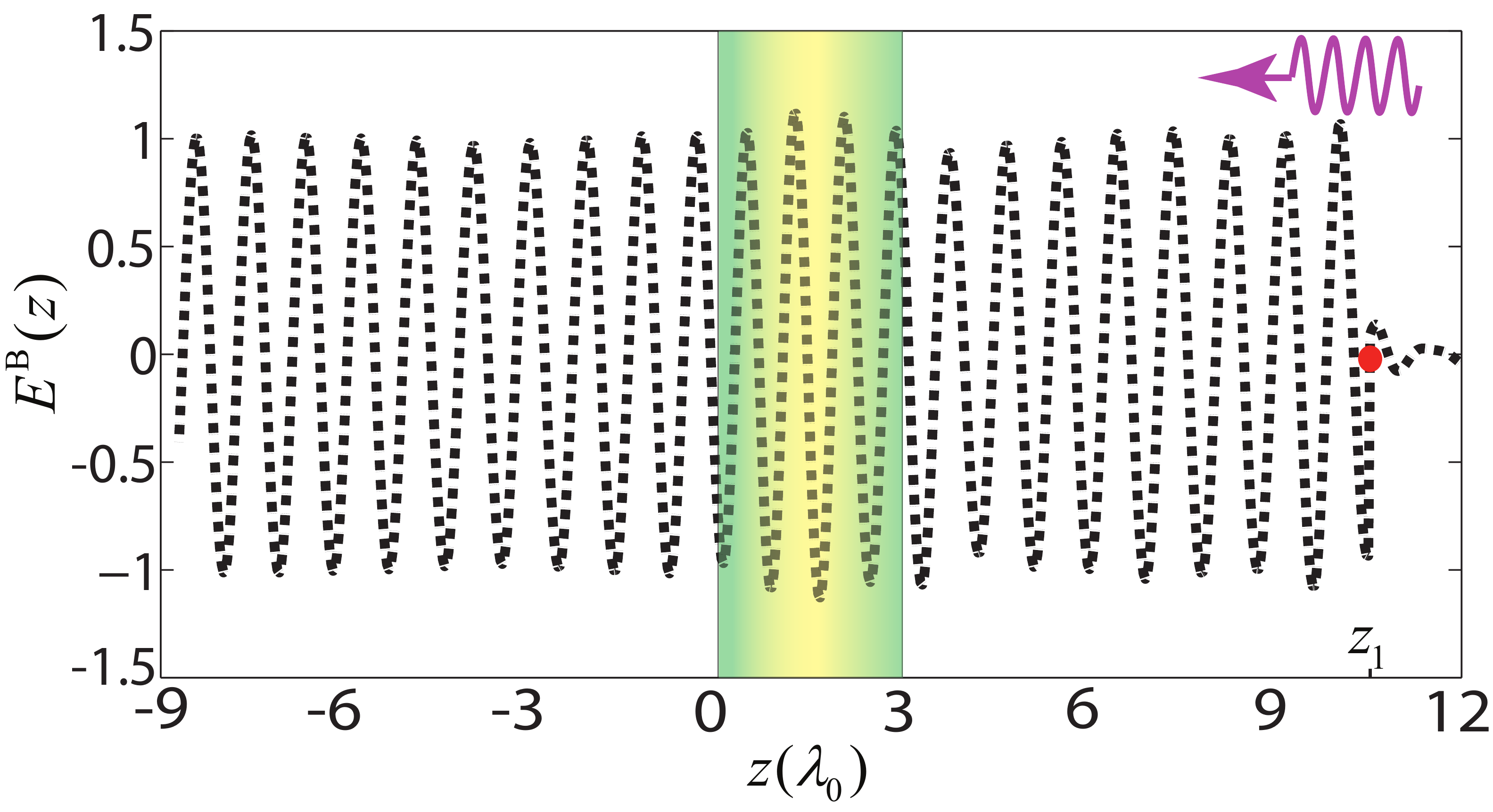}}
		\subfigure[]{\label{Fig:Nonrec_trans_d}
			\includegraphics[width=0.9\columnwidth]{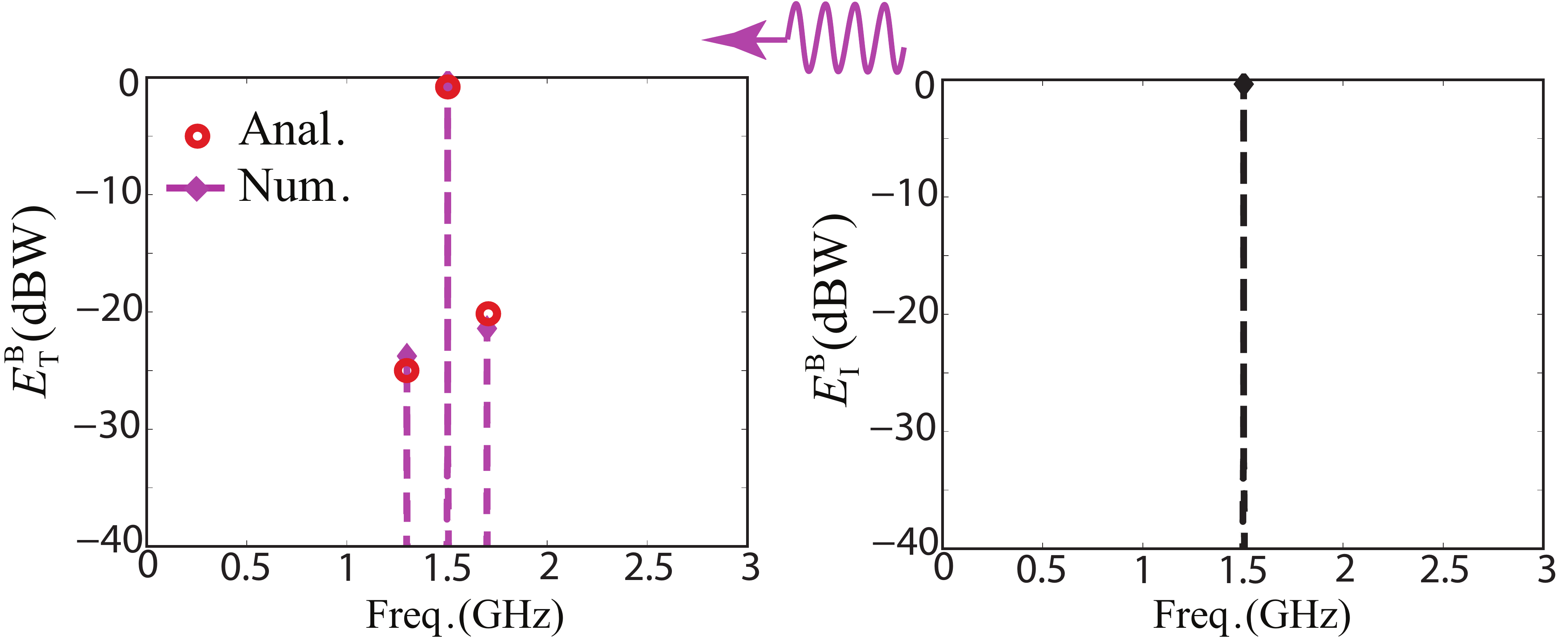}}
		\caption{Analytical and numerical (FDTD) results for the forward and backward problems in the quasisonic condition with parameters $\delta_\epsilon=0.3$, $\omega_0=2\pi\times1.5$~GHz, $\omega_\text{m}=2\pi\times 0.2$~GHz, $L=3\lambda_0$ and $\gamma=0.85$~\cite{Taravati_PRB_2017}. (a)~and (c) FDTD waveforms showing the electric field amplitude for the forward and backward problems, respectively~\cite{Taravati_PRB_2017}. (b)~and (d)~Frequency spectrum of the incident and transmitted fields for the forward and backward problems, respectively~\cite{Taravati_PRB_2017}.}
		\label{Fig:Nonrec_trans}
	\end{center}
\end{figure*}

We next demonstrate the effect of unidirectionality of the ST modulation by investigating the wave transmissions through the STM slab for forward and backward problems. Figs.~\ref{Fig:Nonrec_trans_a} and~\ref{Fig:Nonrec_trans_b} plot, respectively, the time-domain and frequency-domain numerical result for the amplitude of the electric field in the forward problem. Here, the incident wave impinges on the quasisonic STM slab ($\gamma=0.85$, with $\gamma_\text{s,min}=0.867$) possessing sinusoidal STM permittivity of $\delta_\epsilon=0.3$ and uniform permeability $\delta_\mu=0$. It may be observed from these figures that the wave is strongly interacting with the medium, the incident power at $\omega_0$ is effectively transferred to the STHs at $\omega_0\pm n\omega_\text{m}$, $n\ge1$, leaving a weak transmitted wave at the incident frequency $\omega_0$. In contrast, and the incident wave in the backward problem (Figs.~\ref{Fig:Nonrec_trans_c} and~\ref{Fig:Nonrec_trans_d}) passes through the STM slab with weak interaction and minor power exchange with the ST modulation. These plots reveal that the ST medium behaves differently under forward and backward wave incidences, i.e., providing strong STHs in the forward problem, but weak STHs in the backward problem.

\subsection{Superior Performance of Equilibrated ST Modulation}\label{sec:equi}
It has been shown that an equilibrated STM medium ($\delta_\mu=\delta_\epsilon>0$) presents much stronger nonreciprocity in comparison with conventional permittivity STM medium~\cite{Taravati_PRApplied_2018}. Such a strong nonreciprocity is accompanied by a larger sonic condition interval which provides extra design freedom for achieving strong nonreciprocity by a weak pumping strength. It is shown that the width of photonic band gaps in general periodic space-time permittivity- and permeability-modulated media is proportional to the absolute difference between the electric and magnetic pumping strengths $|\delta_{\epsilon}-\delta_{\mu}|$. Conventional space-time permittivity-modulated media provide moderate photonic transitions from the excited mode to its two adjacent modes. However, in an equilibrated space-time-varying medium, strong photonic transitions occur from the excited mode to its four adjacent modes~\cite{Taravati_PRApplied_2018}. 
	
\begin{figure*}
	\begin{center} 
		\subfigure[]{\label{Fig:dispersion_curves_c}
			\includegraphics[width=0.64\columnwidth]{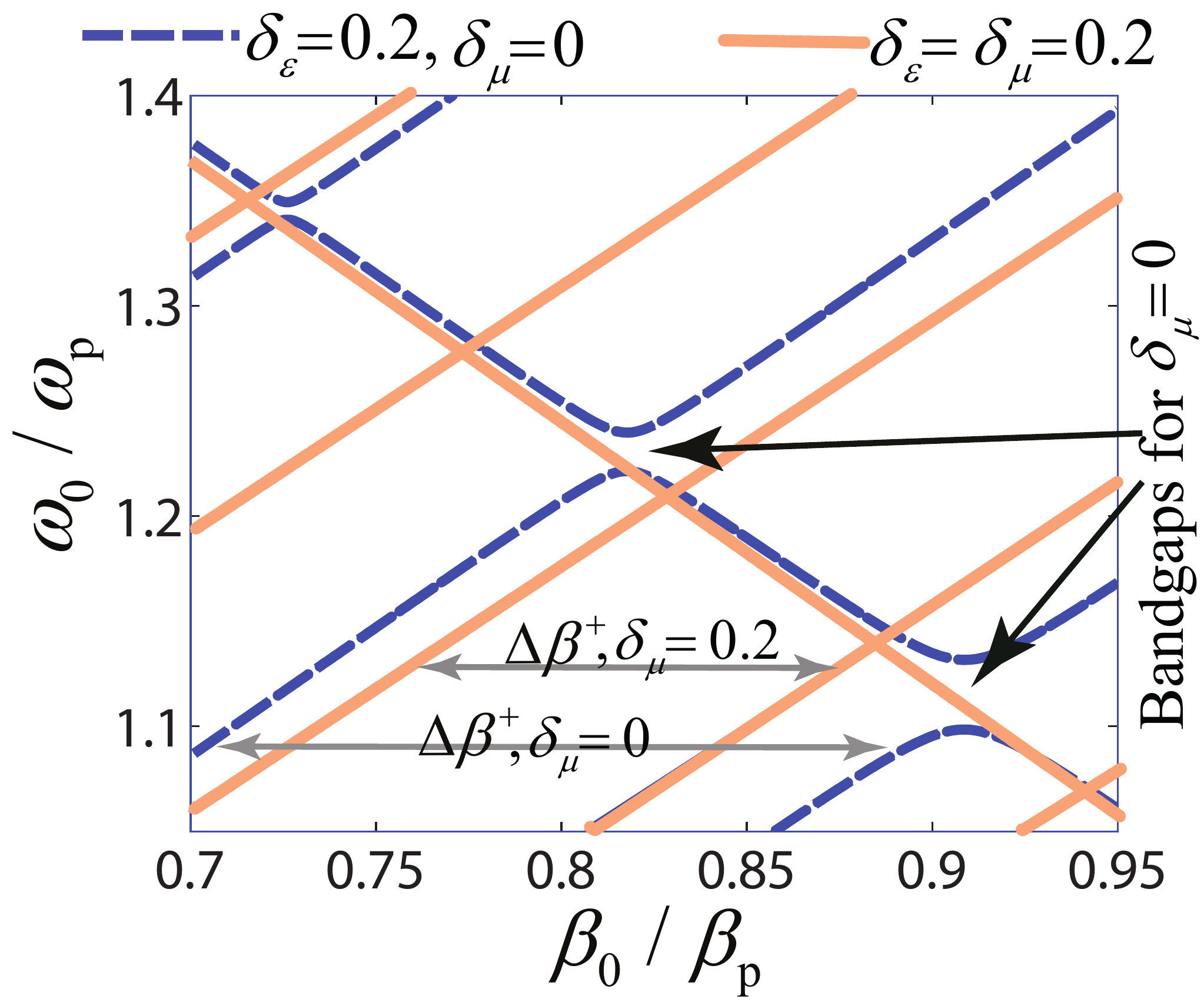}}
			\subfigure[]{\label{Fig:ratio}
			\includegraphics[width=0.6\columnwidth]{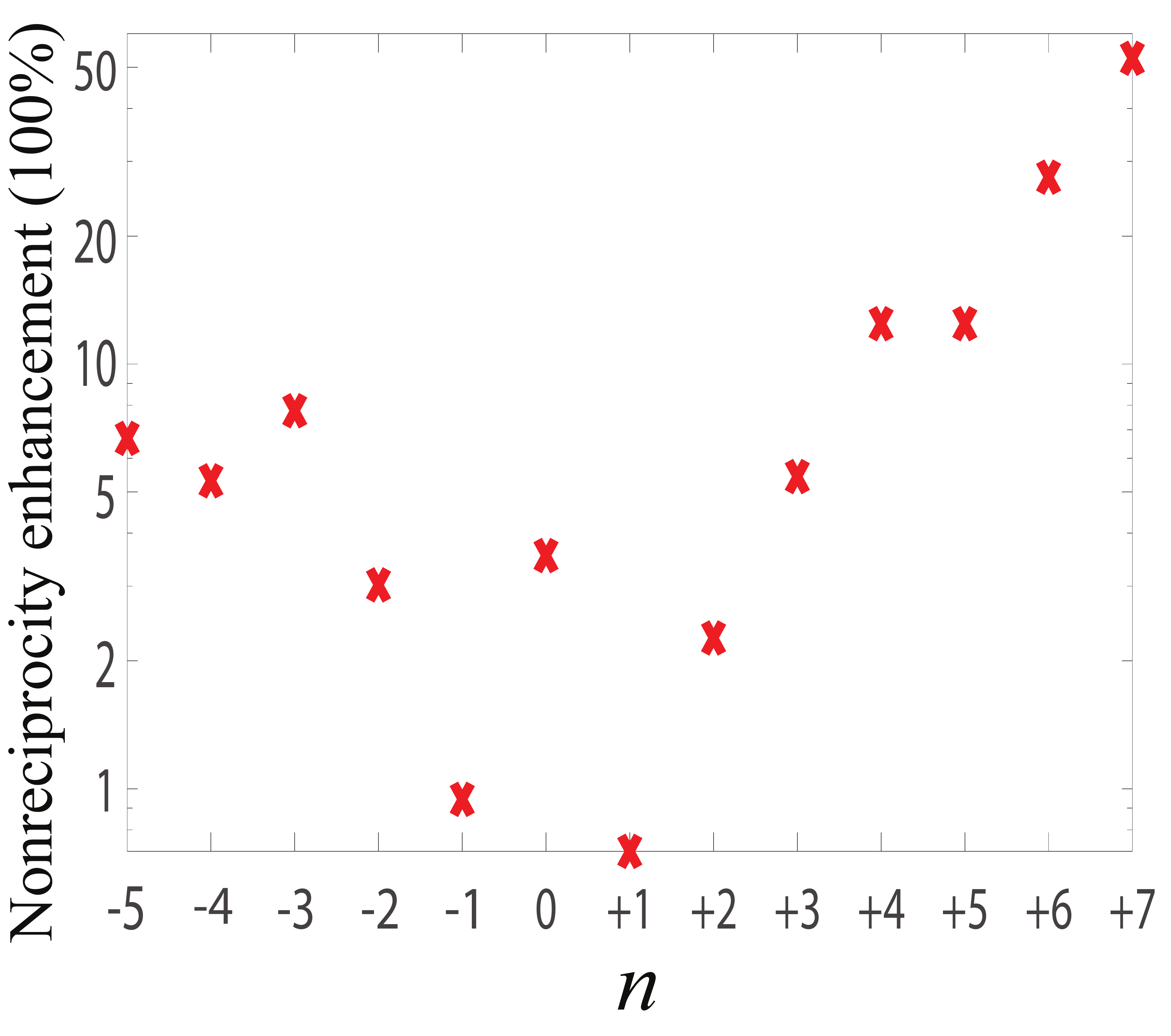}} \\
			\subfigure[]{\label{Fig:refl_conv_a}
			\includegraphics[width=0.6\columnwidth]{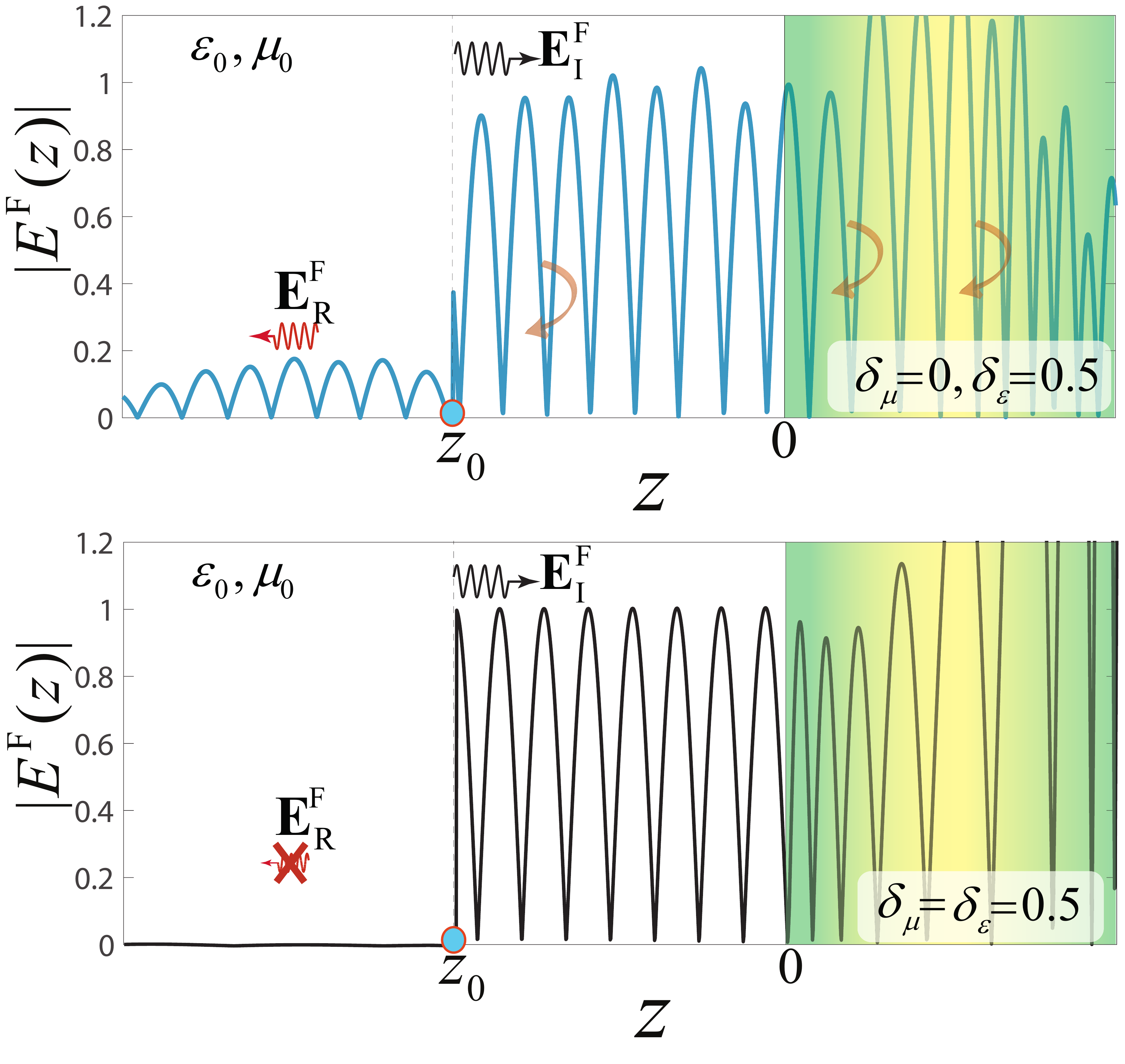}} 
		\subfigure[]{\label{Fig:refl_conv_b}
			\includegraphics[width=0.6\columnwidth]{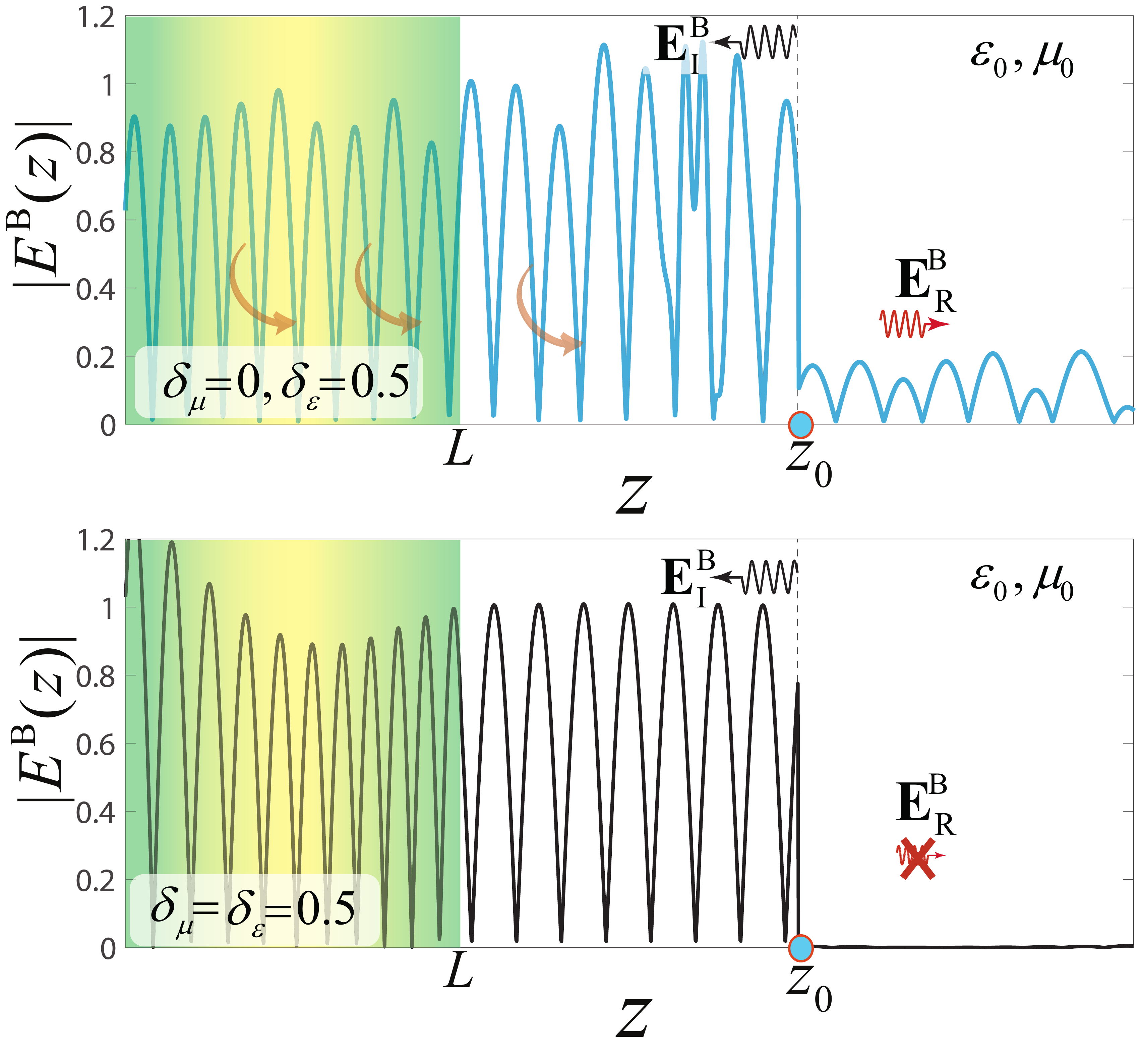}} 
			\subfigure[]{\label{Fig:circ_eq}
			\includegraphics[width=0.72\columnwidth]{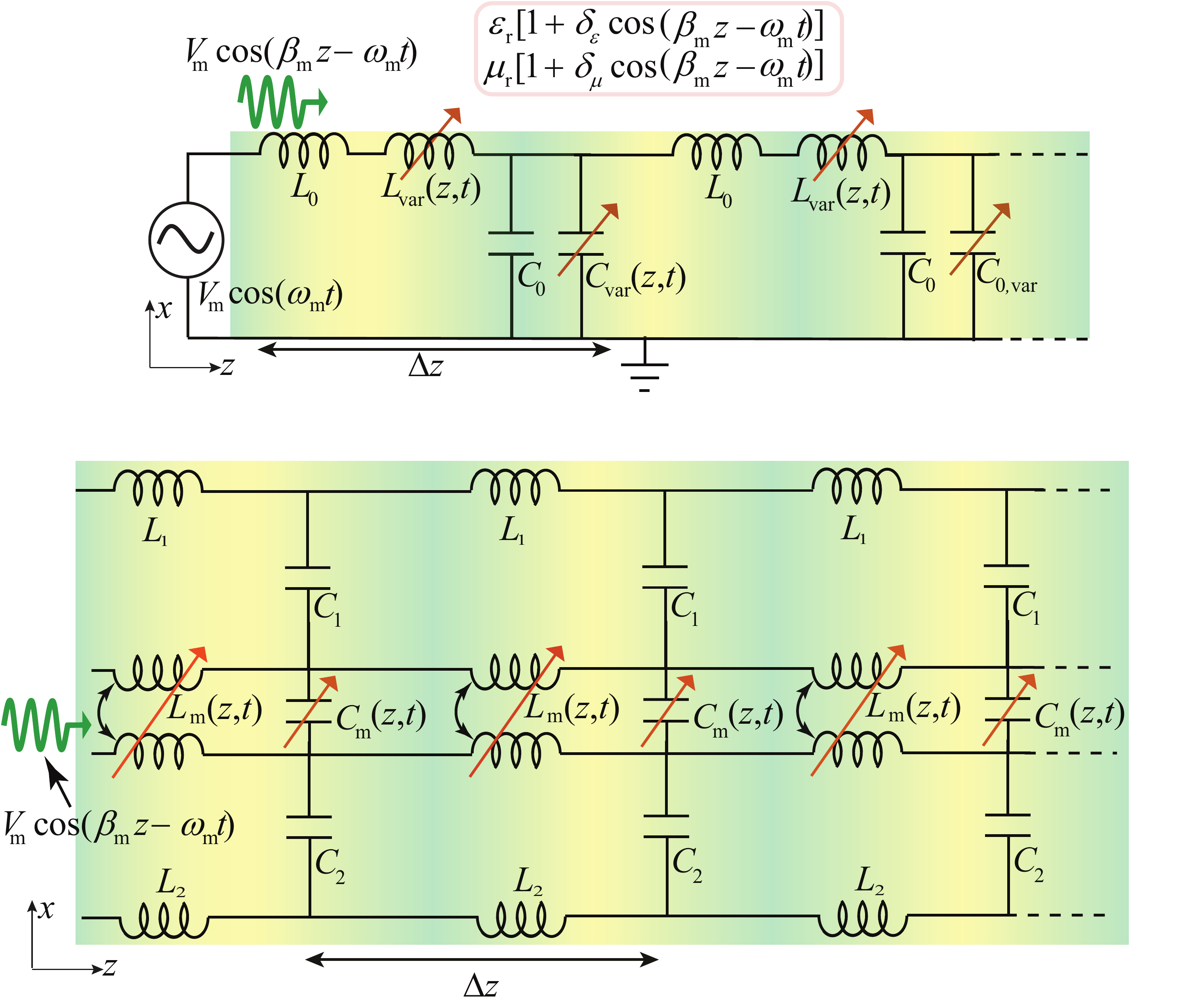} }    
		\caption{Equilibrated sinusoidally STM medium. (a)~Dispersion diagram comparison between the equilibrated ($\delta_\mu=\delta_\epsilon=0.2$, $\gamma_\text{s,l}=0.833$) and non-equilibrated ($\delta_\mu=0$, $\delta_\epsilon=0.2$, $\gamma_\text{s,l}=0.913$) STM media with $\gamma=0.8$~\cite{Taravati_PRApplied_2018}. (b)~Nonreciprocity enhancement of the equilibrated ST slab in comparison with the nonreciprocity provided by the conventional ST slabs~\cite{Taravati_PRApplied_2018}. (c) and (d)~Reflection from conventional and equilibrated STM slabs for the forward problem (c) and the backward problem (d)~\cite{Taravati_PRApplied_2018}. (c)~Circuit model for an equilibrated STM transmission line~\cite{Taravati_PRApplied_2018}. (e) ~Circuit model for an equilibrated STM coupled transmission line~\cite{Taravati_PRApplied_2018}.}
		\label{Fig:wave_scat_b}
	\end{center}
\end{figure*}

Fig.~\ref{Fig:dispersion_curves_c} compares the dispersion diagrams of a non-equilibrated ST permittivity-modulated medium with $\delta_\epsilon=0.2$, $\delta_\mu=0$ and $\gamma=0.8$, with the dispersion diagram of an equilibrated ST permittivity- and permeability-modulated medium with $\delta_\epsilon=\delta_\mu=0.2$ and $\gamma=0.8$. It is obvious from this figure that, the conventional, non-equilibrated, STM medium exhibits photonic band gaps at the ST synchronization points, i.e., at the intersections of the forward and backward STHs which lie on oblique lines. Such band gaps correspond to evanescent waves, where the wave propagates through the medium, and experiences high attenuation and reflection. Nevertheless, by switching on the permeability modulation with the same ST modulation amplitude  of $\delta_\mu=\delta_\epsilon=0.2$, equilibrium occurs in the electric and magnetic properties of the medium, and photonic band gaps are closed, for all values of $\gamma$~\cite{Taravati_PRApplied_2018}. Another interesting phenomenon that we see in Fig.~\ref{Fig:dispersion_curves_c} is as follows. The equilibrium in the medium has led to significant enhancement in the nonreciprocity of the medium, so that the equilibrated medium introduces much stronger nonreciprocity, $\Delta\beta^+(\delta_\mu=0.2)<<\Delta\beta^+(\delta_\mu=0)$. The nonreciprocity (NR) of the STM medium may also be defined in the frequency domain for the $n$th harmonic, as the ratio of the forward transmitted field to the backward transmitted field, i.e., $\text{NR}_n=E_{\text{T},n}^\text{F}/E_{\text{T},n}^\text{B}$.

Fig.~\ref{Fig:ratio} plots the nonreciprocity enhancement provided by the equilibrated STM medium ($\delta_{\epsilon}=0.02$, $\delta_{\mu}=0$) in comparison with the non-equilibrated STM medium ($\delta_{\mu}=\delta_{\epsilon}=0.02$), for $\omega_0=2\pi\times1.5$~GHz, $\omega_\text{p}=2\pi\times 0.2$~GHz, $L=20\lambda_0$ and $\gamma=1$. As we see from this figure, the equilibrium in the electric and magnetic properties of the STM slab has led to significant enhancement in the nonreciprocity of STHs. In particular, the nonreciprocity of the fundamental STH, which is the most important harmonic to be largely isolated~\cite{Taravati_TAP_65_2017,Taravati_PRB_2017,Taravati_PRB_Mixer_2018}, has enhanced more than $360\%$, and the nonreciprocity of most of the STHs has enhanced more than $220\%$.

Since the constitutive parameters of an STM medium vary in both space and time, the intrinsic impedance of the medium depends on both space and time, i.e.,
\begin{equation}\label{eqa:eta1}
\eta(z,t)= \sqrt{\dfrac{\mu(z,t)} {\epsilon(z,t)} }
=\sqrt{\dfrac{\mu_0 \mu_\text{av}[1+\delta_\mu ~\cos(\beta_\text{m}z-\omega_\text{m}t) ]
	}{\epsilon_0 \epsilon_\text{av} [1+\delta_\epsilon ~\cos(\beta_\text{m}z-\omega_\text{m}t) ] }}
\end{equation}
\noindent which represents an ST-varying intrinsic impedance. For $\delta_\mu\neq\delta_\epsilon$,~\eqref{eqa:eta1} corresponds to a non-equilibrated STM medium with local ST reflections ($R(z,t)$) in an infinite number of space points, i.e., $R(z_0,t)=[\eta(z_0+\Delta z,t)-\eta(z_0,t)]/[\eta(z_0+\Delta z,t)+\eta(z_0,t)]$ and in an infinite number of time points, i.e., $R(z,t_0)=[\eta(z,t_0+\Delta t)-\eta(z,t_0)]/[\eta(z,t_0+\Delta t)+\eta(z,t_0+\Delta t)]$. This issue is overcome by an \textit{equilibrated} STM medium ($\delta_\mu=\delta_\epsilon$), where the ST-varying intrinsic impedance in~\eqref{eqa:eta1} reduces to
\begin{equation}\label{eqa:eta2}
\eta(z,t)\big|_{\delta_{\epsilon}=\delta_{\mu}} =\eta_\text{eq}= \eta_0 \eta_\text{av}
\end{equation}
\noindent which is constant and independent of both space and time, yielding zero local ST reflections. Figs.~\ref{Fig:refl_conv_a} and~\ref{Fig:refl_conv_a} compare the wave reflections ($\textbf{E}_\text{R}$) in non-equilibrated and equilibrated STM media, respectively for forward and backward problems. It is obvious that the conventional non-equilibrated STM slab exhibits strong space and time local reflections, whereas the equilibrated STM slab presents no space and time local reflections. 

Fig.~\ref{Fig:circ_eq} illustrates two circuit models for sinusoidally ST permittivity- and permeability-modulated slabs. These circuits are made up of an array of subwavelength-spaced unit-cells, each of which is composed of a variable inductor or variable mutual inductance and a variable capacitor or a variable mutual capacitance. These variable elements are placed, respectively, in series and parallel with the intrinsic inductance and capacitance of the transmission line, and are spatiotemporally modulated using a sinusoidal harmonic wave, i.e., $V_\text{P} \cos(\omega_\text{m} t)$. Interestingly, such a structure with highly enhanced nonreciprocity, shown in Fig.~\ref{Fig:ratio}, requires the same pumped energy as a conventional ST permittivity-modulated structure~\cite{Fan_PRL_109_2012,Wang_TMTT_2014,Taravati_PRB_2017}. 

\section{Applications of ST Modulation}\label{sec:App}

Recently, ST modulation has begun to be used for the realization of a variety of novel and efficient components. This section presents a class of guided and radiated structures that leverage the unique and valuable properties of STM media.

\subsection{Magnet-Free Linear Isolators}\label{sec:Isolator}

\subsubsection{Quasisonic Isolator}\label{sec:QSI}
\begin{figure*}
	\begin{center} 
			\includegraphics[width=2\columnwidth]{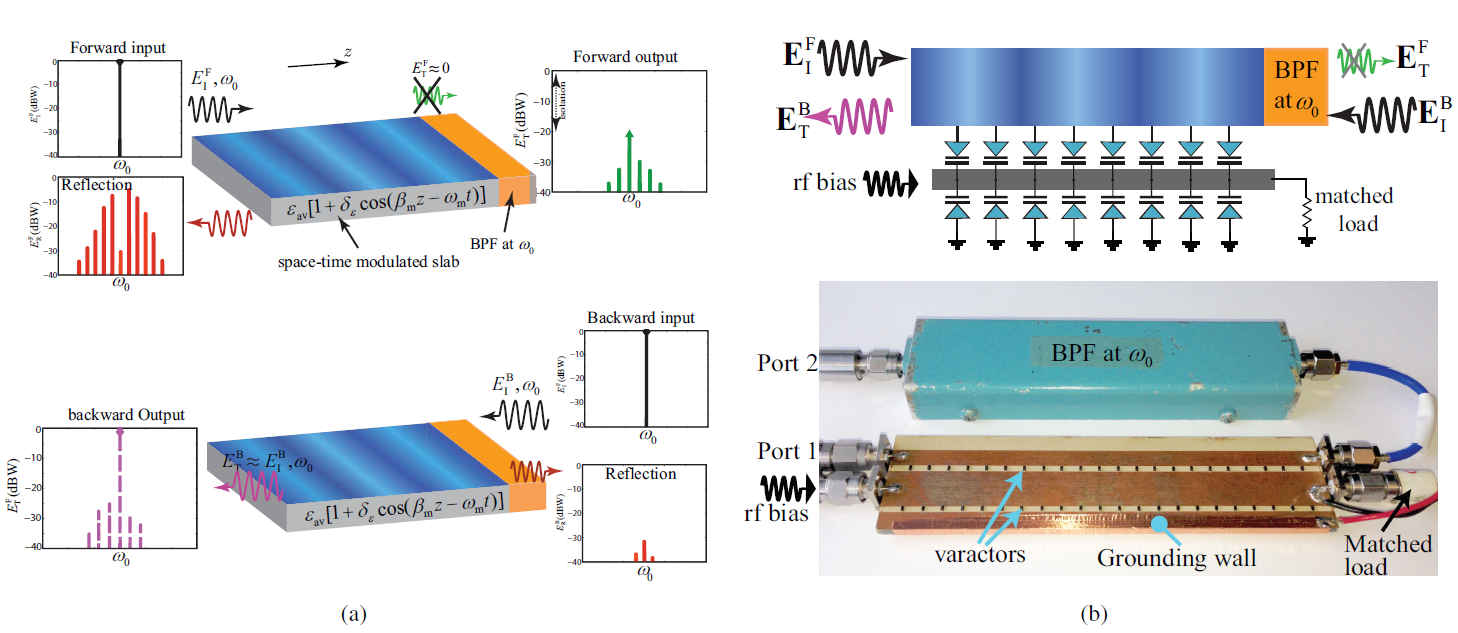} 
		\caption{Quasisonic STM isolator. (a)~Experimental results for the forward and backward problems~\cite{Taravati_PRB_2017}. (b) Circuit model and a photograph of the fabricated isolator~\cite{Taravati_PRB_2017}.}
		\label{Fig:iso}
	\end{center}
\end{figure*}

The strong nonreciprocity of the quasisonic condition is utilized for the realization of a microwave isolator~\cite{Taravati_PRB_2017}. The length ($L$) and the modulation ratio ($\gamma$) are adjusted such that in the forward direction the incident power is efficiently transferred to higher order STHs $\omega_0\pm n \omega_\text{m}$, with $n\ge1$, whereas a small amount of power is transmitted at the fundamental frequency $\omega_0$. In contrast, in the backward direction the STM slab interacts weakly with the incident wave and therefore passes through the slab nearly unaltered. Fig.~\ref{Fig:iso}(a) shows the architecture of the designed quasisonic isolator and the experimental results. In the forward problem, the transmitted wave passes through a bandpass filter (BPF in the figure), with bandpass frequency $\omega_0$, and since most of the power transitted to higher order STHs will be reflected back, as it lies in the the stopband of the filter. However, in the backward direction most of the power is residing at the fundamental frequency $\omega_0$, and hence, passes through the slab with minimal alteration. Thus, the structure operates as an isolator at frequency $\omega_0$. The microwave implementation of this isolator is carried out by emulating the STM medium with the constitutive parameters in~\eqref{eqa:sin_perm}, with $\delta_\mu=0$, by a microstrip transmission line loaded with an array of subwavelength-spaced varactors. The fabricated prototype is shown in Fig.~\ref{Fig:iso}(b), where the varactors are reverse-biased by a DC voltage and are spatiotemporally modulated by an RF bias. Such a structure realizes the STM capacitance
\begin{equation}
C(z,t)= C_\text{av}[1+\delta_C \cos(\beta_\text{m}z-\omega_\text{m}t)].
\end{equation}
where $C_\text{av}$ is proportional to the effective permittivity of the microstrip line and $\delta_C$ corresponds to the average permittivity introduced by varactors. The ST modulation amplitude may be controlled via the amplitude of the modulation RF bias. The experimental results for the electric field amplitude of the incident, transmitted and reflected waves are plotted in Fig.~\ref{Fig:iso}(a). In the forward direction, corresponding to the top of Fig.~\ref{Fig:iso}(a), the transmission level is less than $-20$~dB at the fundamental harmonic, and less than $-30$~dB in other ST harmonics, while the transitted power to higher order harmonics $\omega_0 \pm n\omega_\text{m}$ is reflected by the bandpass filter. In the backward direction, shown in the bottom of Fig.~\ref{Fig:iso}(a), the incident wave is nearly fully transmitted at the fundamental frequency, with a weak transition to higher order STHs with less than $-30$~dB reflection. Hence, the medium provides more than $20$~dB isolation. The insertion loss of the passing direction (backward direction) is about $1.9$~dB, of which $1.2$~dB is due to the bandpass filter insertion loss.

\subsubsection{Isolator Based on Asymmetric Photonic Band Gaps}\label{sec:PBI}
\begin{figure}
	\begin{center}
			\includegraphics[width=1\columnwidth]{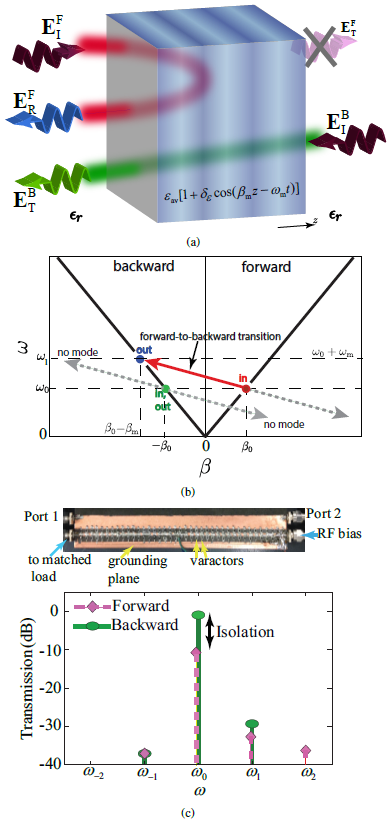}
	\caption{Isolator based on nonreciprocal Bragg reflections in an STM medium with asymmetric electromagnetic band gaps. (a)~Forward problem: propagation of the wave in the electromagnetic band gap of the structure yields complete-reflection. Backward problem: propagation of the wave in the pass-band of the structure yields complete-transmission. (b)~Dispersion diagram~\cite{Chamanara_PRB_2017}. (c)~Photograph of the fabricated structure and the experimental results for nonreciprocal transmission~\cite{Chamanara_PRB_2017}.}
	\label{Fig:asym}
	\end{center}
\end{figure}

Reciprocal media support perfectly horizontal photonic band gaps in the dispersion diagram, which are symmetric with respect to the positive and negative Bloch-Floquet wavenumbers. In the band gaps, the Bloch-Floquet harmonics acquire an imaginary part in their wavenumber and hence become evanescent. Thus, when a wave incident on the structure is modulated at a frequency falling within these band gaps, complex and hence evanescent modes will be excited. In contrast to reciprocal media, STM media provide oblique ST photonic transitions and exhibit asymmetric band gaps~\cite{Taravati_PRB_2017,Chamanara_PRB_2017}. Fig.~\ref{Fig:asym}(a) depicts the wave isolation based on asymmetric photonic band gaps~\cite{Chamanara_PRB_2017}. This approach leverages the ST variation in the permittivity of medium to generate electromagnetic band structures that are asymmetrically aligned with respect to the direction of propagation, as shown in Fig.~\ref{Fig:asym}(b). In the forward problem, the incident wave at the frequency $\omega_0$ excites an evanescent mode, represented by the magenta dot. If the structure is long enough, almost no power reaches the end of it and the wave is fully reflected. In contrast, in the backward problem, the mode marked by the green dot is a propagating mode, which is excited. Therefore, the incident electromagnetic power is transferred to the other side of the structure. Fig.~\ref{Fig:asym}(c) shows an image of the fabricated prototype and the experimental results for the forward and backward transmitted fields. The ST modulation circuit is formed by 39 unit
cells of antiparallel varactors, uniformly distributed along the microstrip line, with the subwavelength period $p = 5$ mm,
corresponding to $p/\lambda_\text{m} \approx  1/19$~\cite{Chamanara_PRB_2017}. Hence, the
structure emulates a medium with the continuous STM permittivity in~\eqref{eqa:Fourier_perm}. The corresponding length at frequency $\omega_0 = 2\pi \times 2.5$ GHz is $L = 6 \lambda_0$, where $\omega_\text{m} = 2\pi \times 0.675$ GHz, $\beta_\text{m} = 415.79$ rad/m, and $\delta_\text{C} =0.15$. The achieved isolation at $\omega_0 = 2\pi \times 2.5$ is more than $10.5$ dB.

\subsubsection{Self-Biased Broadband Isolator Based on Unidirectional ST Coherency}~\label{sec:SBI}
Consider a condition in which the ST modulation and the incident wave share the same temporal frequency $\omega_0$~\cite{Taravati_PRB_SB_2017}. As a consequence, a temporal coherency occurs between them regardless of the direction of the incident wave, and the nonreciprocal operation of the structure is dictated by the spatial coherency difference between the direction of the STM medium and the incident wave. At certain ST modulation phase shifts and ST modulation amplitudes, corresponding to the ST coherency conditions, the structure operates as an isolator. The proposed structure provides broad operation bandwidth and small size, and hence, exhibits superior efficiency compared to previously proposed STM isolators~\cite{Fan_NPH_2009,Fan_PRL_109_2012,Alu_PRB_2015,Fan_APL_2016,Chamanara_PRB_2017,Taravati_PRB_2017}. Such a coherent STM isolator may be realized using a self-biased architecture, where the input signal modulates the structure itself, and it thus operates as a self-biased isolator. Moreover, the proposed isolator is capable of providing transmission gain, as well as introducing nonreciprocal reflection gain. This isolator is formed by a transmission line with the length~$L$ and permittivity
\begin{equation}\label{eq:perm}
\epsilon(z,t)=\epsilon_\text{av}[1+\delta_\epsilon~\cos(\beta_\text{m}z-\omega_0t+\phi_\text{m})].
\end{equation}
\noindent where $\phi_\text{m}$ is the modulation phase, and $\omega_0$ denotes the temporal frequency of both the modulation and incident wave. The closed form solution of the scattered electromagnetic fields may be derived, by considering the transformation $W=\gamma \sqrt{  8\delta_\epsilon}    e^{j(\beta_\text{m} z-\omega_0 t+\phi_\text{m})/2}$ and $t'=t$. Then, the electric field inside the spatiotemporally coherent medium may be expressed in terms of the modified Bessel functions of the first and second kinds~\cite{Taravati_PRB_SB_2017}.

\begin{figure*}
	\begin{center}
		\subfigure[]{\label{Fig:concept_b}
			\includegraphics[width=0.8\columnwidth]{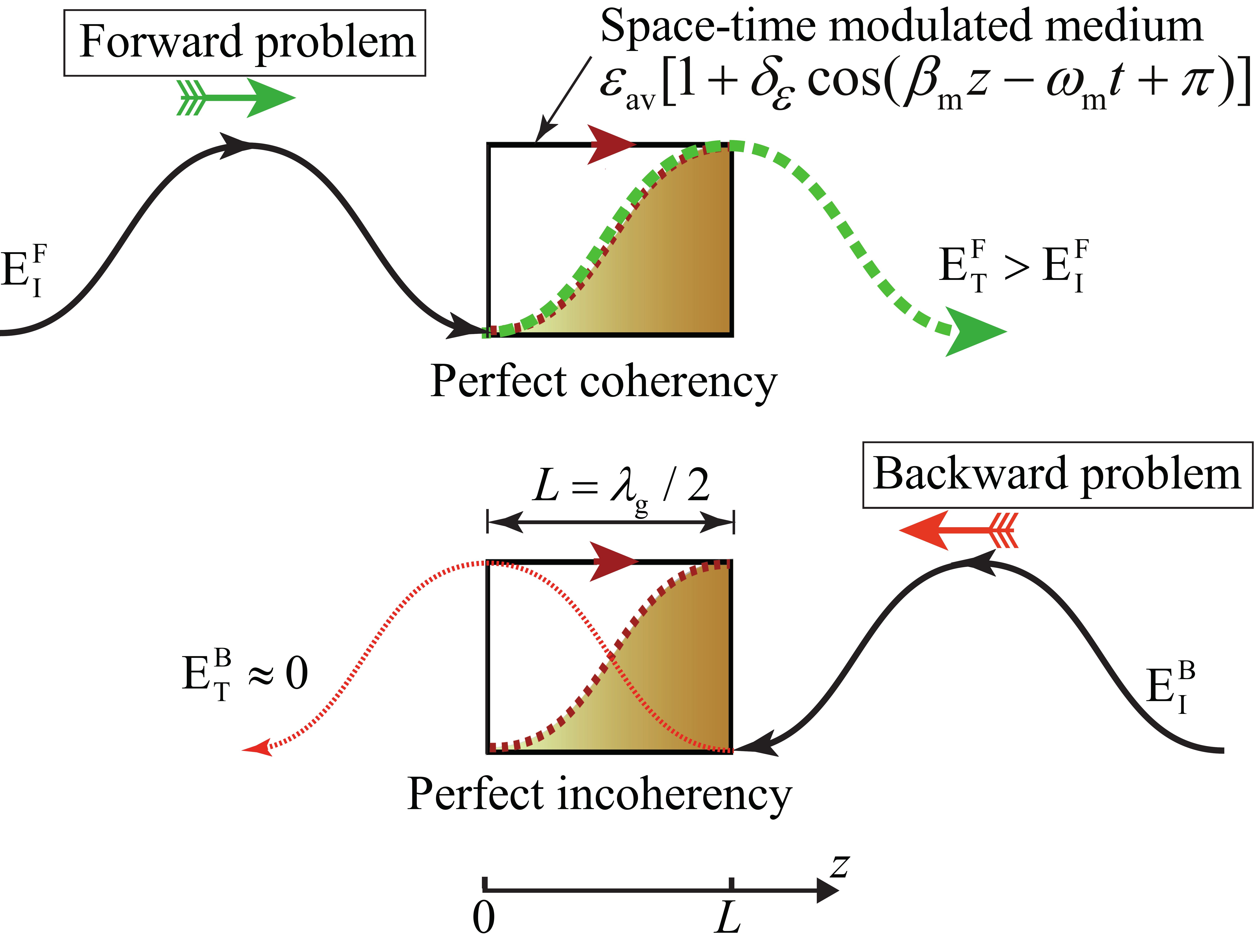} }
		\subfigure[]{\label{Fig:varying_phi_nr_conc}
			\includegraphics[width=0.8\columnwidth]{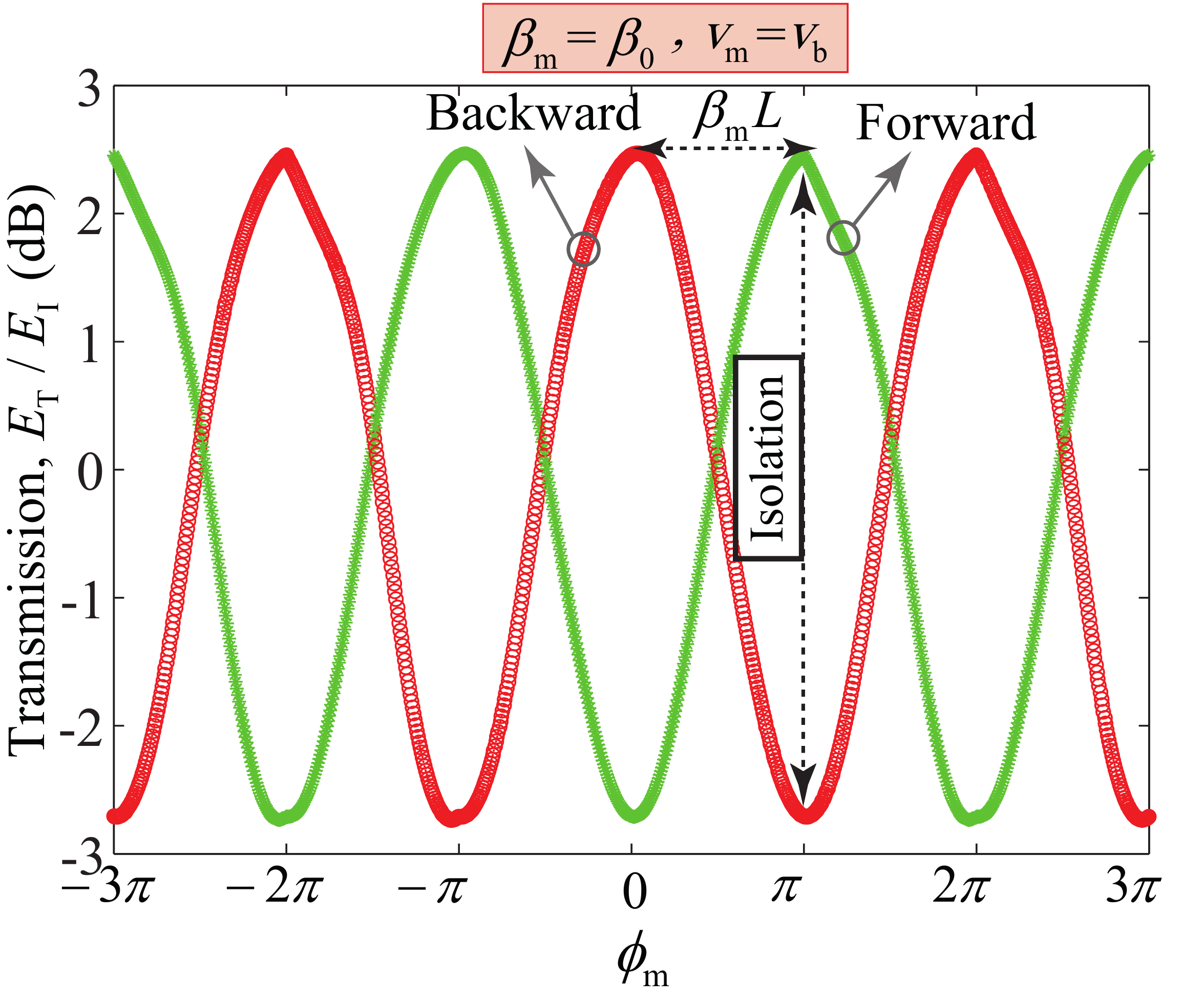}}
		\subfigure[]{\label{Fig:varying_gama}
			\includegraphics[width=0.8\columnwidth]{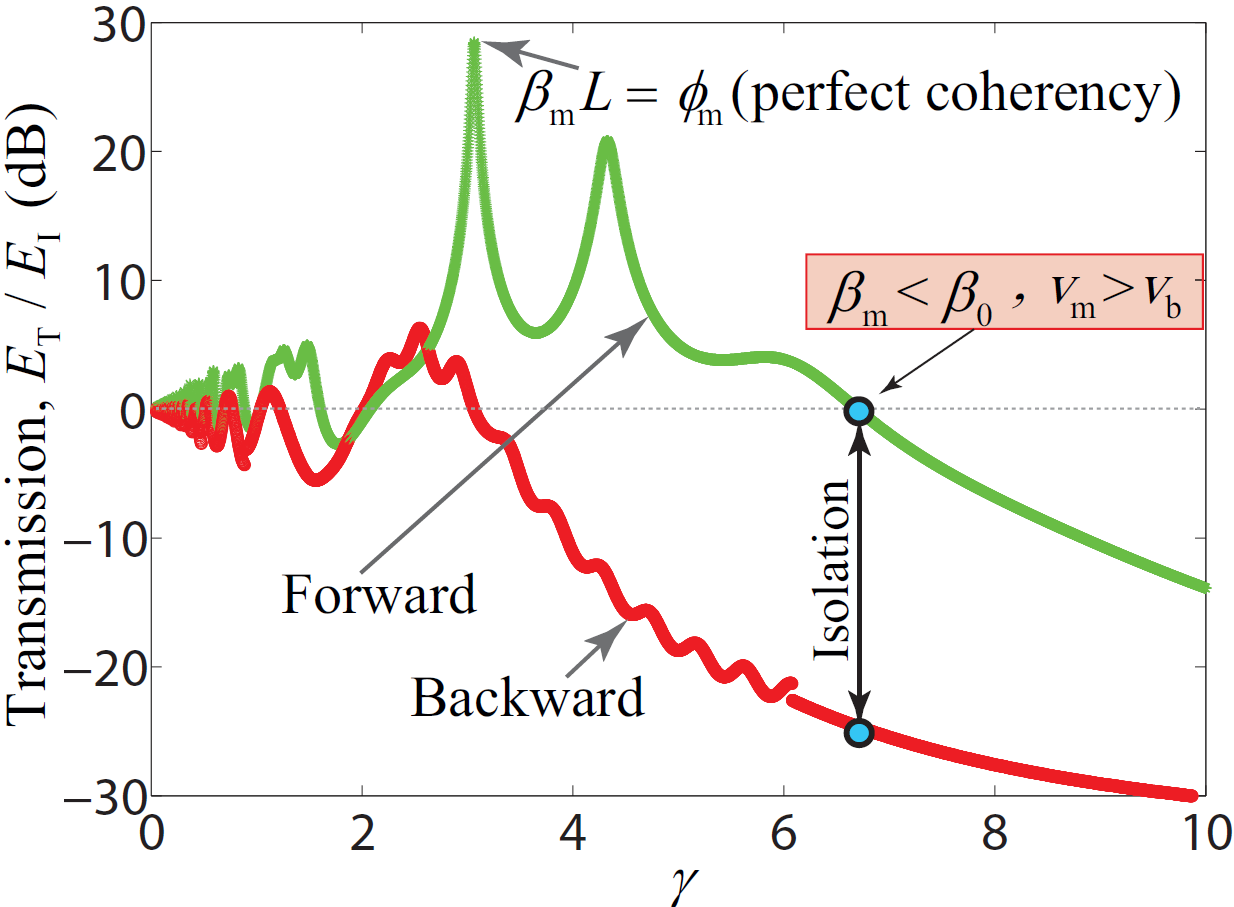} }
			\subfigure[]{\label{Fig:circuit_self}
			\includegraphics[width=1\columnwidth]{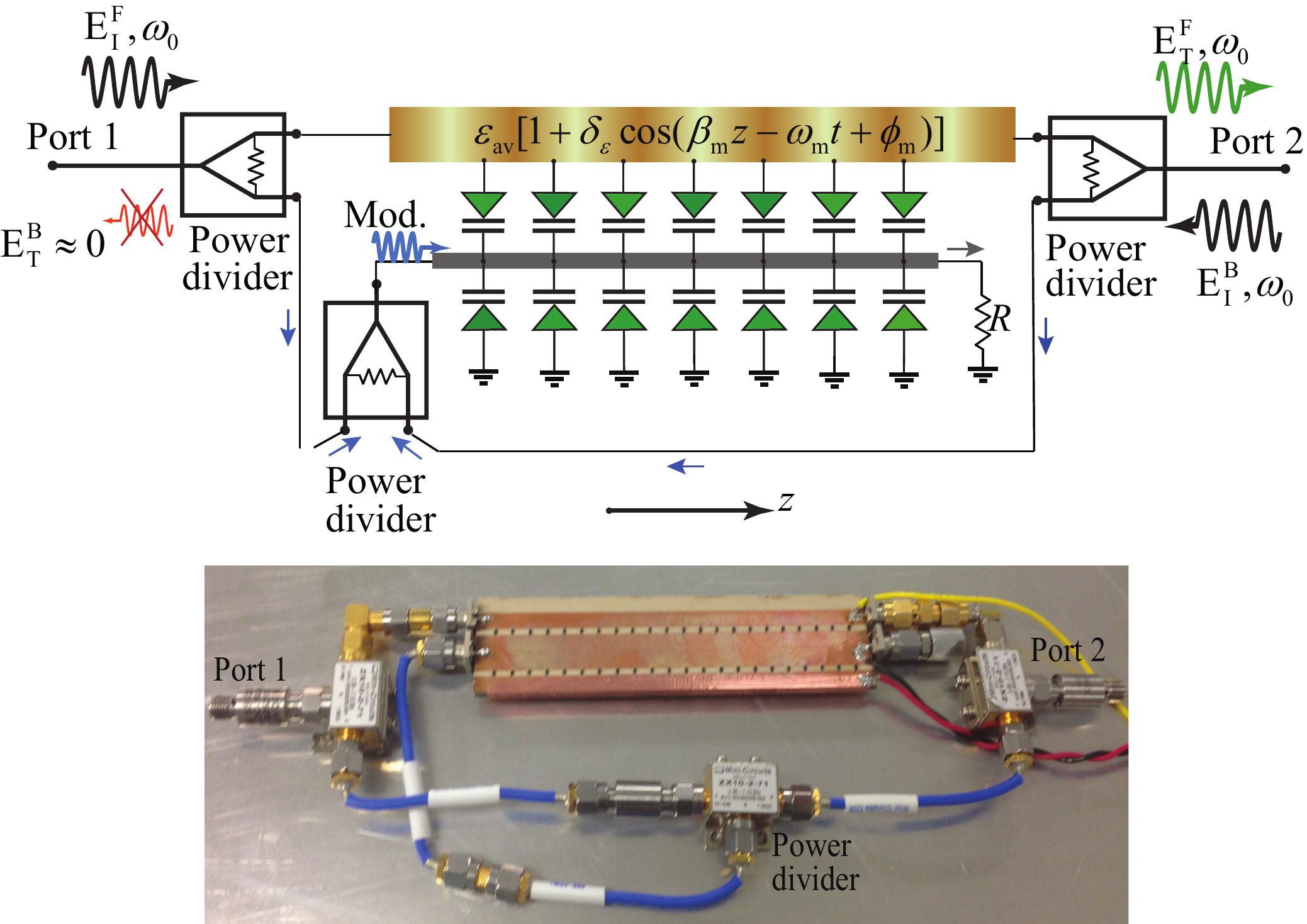}}
		\subfigure[]{\label{Fig:BW}
			\includegraphics[width=0.9\columnwidth]{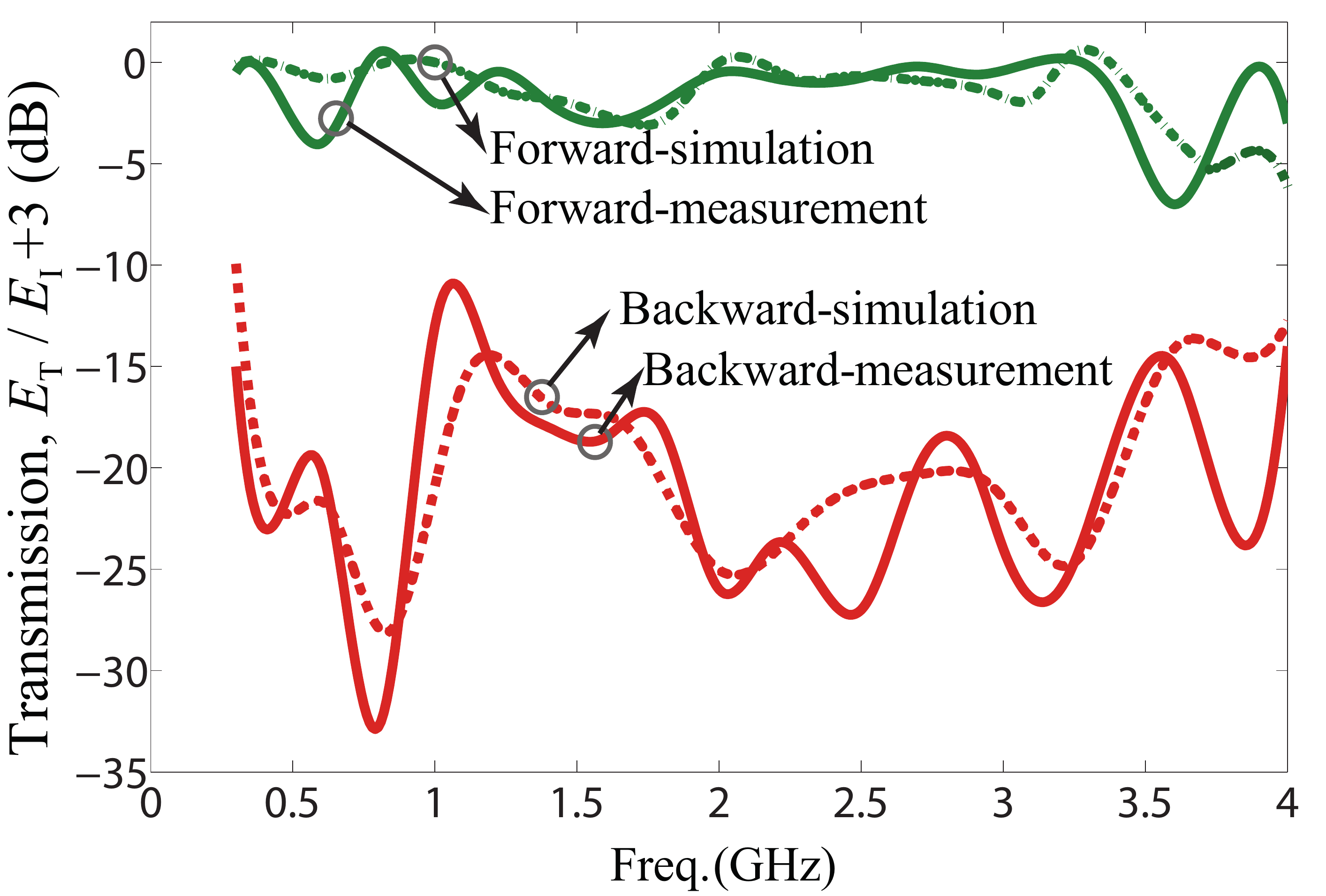} }
		\caption{One-way ST coherency in a STM medium, i.e. $\omega_\text{m}=\omega_{0}$. (a)~Unidirectional perfect space-and-time coherence, i.e., $\beta_\text{m}=\beta_{0}$~\cite{Taravati_PRB_SB_2017}.  
		(b)~Effect of the modulation phase, $\phi_\text{m}$, on the forward and backward transmissions with $\delta_\epsilon=0.15$~\cite{Taravati_PRB_SB_2017}.	(c)~Transmission versus the ST velocity ratio~\cite{Taravati_PRB_SB_2017}. (d)~Circuital schematic and photograph of the fabricated structure~\cite{Taravati_PRB_SB_2017}. (e) Simulation and experimental results~\cite{Taravati_PRB_SB_2017}.}
	\end{center}
\end{figure*}

Fig.~\ref{Fig:concept_b} provides a generic representation of nonreciprocal electromagnetic wave transmission through the unidirectionally perfectly coherent STM medium, with perfect forward coherency and perfect backward incoherency. Fig.~\ref{Fig:varying_phi_nr_conc} plots the closed-form solution results for the forward and backward transmissions versus the modulation phase $\phi_\text{m}$ for $\beta_\text{m}=\beta_0$, $\delta_\epsilon=0.15$, $\epsilon_\text{av}=7.06$, $\omega_0=2\pi\times2$~GHz and with $L=2\lambda_\text{g}$ yielding a maximal contrast between the forward and backward transmissions. However, to achieve a stronger isolation level and arbitrary transmission gain, we consider imperfect space-coherency, where $\beta_\text{m} \neq \beta_0$ ($\gamma\neq 1$), and perfect temporal coherency. Fig.~\ref{Fig:varying_gama} shows that, forward transmission of $E_\text{T}^\text{F}/E_\text{I}=0$ dB with the isolation level of $24.5$ dB may be achieved for $\gamma=6.66$. A forward transmission peak is observed around $\gamma=3.063$ which corresponds to the perfect spatial coherency of the forward transmitted wave, where $\beta_\text{m} L=\phi_\text{m}$. Moreover, a reciprocal transmission occurs at $\gamma \rightarrow \infty$ ($\beta_\text{m} \rightarrow 0$) and $\gamma \rightarrow 0$ ($\omega_\text{0,m}\rightarrow0$)~\cite{Taravati_PRB_2017}, where the structure is undirected. Fig.~\ref{Fig:circuit_self} shows the architecture and an image of the fabricated self-biased isolator using three Wilkinson power dividers, extracting the ST modulation wave from the input wave. Fig.~\ref{Fig:BW} plots the experimental and simulation results for the forward and backward transmissions through the spatiotemporally coherent isolator, where more than $8$~dB isolation (with linear phase response) is achieved across $122\%$ fractional bandwidth. To achieve a higher isolation level and stronger transmitted signal, one may enhance the ST modulation amplitude $\delta_\epsilon$ by amplifying the extracted modulation wave using a transistor-based amplifier.
We shall stress that this device provides the same linear response for a modulated or non-sinusoidal input signal. The dynamic range of the isolator, for more than 15 dB isolation, is about 30 dB, i.e., from $-10$ dBm to $+20$ dBm. There is a danger that the proposed solution for amplifying the extracted modulation using a transistor-based amplifier, permitting operation with small input-signals, could compress the varactors and restrict the maximum power handling and linearity. This could be  prevented by using an automatic gain control (AGC) that controls the gain of the amplifier.
\subsection{STM Magnet-Free Circulator}\label{sec:Circ}
\begin{figure*}
	\begin{center}
        \subfigure[]{\label{Fig:circ_gen}
			\includegraphics[width=0.9\columnwidth]{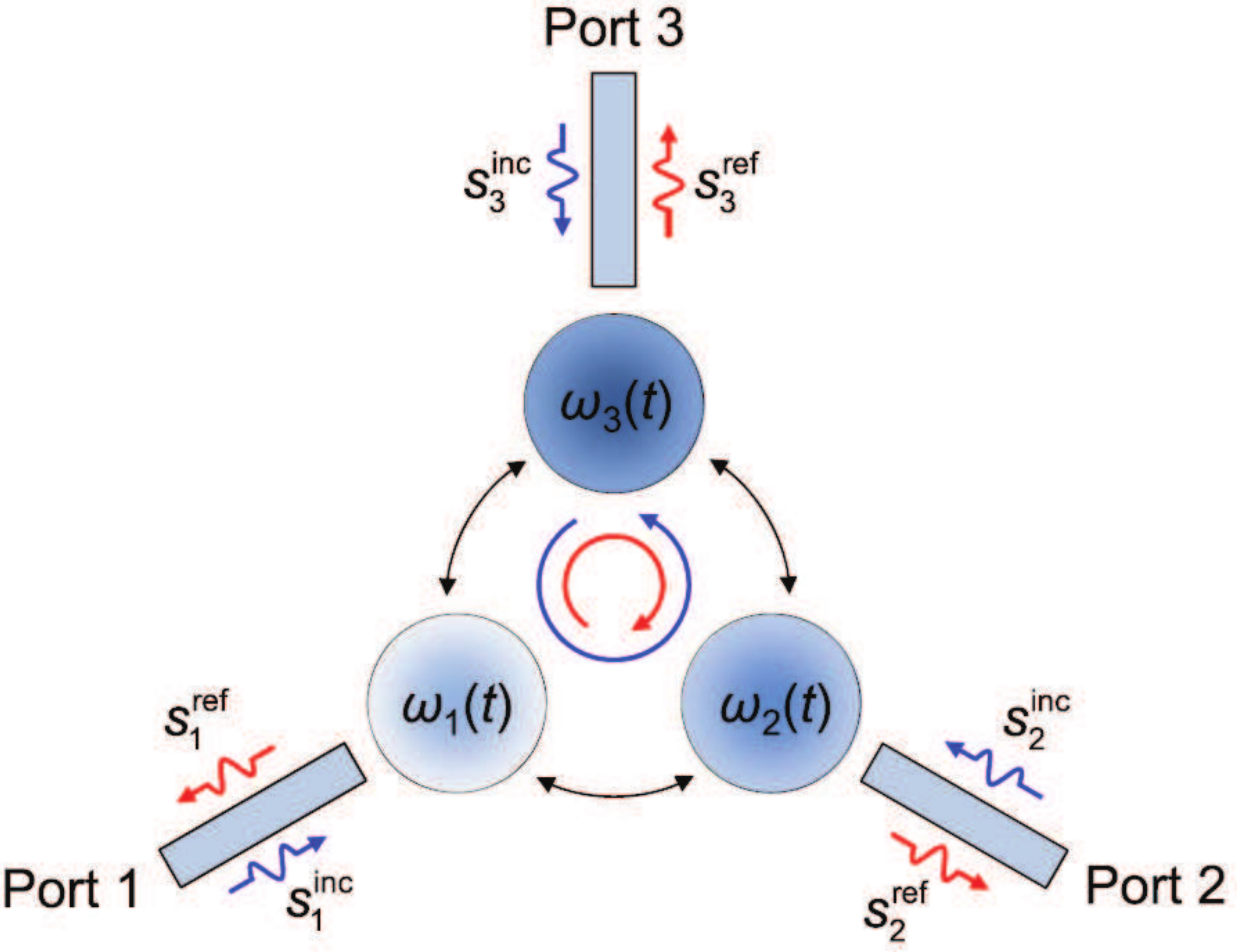} }
		\subfigure[]{\label{Fig:circ_circ}
			\includegraphics[width=1\columnwidth]{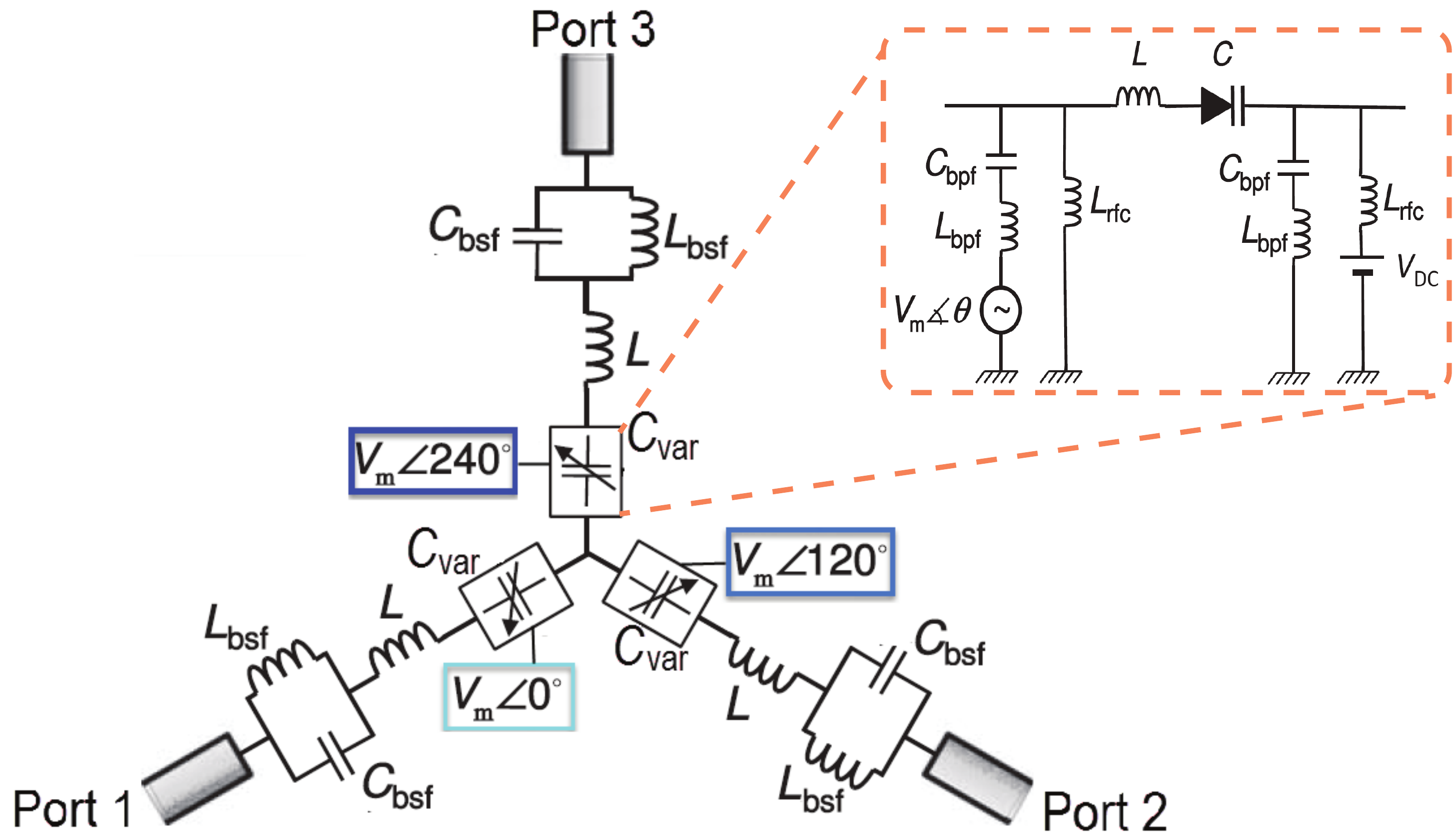}}
		\subfigure[]{\label{Fig:circ_photo}
			\includegraphics[width=0.6\columnwidth]{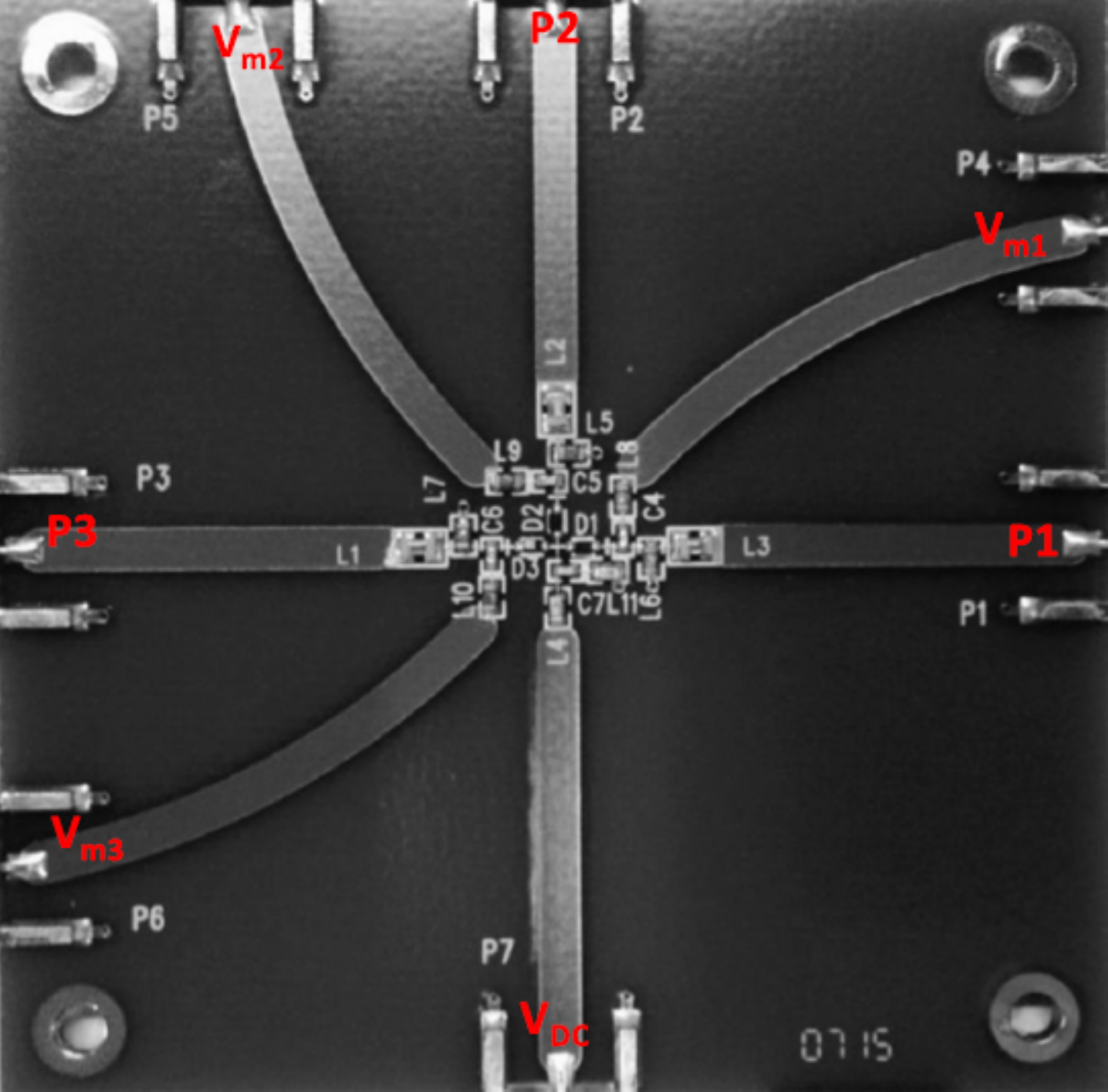} }
		\subfigure[]{\label{Fig:circ_res}
			\includegraphics[width=0.8\columnwidth]{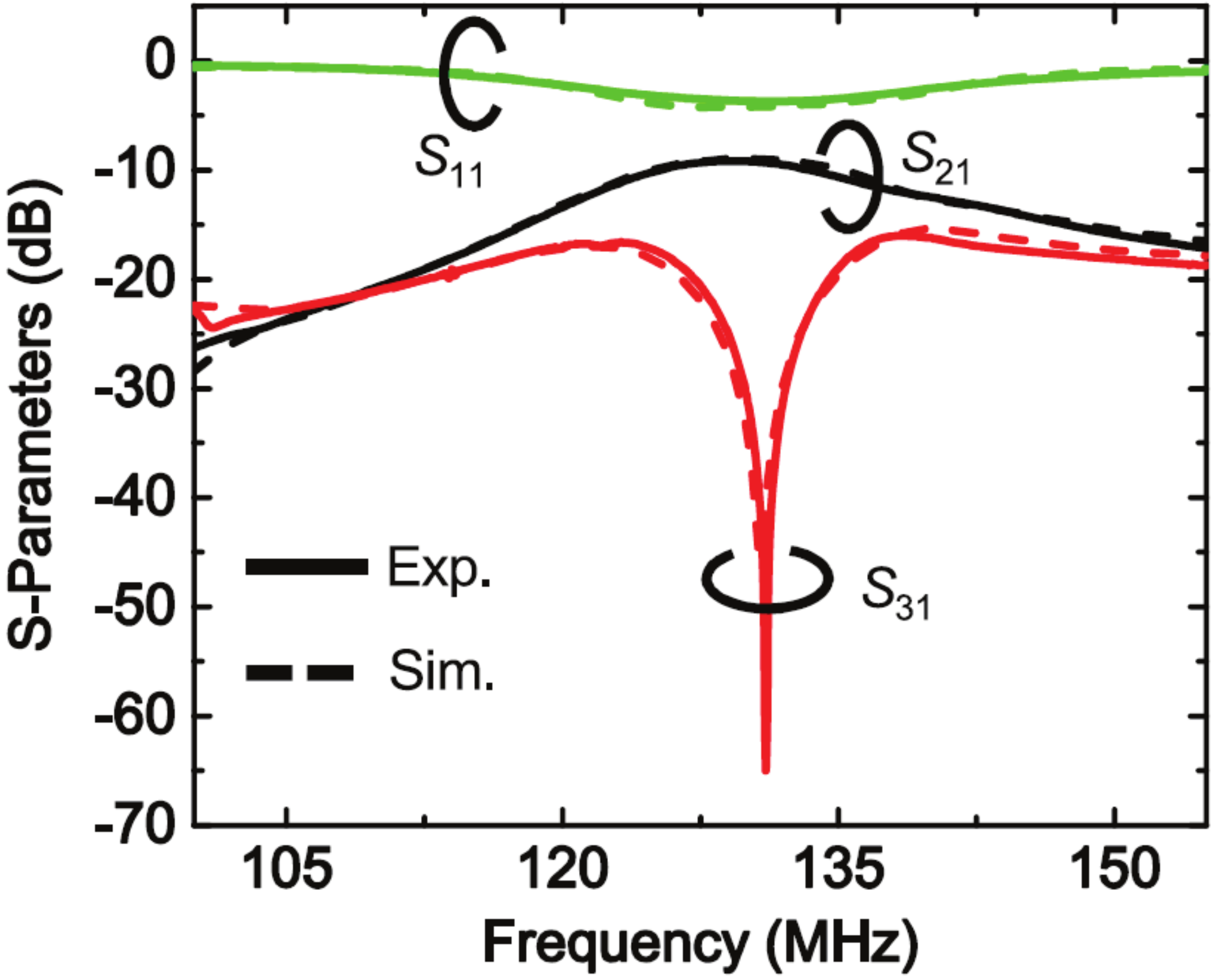}} 
		\caption{Magnetless circulator based on time-modulated coupled resonators. (a) Generic schematic of the circulator~\cite{estep2016magnetless}. (b) Realization using lumped-element varactor-based wye resonator. The leakage of the modulation signal to the external ports is prevented using three parallel bandstop filters ($L_\text{bsf}$ and $C_\text{bsf}$)~\cite{estep2016magnetless}. (c) Photograph of the fabricated prototype~\cite{estep2016magnetless}. (d) Experimental and numerical simulation results~\cite{estep2016magnetless}.} 
	\end{center}
\end{figure*}

Circulators are nonreciprocal three-port devices which are the fundamental units of all full-duplex communication systems. Assuming a clockwise circulator, the signal entering into it (port 1) exits from the next port in a clockwise direction (port 2), where port 3 is isolated. However, if some of the exited light is reflected back to the circulator, it come out of port $3$, where port 1 is isolated. Recently, there has been substantial scientific interest in wave circulation based on ST modulation~\cite{Wang_TMTT_2014,Alu_NP_2014,estep2016magnetless,reiskarimian2017cmos,reiskarimian2018integrated}. STM circulators exhibit superior performance over conventional magnetically biased ferromagnetic circulators and active circulators. An elegant approach for the realization of an STM microwave circulator would consist of STM coupled-resonators, where the effective spin is imparted to the structure by the ST modulation~\cite{Alu_NP_2014,estep2016magnetless}. Such a structure is compatible with integrated circuits and could potentially exhibit good power and noise performance. Fig.~\ref{Fig:circ_gen} shows a generic representation of the STM circulator presented in~\cite{estep2016magnetless}, formed by three identical time-modulated resonators symmetrically coupled to three transmission lines. These three resonators are time-modulated so that the time-dependent resonance frequency of the $i$th resonator reads
\begin{equation}
\omega_i(t)=\omega_0+\delta\omega_\text{m} \cos\left(\omega_\text{m} t+(i-1) \frac{2\pi}{3} \right), \qquad i=1,2,3
\label{eqa:circ_reson}
\end{equation}
where $\omega_0$ is the resonance frequency of the unnmodulated resonators, $\omega_\text{m}$ is the modulation frequency and $\delta\omega_\text{m}$ is the amplitude of the resonance-frequency perturbation~\cite{estep2016magnetless}. Therefore, three identical resonators are time-modulated with the same frequency and amplitude, but phases of $0^\circ$, $120^\circ$ and $240^\circ$. In the absence of the time modulation, the loop of resonators supports two degenerate counter-rotating modes possessing phases $0^\circ$, $\pm 120^\circ$ and $\pm240^\circ$ in the three resonators. The time modulation of resonators results in right-handed and left-handed states, so that exciting the circulator from, for instance, port 1 yields excitation of the right- and left-handed states with the same amplitude and opposite
phases $\phi_\text{R}=-\phi_\text{L}$. It can be shown that by exciting port 1, the signal at port 2 reads $e^{i2\pi/3} e^{i \phi_\text{R}}+e^{-i2\pi/3} e^{i\phi_\text{L}}$ and the signal at port 3 is $e^{i4\pi/3} e^{i\phi_\text{R}}+e^{-i4\pi/3} e^{i\phi_\text{L}}$~\cite{Alu_NP_2014,estep2016magnetless}. Hence, by choosing $\phi_\text{R}=\pi/6$ and $\phi_\text{L}=-\pi/6$ a constructive interference at port 2 and destructive interference at port 3 is achieved, yielding an appropriate signal circulation. Fig.~\ref{Fig:circ_circ} shows the experimental realization of this STM circulator at microwave frequencies based on a loop of lumped LC resonators, varactors and wye resonators. Fig.~\ref{Fig:circ_photo} shows a photograph of the fabricated prototype, and Fig.~\ref{Fig:circ_res} plots the simulation and experimental results. The microwave version of this circulator may be realized using distributed elements. In addition,~\cite{estep2016magnetless} presents a microwave version of the proposed circulator based on distributed elements for wireless-communications band $2.2$ GHz, where theoretical and simulation results are presented.
\subsection{STM Transceiver Front-Ends}\label{sec:LWA}
Thanks to the extraordinary properties of STM media, various multifunctionality components have been recently reported. These components are appropriate for the front-end of RF communication systems, where a single STM structure simultaneously may provide the tasks of frequency up-/down-conversion, duplexing, antenna and filtering process.
\subsubsection{Broadband low-loss nonreciprocal component}\label{sec:WANG}
\begin{figure*}
	\centering
			\centering{\includegraphics[width=2\columnwidth] {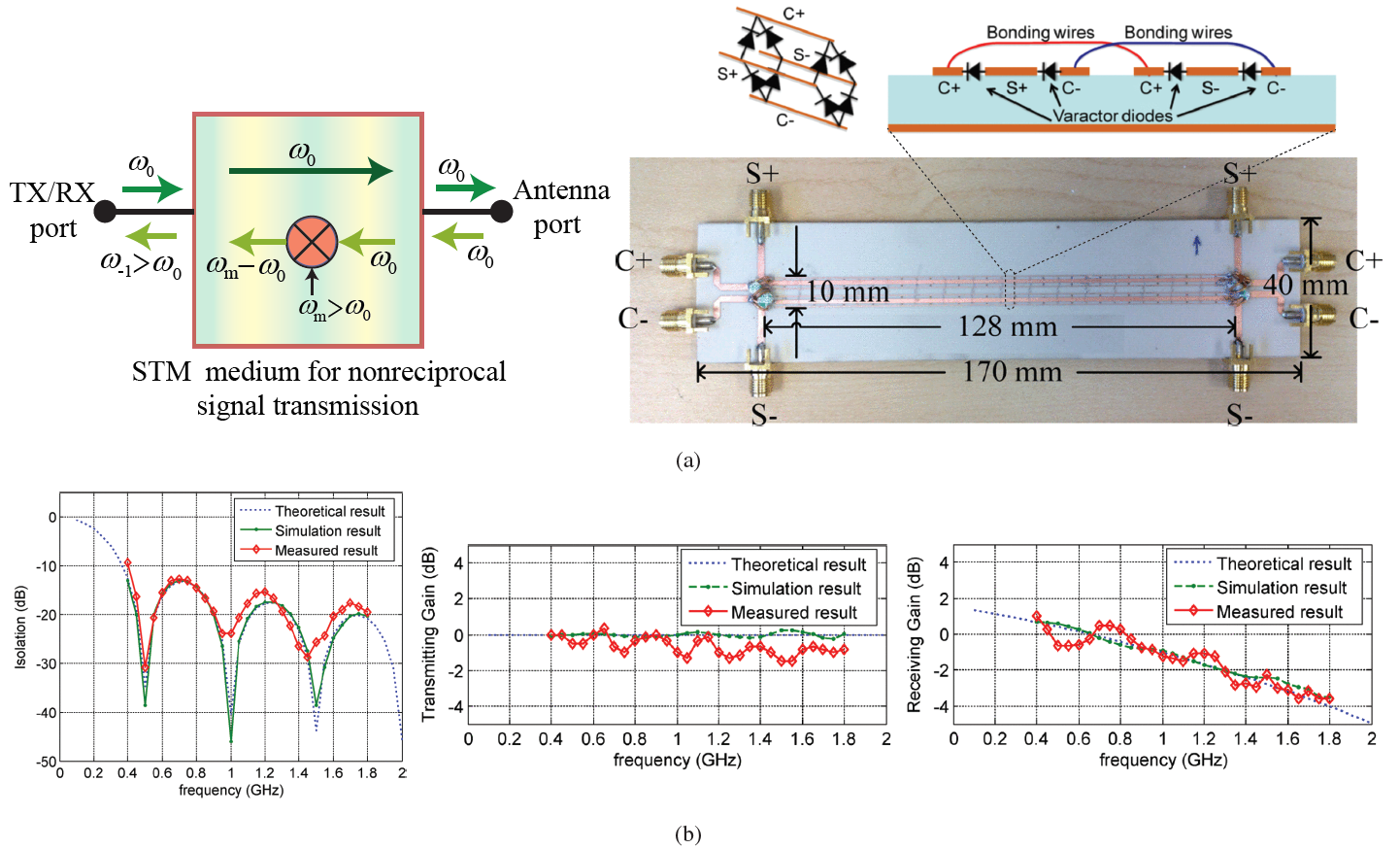}}  
	\caption{ST modulation yielding broadband and low-loss signal circulation. (a) Circuital representation and an image of the fabricated prototype. In the down-link, up-conversion from $\omega_0$ to $\omega_{-1}=\omega_\text{m}-\omega_0$ occurs, so that additional down-conversion to $\omega_0$ using an external mixer is required~\cite{qin2014nonreciprocal}. (b) Theoretical, simulation and experimental results for isolation ($S_{31}$), transmission gain ($S_{21}$), and receive gain ($S_{32}$)~\cite{qin2014nonreciprocal}.}
	\label{Fig:Wang}
\end{figure*}

Fig.~\ref{Fig:Wang}(a)~\cite{qin2014nonreciprocal} shows an STM structure which is designed for broadband and low-loss nonreciprocal signal transmission and reception. The operation of this device is as follows. The temporal frequency of the ST modulation is larger than that of the input signal, i.e., $\omega_\text{m}>\omega_0$. In the transmit mode (left-to-right, backward problem), the input signal at $\omega_0$ passes through the structure with nearly no alteration in the frequency and amplitude. However, in the receive mode (right-to-left, forward problem), the incoming signal from the antenna at $\omega_0$ is up-converted to $\omega_{-1}=\omega_\text{m}-\omega_0$. Hence, this device can be used as a circulator if the TX/RX port is connected to a diplexer, with two pass bands at $\omega_0$ and $\omega_{-1}$ separating the TX and RX paths. In addition, to cancel out the up-conversion in the down-link, an external mixer may be used in the RX port side for down-conversion from $\omega_{-1}$ to $\omega_0$. The complete analysis of the structure and its noise performance is provided in~\cite{qin2014nonreciprocal}.\\
\indent This nonreciprocal component is designed at microwave frequencies using two pairs of differential transmission lines loaded with reverse-biased varactors in a double-balanced fashion. They support separate traveling of the modulation signal and the main signal, where $C+$ and $C-$ denote the modulation (carrier) transmission lines, and $S+$ and $S-$ represent the main signal transmission lines. Due to
the symmetry, these two transmission-line pairs are isolated from each other, where each transmission line pair is on the virtual ground of the other one~\cite{qin2014nonreciprocal}. Fig.~\ref{Fig:Wang}(b) plots the theoretical, simulation and experimental results for the isolation ($S_{31}$), transmission gain ($S_{21}$), and receive gain ($S_{32}$)~\cite{qin2014nonreciprocal} provided by the nonreciprocal component in Fig.~\ref{Fig:Wang}(a). The results demonstrate a broadband operation, that is, from $\omega_0=2\pi\times 0.4$~GHz to $2\pi\times 1.8$~GHz.
\subsubsection{Nonreciprocal transmitter}\label{sec:ALU}
\begin{figure*}
	\centering
		\subfigure[]{\label{Fig:Alu}
			\centering{\includegraphics[width=1.9\columnwidth] {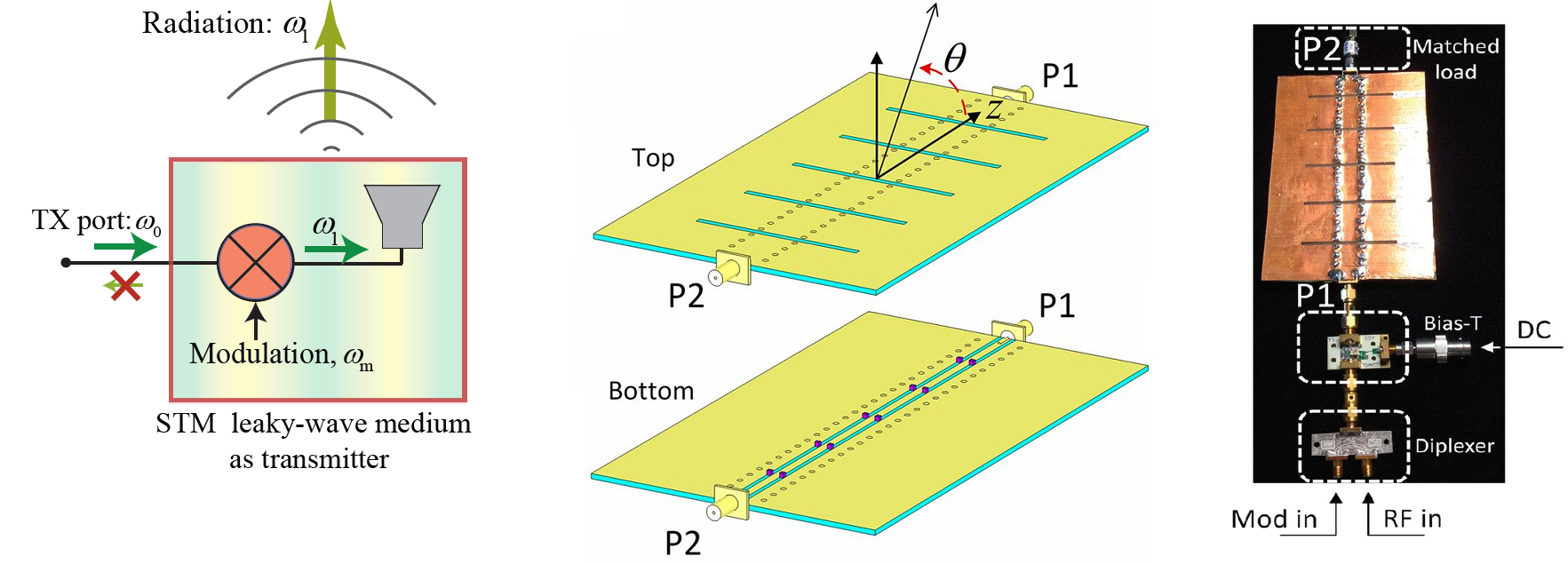}}  }  
		\subfigure[]{\label{Fig:Alu_res}
			\centering{\includegraphics[width=1.9\columnwidth] {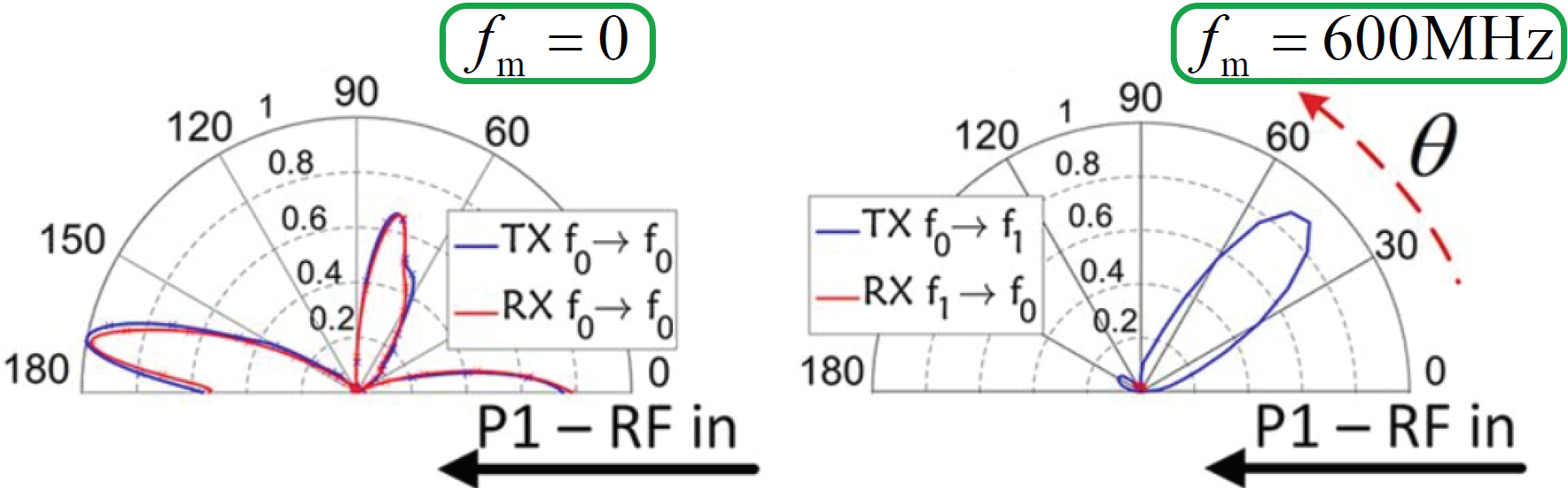}}  }  
	\caption{STM leaky-wave medium casting as a nonreciprocal mixer-antenna. (a)~Circuit representation and an image of the fabricated prototype~\cite{Alu_PNAS_2016}. (b)~Experimental results for (left-side) Unmodulated structure, $f_\text{m}=0$, and (right-side) Modulated structure, $f_\text{m}=600$~MHz~\cite{Alu_PNAS_2016}.}
	\label{Fig:***}
\end{figure*}

\indent The ST modulation may be realized in conjugation with the leaky-wave modes. As a consequence, in addition to the frequency mixing and nonreciprocal wave transmission, a leaky-wave radiation will be enabled, leading to  nonreciprocal mixer-antenna structures~\cite{Taravati_APS_2015,Alu_PNAS_2016,Taravati_TAP_65_2017}. The ST modulation is well-integrated with the leaky-wave structures since both the ST modulation and leaky-wave radiation are based on progressive change in the traveling wave. In particular, the STM leaky-wave structure has been proposed for the front-end of transmitters providing frequency up-conversion and radiation~\cite{Taravati_APS_2015,Alu_PNAS_2016}. Fig.~\ref{Fig:Alu} depicts the application of the STM leaky-wave medium to the front-end of a transmitter~\cite{Alu_PNAS_2016}, where a desired up-conversion and leaky-wave radiation is achieved in the transmit mode. In this scheme, the radiation is enabled by periodic interruption (by thin radiation apertures slots) in a coplanar waveguide (CPW). The voltage-dependent capacitors are placed right below the slots to control the
propagation phase. The modulation control is achieved using a diplexer that combines the main RF signal and the modulation signal in a single port connected to the antenna via a Bias-T to superimpose the DC bias~\cite{Alu_PNAS_2016}.\\
\subsubsection{Full-duplex transceiver}\label{sec:TARAVATI}
\begin{figure*}
	\centering
		\includegraphics[width=2\columnwidth]{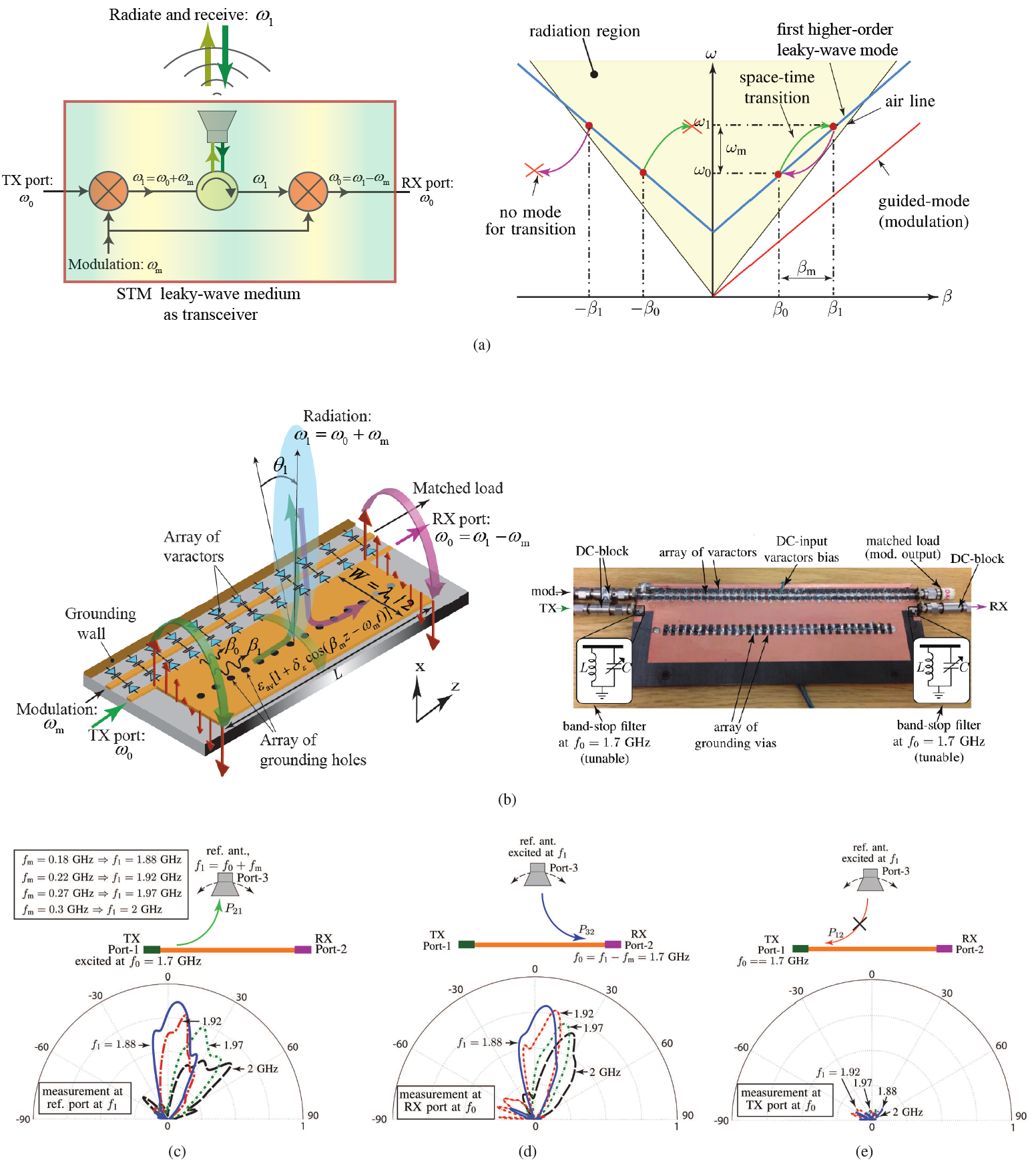}
	\caption{STM mixer-duplexer-antenna leaky-wave system. (a)~STM leaky-wave medium as a full transceiver composed of up-converter, duplexer, leaky-wave antenna and down-converter~\cite{Taravati_TAP_65_2017}. (b)~Realization of the space-time modulated mixer-duplexer-antenna system based on the half-wavelength microstrip leaky-wave antenna~\cite{Taravati_TAP_65_2017}. (c),(d),(e)~Experimental results for the scanning at $\theta_1=\{4, 11.5, 18, 24.5\}^\circ$ corresponding to $f_\text{1}=\{1.88, 1.92, 1.97, 2\}$~GHz, i.e., (c)~up-link radiated power, (d)~down-link received power at the RX port, and (e)~down-link received power at the TX port~\cite{Taravati_TAP_65_2017}.}
	\label{Fig:ST_LW_med}
\end{figure*}

\indent The ST modulation has been integrated with the leaky-wave medium in~\cite{Taravati_TAP_65_2017}, for the realization of a full transceiver front-end, composed of up-converter, duplexer, leaky-wave antenna and down-converter. Fig.~\ref{Fig:ST_LW_med}(a) shows the operation principle of the STM mixer-duplexer-antenna leaky-wave structure. Such a structure operates based on the up-link and down-link ST transitions (shown in Fig.~\ref{Fig:ST_LW_med}(a)), as up-conversion and down-conversion mixing operations, and separation of the up-link and down-link paths as a duplexing operation. Moreover, it provides ST frequency beam scanning at fixed input frequency by virtue of the modulation parameters (here the modulation frequency). In the up-link, a signal with frequency $\omega_\text{0}$ is injected into the transmitter port and propagates in the $+z$ direction, corresponding to the right-hand side of the dispersion diagram in Fig.~\ref{Fig:ST_LW_med}(a). As a result of mixing with the periodic modulation, this signal experiences progressive temporal frequency up-conversion from $\omega_\text{0}$ to $\omega_\text{1}=\omega_\text{0}+\omega_\text{m}$ along with spatial frequency transition from $\beta_\text{0}$ to $\beta_\text{1}=\beta_\text{0}+\beta_\text{m}$. 

In the down-link, the signal with frequency $\omega_\text{1}$ is impinging on the structure under the same angle $\theta_1$ and picked up by the structure. In conformity with the phase matching, the signal can only propagate in the direction of the modulation, i.e., in the $+z$ direction, corresponding to the right-hand side of the dispersion diagram in Fig.~\ref{Fig:ST_LW_med}(a), and experiences progressive spatial and temporal frequency transitions (down-conversion), from $\beta_\text{1}$ to $\beta_\text{0}=\beta_\text{1}-\beta_\text{m}$ and from $\omega_\text{1}$ to $\omega_\text{0}=\omega_\text{1}-\omega_\text{m}$, respectively, while propagating toward the receiver port. No ST transition is allowed for signal propagation in the backward ($-z$) direction (left-hand side of the dispersion diagram in Fig.~\ref{Fig:ST_LW_med}(a)) due to the unavailability of leaky modes at the points $(-\beta_0+\beta_\text{m},\omega_0+\omega_\text{m})$ from $(-\beta_0,\omega_0)$ for up-conversion and $(-\beta_1-\beta_\text{m},\omega_0)$ from $(-\beta_1,\omega_0+\omega_m)$ for down-conversion. The radiation angle (for both transmit and receive at frequency $\omega_1$) is denoted by $\theta_\text{1}$, which is given by~\cite{Taravati_TAP_65_2017} 
\begin{equation}
\theta_\text{1}=\sin^{-1} \left( \frac{ \beta_1(\omega)}{ k_{01}} \right)=\sin^{-1} \left( \frac{ c(\beta_0+\beta_\text{m})}{\omega_\text{0}+\omega_\text{m}} \right),
\label{eqa:disp_unm}
\end{equation}
where $k_{01}=\omega_1/c$ is the effective wavenumber at the frequency $\omega_1$. According to Eq.~\eqref{eqa:disp_unm}, the radiation angle depends on both input and modulation frequencies, $\omega_\text{0}$ and $\omega_\text{m}$. Hence, we take advantage of this property and achieve space scanning by varying the modulation frequency, i.e., $\theta_\text{1} \left(\omega_\text{m} \right)$, where the input frequency $\omega_\text{0}$ is fixed. Beam scanning can be achieved at a fixed input frequency, $\omega_\text{0}$, simply by varying the modulation frequency, $\omega_\text{m}$.

Fig.~\ref{Fig:ST_LW_med}(b) shows the details of the fabricated prototype. The antenna system supports two modes, i.e., the narrow strip corresponds to a dominant even mode, which provides a guided-mode path for the RF signal modulating the varactors, and the wide strip which represents a higher order odd ($EH_1$) mode, providing the leaky-mode channel for radiation. Fig.~\ref{Fig:ST_LW_med}(c) plots the experimental result for the normalized radiated power of the up-link frequency beam scanning. The input frequency is set at $f_\text{0}=1.7$~GHz, where the modulation frequency set reads $f_\text{m}=\{0.18, 0.22, 0.27, 0.3\}$~GHz, corresponding to radiation frequencies $f_\text{1}=\{1.88, 1.92, 1.97, 2\}$~GHz and radiation beam angles of $\theta_1=\{4, 11.5, 18, 24.5\}^\circ$. Figs.~\ref{Fig:ST_LW_med}(d) and~\ref{Fig:ST_LW_med}(e) plot the experimental results for the normalized received power at the receive and transmit ports for down-link frequency beam scanning. The input frequency at the reference antenna port varies as $f_\text{1}=\{1.88, 1.92, 1.97, 2\}$~GHz, corresponding to $f_\text{m}=\{0.18, 0.22, 0.27, 0.3\}$~GHz, for a fixed received signal $f_\text{0}=f_1-f_\text{m}=1.7$~GHz. The isolation achieved at specified radiation angles corresponds to $\{31.5, 12.3, 9.2, 15.5\}$~dB.
\subsection{Pure Frequency Mixer Based on Aperiodic ST modulation }\label{sec:PFM}
Frequency mixing is the key operation in telecommunication circuits and systems which is conventionally achieved by nonlinear components, e.g., diodes and transistors, or by linear time-varying circuits where the nonlinearity is due to the parasitic effects of the diode or switch elements. The nonlinear response of these nonlinear components results in generation of an infinite number of undesired mixing products. Periodic ST modulation provides the required energy and momentum for transition from the fundamental temporal frequency to an infinite number of ST frequency harmonics~\cite{Oliner_PIEEE_1963,Taravati_PRB_2017,Taravati_thesis}. A general technique may be used for pure frequency mixing in STM media, where harmonic photonic transitions in temporally periodic systems are prohibited by tailored photonic band gaps introduced by the engineered spatial-aperiodicity of the structure. In contrast to conventional periodic uniform STM media, here the medium is aperiodic in space and therefore exhibits an aperiodic dispersion diagram and photonic band gaps tailored in a way to provide photonic band gaps at undesired ST mixing products, and, hence, introduces a pure frequency mixing. In addition, the frequency bands of the mixer may be tuned via ST modulation parameters. The proposed mixer takes advantage of the aperiodic spatially varying structures which are capable of providing a variety of electromagnetic responses that are unattainable by uniform structures. Such structures may be used for the realization of highly efficient electromagnetic systems~\cite{Taravati_IET_2012,Zutter_1992,Schutt_1992,Plant_MWCL_2007,Taravati_IET_7_2013,Shikova_2004,Taravati_IJRMCAE_2013,Laso_MWCL_2011,Taravati_MWCL_2016,Khalaj_PIER_2006,Taravati_JEMWA_2012,Laso_MTT_60_2012,Taravati_JEMWA_BLC_2012,Plant_TMTT_2007,Taravati_IJRMCAE_23_2013}. 

\begin{figure*}
	\begin{center}
			\includegraphics[width=2\columnwidth]{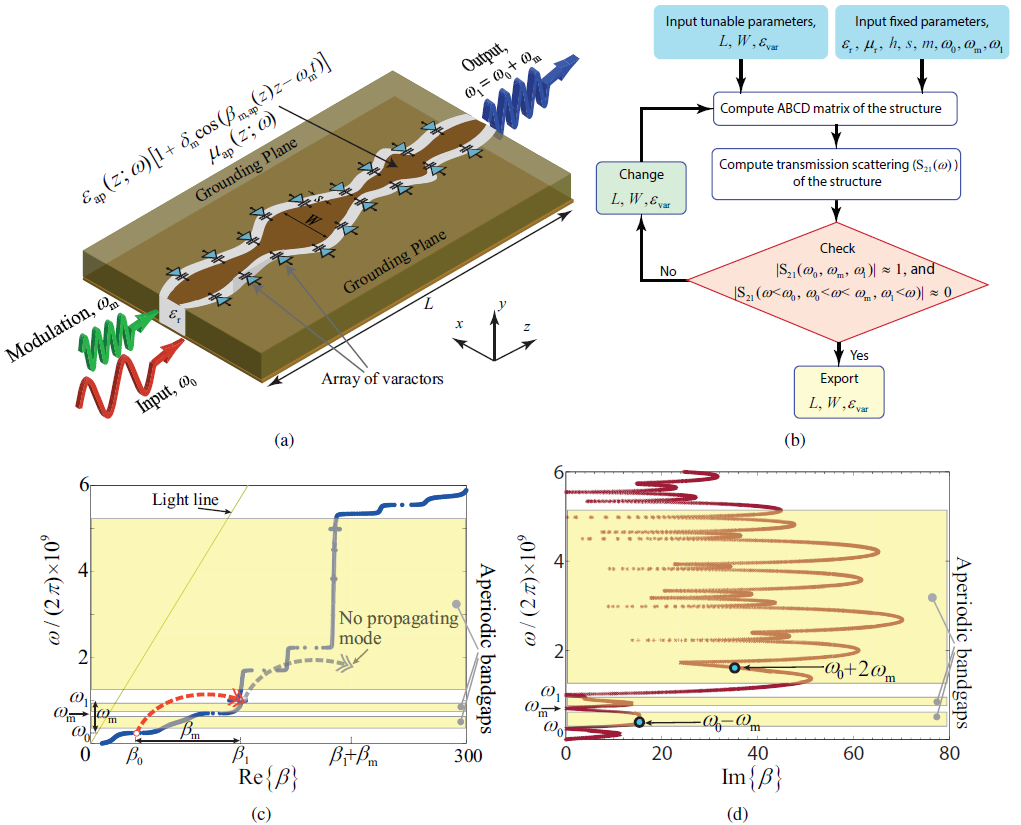}
		\caption{STM pure frequency mixer. (a)~Generic representation of the spatially aperiodic STM mixer yielding pure frequency conversion from $(\beta_0,\omega_0)$ to $(\beta_1,\omega_1)$~\cite{Taravati_PRB_Mixer_2018}. (b) Optimization procedure for mixer synthesis. (c)~and (d)~Real and imaginary parts of the wavenumber, where aperiodic photonic band gaps prohibit harmonic transitions to undesirable ST harmonics~\cite{Taravati_PRB_Mixer_2018}.}
		\label{Fig:mixer}	
	\end{center}
\end{figure*}

Fig.~\ref{Fig:mixer}(a) shows the schematic of an aperiodic STM pure frequency mixer, based on a nonuniform conductor-backed coplanar waveguide (CPW) loaded with an array of subwavelength-spaced varactors. The structure assumes the thickness $L$, a spatially aperiodic temporally periodic permittivity, i.e.,
\begin{subequations}\label{eqa:ap_constit}
		\begin{equation}\label{eqa:perm}
		\epsilon(z,t;\omega)=\epsilon_{0} \epsilon_\text{ap}(z;\omega) \left[1+\delta_\text{m} \cos(\beta_\text{m,ap}(z;\omega) z-\omega_\text{m} t) \right],
		\end{equation}
and a spatially aperiodic permeability
		\begin{equation}\label{eqa:permeab}
		\mu(z;\omega)=\mu_{0} \mu_\text{ap}(z;\omega),
		\end{equation}
\end{subequations}
where $\epsilon_\text{ap}(z;\omega)$ and $\mu_\text{ap}(z;\omega)$ are aperiodic functions of space and frequency, and $\beta_\text{m,ap}(z)$ and $\omega_\text{m}$ respectively denote the spatial frequency and temporal frequency of the pump wave. Such an STM mixer is represented by average spatially variant phase velocity $v_\text{ap}(z;\omega)=c/\sqrt{\epsilon_\text{ap}(z;\omega)\mu_\text{ap}(z;\omega)}$; the pump wave shows an aperiodic space-dependent spatial frequency of $\beta_\text{m,ap}(z;\omega)=\omega_\text{m}/v_\text{ap}(z;\omega)$, where the subscript ``ap" denotes ``aperiodic" spatial variation of these functions. The input wave at $\omega_0$ is injected into the aperiodic STM medium in Fig.~\ref{Fig:mixer}(a), and then, propagates inside the medium along the $+z$ direction and experiences a photonic transition to $\omega_n=\omega_0+n\omega_\text{m}$. However, with a proper spatial-aperiodicity all the undesired ST mixing products can be significantly suppressed by virtue of aperiodic photonic band gaps, leading to a pure transition from $\omega_0$ to $\omega_1=\omega_0+\omega_\text{m}$.

The spatial variation of the constitutive parameters, $\epsilon_\text{ap}(z;\omega)$ and $\mu_\text{ap}(z;\omega)$, is designed through a synthesis method which provides the specified dispersion diagram with proper photonic band gaps, so that $|S_{21}(\omega_0,\omega_\text{m},\omega_1)|=1$ and $|S_{21}(\omega<\omega_0,\omega_0 <\omega< \omega_\text{m},\omega_1<\omega)| \rightarrow 0$. The synthesis procedure, shown in Fig.~\ref{Fig:mixer}(b), is based on optimization of the nonuniform spatial profile of the CPW and the average permittivity of the varactors, $\epsilon_\text{var}$. In Fig.~\ref{Fig:mixer}(b), $W$ represents the width of the middle line, $s$ is the gap between the middle line and the grounding plane and $h$ denotes the height of the substrate. Figs.~\ref{Fig:mixer}(c) and~\ref{Fig:mixer}(d) plot the optimized analytical dispersion diagram, i.e., the real and imaginary parts of the wavenumber $\text{Re}\{\beta(\omega)\}$ and $\text{Im}\{\beta(\omega)\}$, for the unbounded aperiodic dispersion-tailored STM mixer with the optimized permittivity and permeability. It may be seen from these two figures that periodic transitions to undesired STHs are prohibited (Fig.~\ref{Fig:mixer}(c)) by virtue of aperiodic photonic band gaps, where $\partial \text{Re}\{\beta(\omega)\}/\partial \omega \rightarrow 0$ and $\text{Im}\{\beta(\omega)\}>>0$. Thanks to the tailored spatial aperiodicity of the structure, except $\omega_{0}$ and $\omega_1$, all the time harmonics lie in the band gap of the structure and hence will not mature, but rather propagate as weak waves and are attenuated. Figs.~\ref{Fig:mixer}(c) and~\ref{Fig:mixer}(d) show the suppression level at different band gaps for undesired harmonics. It should be noted that, except for the input wave at $\omega_{0}$, the power level of all the space-time harmonics is equal to zero at the input of the structure, i.e., at $z = 0$. As a result, the band gaps are to prohibit growth of the power of the undesired harmonics, rather than attenuation of the strong waves. Consequently, even a weak or moderate attenuation in a band gap may lead to strong suppression of the corresponding space-time harmonic.

\begin{figure*}
	\begin{center}
		\subfigure[]{\label{Fig:photo_cpw}
			\includegraphics[width=1.05\columnwidth]{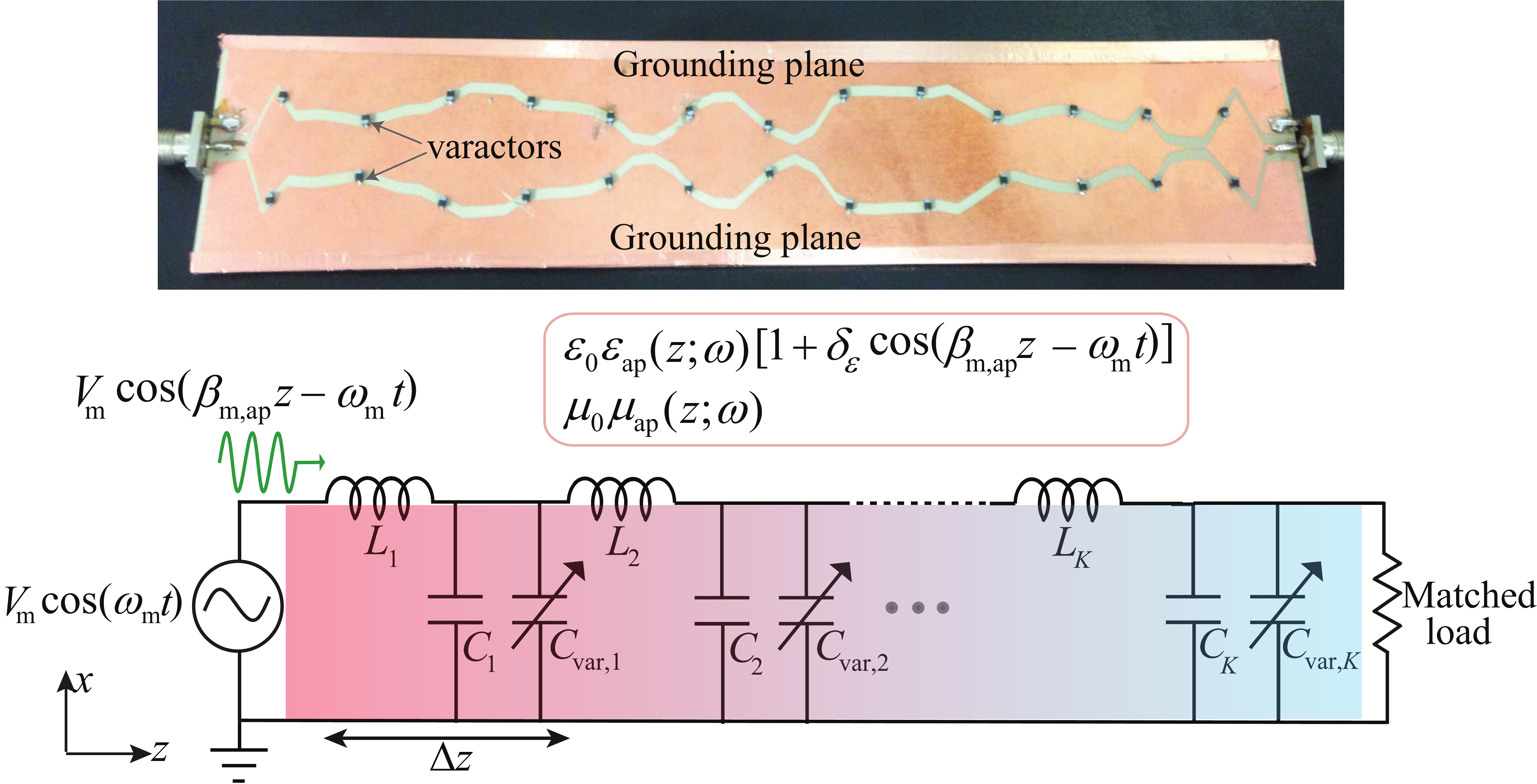}} 
		\subfigure[]{\label{Fig:exp_setup}
			\includegraphics[width=0.9\columnwidth]{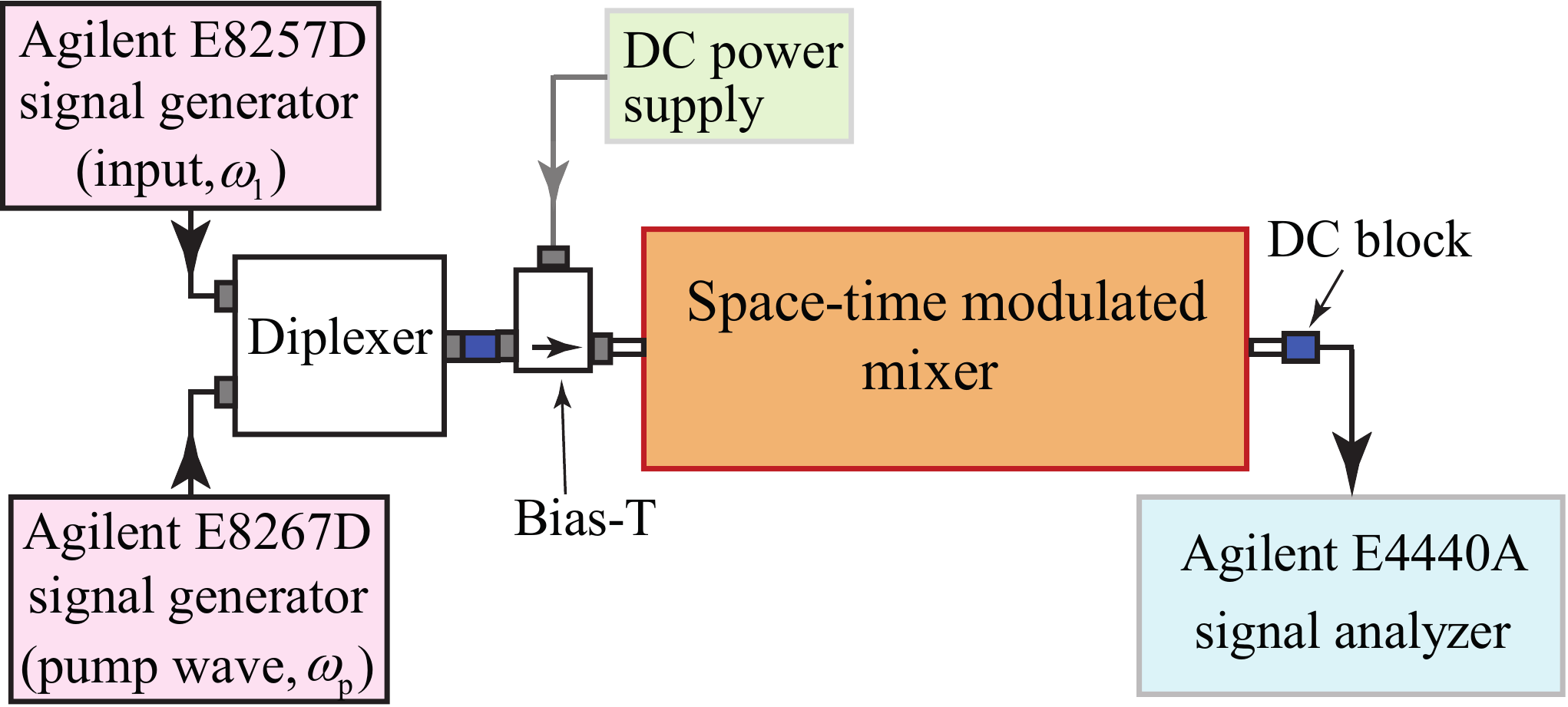}} 		
		\subfigure[]{\label{Fig:exp_res}
			\includegraphics[width=0.9\columnwidth]{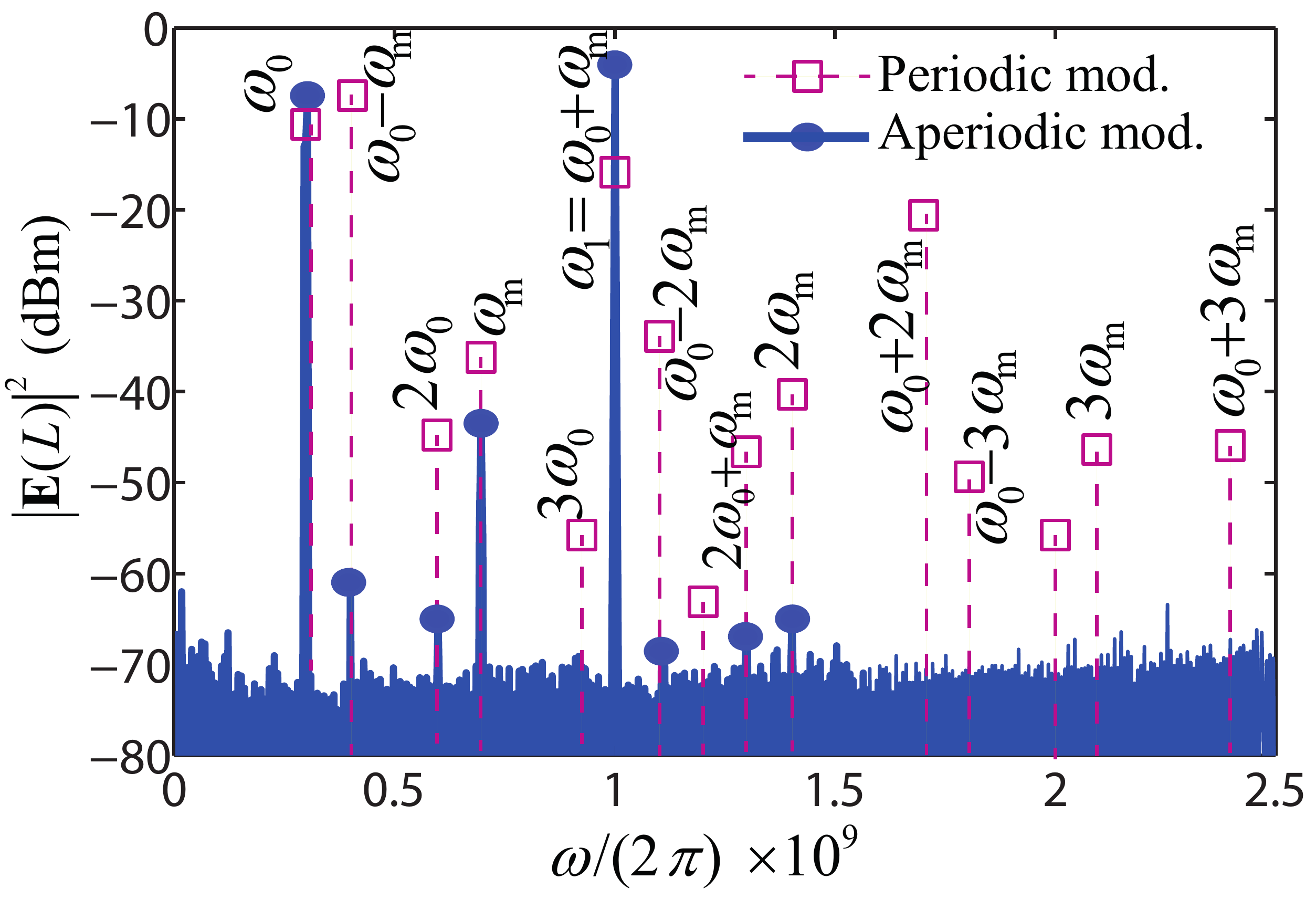} }
		\caption{STM pure frequency mixer. (a)~Experimental implementation using an array of STM varactors distributed on top of an aperiodic spatially variant conductor-backed coplanar waveguide~\cite{Taravati_PRB_Mixer_2018}. (b)~Experimental set-up. (c)~Experimental comparison of the output spectra of the periodic and the dispersion-tailored aperiodic STM structures, wherein the dispersion-tailored aperiodic STM mixer exhibits negligible power levels of the unwanted harmonics~\cite{Taravati_PRB_Mixer_2018}.}
		\label{Fig:disp_exp}	
	\end{center}
\end{figure*}

Fig.~\ref{Fig:photo_cpw} shows the experimental implementation of the aperiodic STM waveguide mixer with constitutive parameters in~\eqref{eqa:ap_constit}, considering the experimental set-up in Fig.~\ref{Fig:exp_setup}. Fig.~\ref{Fig:exp_res} plots the experimental results for the output spectra of the periodic and aperiodic STM frequency mixers. This plot shows that the aperiodic dispersion-tailored STM mixer has significantly suppressed undesired STHs, while providing the required energy and momentum for a pure transition from the input frequency $\omega_0$ to the desirable harmonic at $\omega_0+\omega_\text{m}$. 

\begin{table*}
	\centering
	\caption{Advantages and disadvantages of the STM microwave components presented in Sec.~\ref{sec:App}} 
	\label{tab:table1}
	\begin{tabular}{ |m{2cm}| |m{2.6cm}|m{2cm} |m{2.5cm}|m{2cm}|  }
		\hline
		STM components&  ISOLATORS: \qquad i)~Sec.~\ref{sec:QSI}~\cite{Taravati_PRB_2017};\qquad   ii)~Sec.~\ref{sec:PBI}~\cite{Chamanara_PRB_2017}; \qquad  iii)~Sec.~\ref{sec:SBI}~\cite{Taravati_PRB_SB_2017}& CIRCULATOR: Sec.~\ref{sec:Circ}~\cite{estep2016magnetless}& FRONT-ENDS: i)~Sec.~\ref{sec:WANG}~\cite{Wang_TMTT_2014}; ii)~Sec.~\ref{sec:ALU}~\cite{Alu_PNAS_2016}; iii)~Sec.~\ref{sec:TARAVATI}~\cite{Taravati_TAP_65_2017} & PURE FREQ. MIXER: Sec.~\ref{sec:PFM}~\cite{Taravati_PRB_Mixer_2018}\\
		\hline 	\hline
		%
		%
		Size& i)~Moderate; \qquad ii)~Large;\qquad  \qquad iii)~Small &Small & i)~Moderate,\qquad  ii)~Small,\qquad   iii)~Small & Moderate\\\hline
		Implementation complexities& Variable capacitors (or inductors) are required &Variable capacitors (or inductors) are required & Variable capacitors (or inductors) are required& Variable capacitors (or inductors) are required \\\hline
		Frequency bandwidth&   i)~Fair;\qquad\qquad ii)~Narrow;\qquad \qquad iii)~Ultra-wide&      Narrow& i)~Broadband; \qquad ii)~Narrow; \qquad iii)~Narrow& Narrow\\
		\hline
		Tunable& Yes    &Yes&Yes& Yes \\\hline
		External RF bias required& Yes,\qquad  Yes, \qquad No &Yes&Yes& Yes \\\hline
		Integrable with circuit technology& Yes    &Yes&Yes& Yes 
		\\\hline
		Advantages over conventional techniques&  magnetless, linear, lightweight, (\cite{Taravati_PRB_SB_2017}: Self-biased and broadband) &Magnetless, linear, lightweight &Magnetless, linear and lightweight & No undesired mixing products, power efficient  
		\\\hline	
	\end{tabular}
\end{table*}

\section{Outlook and Conclusion}
Space-time-modulated (STM) structures represent an emerging class of dynamic and directional materials. ST modulation is a promising new paradigm for nonreciprocity and frequency generation, ranging from acoustics and radio frequencies to terahertz and optics, and is compatible with high frequency techniques~\cite{Fan_PRL_109_2012,serranomagnetic}. This paper presented a global perspective on the principles and applications of space-time-modulated (STM) media. It is shown that magnet-free, wide-bandwidth, linear, and highly versatile microwave nonreciprocal systems may be realized by leveraging unique features of an STM medium. ST modulation represents a candidate for the realization of high efficiency components for 5G frequency bands, particularly 3.5~GHz and 26/28~GHz. 
In addition, STM media can be implemented in conjunction with advanced space-time multiple-antenna systems~\cite{mehran2013wireless}, which are promising candidates for enhanced-efficiency 4G, 5G and forthcoming generations of wireless communication systems, and may be adopted with link-level ST modulation schemes~\cite{nikitopoulos2017space} for the realization of versatile communication systems. Table~\ref{tab:table1} summarizes the advantages and implementation complexities of the STM microwave components presented in this paper.

\newpage
\phantom{.}
\newpage
\phantom{.}
\newpage

\bibliographystyle{IEEEtran}
\bibliography{Taravati_Reference}
\end{document}